**Title: Cooperation among an anonymous group protected Bitcoin during failures of decentralization**

**Authors:**
Alyssa Blackburn[1,2,3], Christoph Huber[4], Yossi Eliaz[3,5*], Muhammad S. Shamim[1,2,3,6*], David Weisz[1,2,3], Goutham Seshadri[1], Kevin Kim[1], Shengqi Hang[1], and Erez Lieberman Aiden[1,2,3,7#]

**Affiliations:**
1 The Center for Genome Architecture, Baylor College of Medicine, Houston, TX 77030, USA
2 Department of Molecular and Human Genetics, Baylor College of Medicine, Houston, TX
3 Center for Theoretical Biological Physics, Rice University, Houston, TX 77030, USA
4 Institute for Markets and Strategy, WU Vienna University of Economics and Business, Vienna, Austria
5 Department of Physics, University of Houston, Houston, TX 77204
6 Medical Scientist Training Program, Baylor College of Medicine, Houston, Texas, USA
7 Department of Computer Science, Rice University, Houston, TX 77030, USA
* These authors contributed equally
# Corresponding author
Correspondence to: erez@erez.com (E.L.A.)

**Abstract:**
Bitcoin is a digital currency designed to rely on a decentralized, trustless network of anonymous agents. Using a pseudonymous-address-linking procedure that achieves >99% sensitivity and >99% specificity, we reveal that between launch (1/3/2009), and when the price reached $1 (2/9/2011), most bitcoin was mined by only sixty-four agents. This was due to the rapid emergence of Pareto distributions in bitcoin income, producing such extensive resource centralization that almost all contemporary bitcoin addresses can be connected to these top agents by a chain of six transactions. Centralization created a social dilemma. Attackers could routinely exploit bitcoin via a "51% attack", making it possible for them to repeatedly spend the same bitcoins. Yet doing so would harm the community. Strikingly, we find that potential attackers always chose to cooperate instead. We model this dilemma using an N-player Centipede game in which anonymous players can choose to exploit, and thereby undermine, an appreciating good. Combining theory and economic experiments, we show that, even when individual payoffs are unchanged, cooperation is more frequent when the game is played by an anonymous group. Although bitcoin was designed to rely on a decentralized, trustless network of anonymous agents, its early success rested instead on cooperation among a small group of altruistic founders.

**One Sentence Summary:**
Combining data leakage from the bitcoin blockchain with economic experiments, we study responses to a social dilemma in a group of anonymous individuals.

**Main Text:**
Cooperation, critical for most forms of social organization, entails individual sacrifice so that others can derive a benefit. Anonymity can interfere with the cooperative mechanisms of reciprocity, relatedness and reputation and is thus believed to reduce cooperation in general (*1*).

However, little data exists on economic behavior among anonymous players and groups in a high-stakes, real-world setting.

The recent emergence of bitcoin, which was developed to serve as an anonymous digital currency, provides an example of an economic community with built-in identity protections. (See Fig S23.) Bitcoin comprises a public, tamper-resistant ledger of transactions (the ‘blockchain’) involving a digital asset (‘bitcoin’), and a procedure for updating the ledger and thereby creating new bitcoin (‘mining’) (*2*). Although the value of one bitcoin at launch (Jan 3rd 2009) was zero, it appreciated rapidly, achieving parity with the US dollar twenty five months later (*3*), then \$46,257 exactly ten years after price parity, and \$31,309 as of June 6th, 2022. Ten years post parity with the US Dollar, the value of circulating bitcoin (\$1.03T USD) exceeded all but three national currencies: the American dollar (\$2.12T), the Euro (\$1.74T), and China’s RenMinBi (\$1.32T) [Supplementary Notes].

The blockchain’s contents could provide an ideal resource for examining behavior in an anonymous economic community. But such analyses are difficult in practice, precisely because bitcoin was designed to be anonymous. More precisely, all bitcoin transactions are pseudonymous: individual users can employ an unlimited number of bitcoin addresses (bitcoin’s substitute for user identity) (*2*) (*4*). These identity masking features make bitcoin appealing for illegal activity, which has been estimated to account for 46% of transactions (*5*). For example, Special Counsel Robert Mueller’s report described how Russian operatives used bitcoin when they hacked the Democratic National Committee (*6*).

However, bitcoin’s identity-masking can be compromised by address-linking—determining that multiple addresses are controlled by one agent. Pioneering studies of this type, in the literature (*7-14*) and online (*15-19*) have, for example, estimated that the community included ~75,000 participants when bitcoin achieved dollar parity (*13*).

Yet current address-linking methods fail to detect most links [Supplementary Notes]. As such, a single agent in the community may correspond to thousands of reconstructed clusters, making analysis difficult. An ideal method would reliably partition addresses among agents by detecting overwhelmingly true linkages (high sensitivity) with minimal errors (high specificity). Of course, this would compromise bitcoin’s anonymity, because determining one of an agent’s addresses would reveal the rest.

Here, we use new address-linking strategies to achieve both high specificity (>99%) and high sensitivity (>99%), enabling us to study the bitcoin community in detail between launch (January 3rd 2009) and parity with the US Dollar (February 9th 2011). The end of this period was also punctuated by the launch of the Silk Road, an online, bitcoin-based, black market. This interval captures bitcoin’s transition from a digital object with no value to a functional monetary system.

Using these address-linking techniques, we show that the bitcoin blockchain makes it possible to explore the socioeconomic behavior of bitcoin’s participants. We find that, in line with the findings of Vilfredo Pareto (*20*) in 1896 (and subsequent studies of many national economies), wealth, income, and resources in the bitcoin community were highly centralized. This threatened bitcoin’s security, which relies on decentralization, routinely enabling agents to perform a 51% attack that would allow double-spending of the same bitcoins. The result was a social dilemma for bitcoin's participants: whether to benefit unilaterally from attacks on the currency, or to act in the interest of the collective. Strikingly, participants declined to perform a 51% attack in every case, instead choosing to cooperate.

We model this social dilemma via an N-player generalization of the classic centipede game, in which a group of anonymous players have a one-time opportunity to exploit an appreciating good for short-term benefit. Though immediate exploitation is the sole subgame-perfect Nash equilibrium [Methods], we find that experimental subjects learn to avoid this strategy, and instead tend to cooperate the vast majority of the time – even when they cannot possibly benefit. Surprisingly, cooperation is significantly higher in an N-player centipede game than a classic 2-player centipede game with identical payoffs, suggesting that players in an anonymous group want to avoid spoiling the collective outcome.

Our findings highlight how cooperation can arise among anonymous, self-interested agents, giving rise to new types of economic and social organization.

**We construct an accurate catalog of agents in the early bitcoin community by identifying pseudonymous addresses controlled by the same agent.**

We began by defining a graph whose nodes are all bitcoin addresses. We then added links between the nodes using four bitcoin address-linking techniques. Two of these techniques exploit how the bitcoin mining software generated apparently-meaningless strings, which were used as part of bitcoin's cryptographic protections against forgery. In fact, there are extensive correlations between the apparently-meaningless strings associated with a single user. The other two techniques exploit insecure user behaviors, such as the use of multiple addresses to pay for a single transaction, that make it possible to link addresses based on transaction activity. Finally, we enumerated the connected components of the graph, which correspond to the individual agents. The address-linking procedure is described in detail in the Methods section.

The resulting catalog of agents is shown in Figure 1. See also Figures S8, S9, and S10.

We compared this catalog to ground truth data generated by scraping bitcoin discussion forums and looking for posts indicating the owner of an address. There were 2,982 cases in which two addresses in our catalog were tagged with the same agent in the ground truth data; 2,978 (99.87%) were assigned to the same agent in our catalog. There were also 13,308 cases in which two addresses in our agent list corresponded to two different agents in the ground truth data; 13,221 (99.35%) were assigned to different agents in our catalog. These findings confirm the accuracy of our agent catalog [Methods].

As expected, the largest agent (dubbed "Agent #1," ₿1,108,550; Net Present Value: $34.7B) was associated with Satoshi Nakamoto, pseudonym for bitcoin's creator. The next five largest agents mined ₿188,150 (Agent #2, PV: $5,890,788,350), ₿120,850 (Agent #3, PV: $3,783,692,650), ₿73,200 (Agent #4, PV: $2,291,818,800), ₿70,000 (Agent #5, PV: $2,191,630,000), and ₿64,850 (Agent #6, PV: $2,030,388,650). See Fig 1.

These analyses could potentially facilitate deanonymization. For example, several of the agents were imprisoned for crimes involving bitcoin. One is closely associated with Ross Ulbricht (Agent #67), also known as “DreadPirateRoberts”. Ulbricht created the Silk Road, an online black market. Another is Michael Mancil Brown (Agent #19, AKA “Dr. Evil”). Brown claimed to possess tax returns from then-Presidential candidate Mitt Romney, offering them in exchange for a ransom of $1,000,000 in bitcoin. Had both simply held on to their bitcoin, that amount would today be worth $169,068,600 and $901,699,200, respectively. Since our goal was to explore socioeconomic behavior in the early bitcoin community, rather than to deanonymize agents, we do not report any other identities.

**Sixty-four agents mined most of the bitcoin prior to parity with the US dollar.**

Between launch and dollar parity, most of the bitcoin was mined by only 64 agents, collectively accounting for ₿2,676,800 (PV: $84 billion). This is 1000-fold smaller than prior estimates of the size of the early Bitcoin community (75,000) (*13*). In total, our list included 210 agents with a substantial economic interest in bitcoin during this period (defined as agents that mined bitcoin worth >$2,000 at the time.) It is striking that the early bitcoin community created a functional medium of exchange despite having very few core participants.

**Within months of launch, bitcoin mining income fits a Pareto distribution.**

Since the work of Vilfredo Pareto (*20*), the distribution of top incomes in a population—specifically, the fraction of individuals whose income exceeds a certain threshold—has been modeled by a power law (*21, 22*) (*23, 24*). Pareto distributions among top earners have been observed consistently around the world (*24-29*), reflecting the fact that the centralization of wealth and resources is ubiquitous in economic communities.

We plotted the distribution of bitcoin mining income on log-log axes during six time intervals between the launch of bitcoin (January, 3, 2009) and when it achieved parity with the US dollar (February, 9, 2011). A power law was visible in each interval. The presence of this distribution in Interval 1 implies the emergence of Pareto distributions within four months of bitcoin’s launch. (See Fig 2.) This underlines the degree of centralization of the bitcoin blockchain, and highlights that Pareto distributions can emerge extremely rapidly.

**Changes in the Pareto exponents reflect the dynamics of the emerging bitcoin community.**

The Pareto exponent—which manifests as the slope of the income distribution on a log-log plot—changed over time, reflecting changes in the level of income inequality.

During the first four intervals, the Pareto exponents (Interval 1: 0.47, Interval 2: 0.53, Interval 3: 0.64, and Interval 4, 0.95) were much lower than those typically seen in studies of national income distribution (US 2003: 1.22 (*25*), UK 1996: 1.9 (*26*), Forbes Top 400 1996: 1.36 (*28*), Japan 1998: 2 (*27*)). This indicates vastly more inequality than a typical economy. In fact, nearly all the computational power of the network, and nearly all the mining income, was associated with Satoshi Nakamoto.

During the two middle intervals, other agents joined the community and replaced Satoshi Nakamoto at the top of the income distribution. Nevertheless, nearly all wealth was held by a tiny group: 70% of bitcoin income during the first four intervals was generated by thirty-three agents.

During the last two intervals, the Pareto exponents (Interval 5: 1.26, Interval 6: 1.19) increase markedly, falling in line with the values that are typically seen for national economies. This is consistent with the entry of thousands of small stakeholders to the bitcoin community, most holding tiny amounts of bitcoin as compared to their predecessors. We refer to this group as the "early adopters," and to those who joined earlier as "founders." Accounting for 70% of all bitcoin generated during all 6 intervals requires over a thousand agents. The increasing number of small stakeholders correlates with bitcoin's appreciation from <1¢ to $1. See Fig 2, S16.

**Bitcoin's founders were collaborating on the technical challenge of deploying bitcoin, whereas the early adopters were more focused on using it for transactions.**

Founders were more likely to socialize with one another than early adopters. Each one was more likely to have posted on a bitcoin-related forum (10% vs 1%, p=0.01), and they collectively posted 18 times more often. See Fig S13.

We used topic modeling to discover abstract "topics" reflected in posts to bitcointalk.org, a forum launched during period 3 that became the principal venue for discussing bitcoin during the target period.

Posts by the founders were more likely (21.9% vs 8.9%) to relate to two topics: compiling executable files (associated with terms like "compile", "makefile", and "build") and connecting to the peer-to-peer network ("connections," "port," and "ip"). These topics relate to the core functions of the bitcoin software: the creation of the bitcoin client and its use in connecting to the peer-to-peer network.

By contrast, the posts of early adopters were more likely (23.9% vs 13.7%) to reflect economic activity ("auction", "paypal", and "sell"). In intervals 5 and 6, languages other than English also begin to appear, reflecting the community's growing international presence [Methods]. See Figure 3.

It is interesting to compare the above dynamics to the timeline of bitcoin's adoption.

During the first four intervals, the use of bitcoin to mediate meaningful economic transactions (vs. to test the technology) was minimal. The first known purchase using bitcoin occurred on May 27, 2010, immediately prior to the beginning of Interval 5. In this well-documented transaction, Laszlo Hanyecz paid ₿10,000 for two large pizzas (*30*). By the end of Interval 6, bitcoin's value had achieved parity with the US dollar (as noted above), and the Silk Road online marketplace was about to launch.

Thus, the arrival of thousands of early adopters during Intervals 5 and 6, and the shift in focus from implementation to use, were simultaneous with the appreciation of bitcoin and its maturation into a functional medium of exchange.

**Nearly all bitcoin addresses fall within six degrees of separation of the top 64 agents.**

To explore the role of the top 64 agents in the broader bitcoin community, we constructed a contemporary bitcoin transaction network. Nodes are bitcoin addresses. Edges reflect all transactions between launch and December 2017: 346 million addresses and 283 million transactions.

Most addresses present in the bitcoin blockchain (91.9%) are separated from one of the 64 top agents by a chain of four or fewer transactions. Nearly all (99.3%) are separated by six or fewer. This is in line with classic findings on other human social networks (*31-39*). See Figure 4.

The structure of the bitcoin transaction network may have consequences for bitcoin's anonymity today. For instance, should a law enforcement agency determine the identity of the 64 top agents, they could de-anonymize nearly any target bitcoin address by tracing at most 6 transactions [Methods].

**Bitcoin was designed to rely on a decentralized, trustless network of anonymous actors, but its early success rested on cooperation among a small group of altruistic agents.**

Bitcoin's design relies on a large, decentralized network of participants to prevent fraud. If a single agent or group of colluding agents controls most of the computational resources used to mine bitcoin (i.e., to validate transactions), this scheme fails, and the attacker(s) can validate illegitimate transactions. (For instance, attackers could repeatedly spend the same bitcoin.) This is called a 51% attack. Such attacks have been observed in many cryptocurrencies (*40*).

Estimates of the overall number of participants in the bitcoin community – 75,000 (*13*) or 64 (above) – exaggerate the degree of decentralization at any given point in time. This is partly because some agents have far more computational resources than others. In addition, a smaller agent's resources may be concentrated during a short period, allowing them to dominate the network during that period.

For instance, between bitcoin's launch and December 1st, 2009, eleven months later, Satoshi Nakamoto frequently controlled the majority of the computational resources and could perform unilateral 51% attacks (See Figure 5A). This is fully in line with what is already known (*15*).

Surprisingly, we found that other agents could also perform 51% attacks.

For example, three other agents exhibited streaks in which they mined at least six blocks in a row, suggesting that they controlled nearly all of the network's computational resources during the streak. Since the typical time between blocks is ~10 minutes, this implies that these agents could reliably perform a 51% attack over a period of an hour or more. The first of these streaks, by Agent #3, took place on August 18th, 2009, during an interval when Satoshi Nakamoto's hardware appears to have been offline [Supplementary Notes]. The most protracted vulnerability was in early October 2010, when Agent #2 could have performed a 51% attack during five six-hour periods (Figure 5B). Agent #2 was among the first users to accelerate the bitcoin-transaction-validation process (mining) by employing general-purpose graphical processing units [Supplementary Notes].

Of course, the ability to mine every single block in a time interval is not a prerequisite for a majority attack. In fact, we noted that approximately 4.6% of blocks (3903) were contained in intervals of six blocks or more in which a single agent (other than Satoshi Nakamoto) mined most of the blocks, making the blockchain susceptible to majority attack on those occasions. The first such streak was on July 28th, 2009. The last took place on December 25th, 2010. Twenty-eight different agents were associated with such streaks (including Satoshi Nakamoto makes the number twenty-nine). See Figure S15.

Bitcoin was also susceptible to bilateral collusion. There were ninety-nine streaks in which two agents (excluding Satoshi Nakamoto) mined at least six blocks in a row, implying that, were they to collude, they could very reliably perform a 51% attack for an hour or more. The first such streak was on July 25th, 2009. The last took place on January 23rd, 2011, mere weeks before bitcoin achieved parity with the US dollar. Over 100 agents participated in bilateral streaks.

More generally, we found that many longer time intervals (days or weeks) in which no agent had majority control contained shorter time intervals (hours) in which a single agent did have majority control. See Fig S14.

As a measure of bitcoin's decentralization, we compared the empirical frequency of streaks (excluding Satoshi Nakamoto) with theoretical predictions for an idealized network comprising $P$ miners with identical computational resources. This yielded an "effective population size," $P$, for bitcoin's decentralized network: 4.9, or roughly five agents [Methods]. The results were similar whether we considered streaks where a single agent mined all of the blocks, streaks where a single agent mined most of the blocks, or streaks where two agents mined all of the blocks. The network was less prone to streaks than if two agents continually flipped a fair coin for the ability to mine the next block, but less secure than if six agents rolled a fair die (Figure 5C).

Taken together, these data demonstrate that bitcoin, during the period of interest, was often vulnerable to 51% attacks.

This created a social dilemma for bitcoin's participants. As bitcoin's price neared $1, a 51% attack could become increasingly lucrative for the attacker. Yet such attacks would rapidly undermine Bitcoin's credibility. Strikingly, we find no evidence that Satoshi Nakamoto, Agent #2, Agent #3, or any other agent or colluding group, performed a 51% attack on bitcoin.

Rather than relying exclusively on a decentralized, trustless network of anonymous actors, bitcoin depended on altruistic behavior by a group of anonymous agents.

**We model the social dilemma created by bitcoin's intermittent centralization via an N-player Centipede game where a group of anonymous players has the opportunity to exploit an appreciating good.**

To better understand the altruistic behavior described above, we modeled the dilemma faced by bitcoin's participants as an N-player Centipede game. In this game, a generalization of the classic 2-player Centipede game (*41*), a group of players take turns controlling a commodity, such as a cryptocurrency, that is appreciating over time. In each round, a player (corresponding to a potential 51% attacker) is given two options. One option is to cooperate (i.e., "PASS" on the opportunity to attack). If so, the round ends, the commodity's value is multiplied by $d$, and a new player is selected. The other option is defection (i.e., "TAKE" the opportunity to perform a 51% attack, damaging bitcoin's credibility.) This choice ends the game. Everyone who has played so far must sell off their commodity, gaining a payoff $\pi = d^{n-1}$, where $n$ is the round when defection took place. The defector benefits by gaining $b*\pi = b*d^{n-1}$, where $b > d$. (Those who have not played get nothing.) The game also ends after a pre-specified number of rounds, $N$, even if no one has defected. If so, each of the players gets $\pi = d^N$. This game is depicted in Fig 6.

N-player Centipede is identical to 2-player Centipede, except insofar as the former adds a new, anonymous player in each new round, whereas the latter has two players who alternate rounds, enabling direct reciprocity (*42*). Both have a single subgame-perfect Nash equilibrium, which is to defect immediately. This is because the player in the last round is better off defecting (thereby gaining a payoff of $b*d^{n-1}$, rather than cooperating, and gaining $d^n$.) Knowing this, the second-to-last player is also better off defecting (gaining $b*d^{N-2}$ vs. $d^{N-1}$), and so on. Thus, backwards induction renders cooperation impossible among fully rational players.

Of course, it is impossible to fully incorporate the beliefs and possible strategies of an anonymous group, such as the bitcoin community, in a finitely-specified game. For instance, the above model assumes agents believed that their actions might affect bitcoin's success, just as 51% attacks have damaged other cryptocurrencies (*43*). If, by contrast, all agents were certain that bitcoin would become a stable global currency regardless of their choices, they would be more likely to cooperate, providing an alternative explanation for our observations.

**Experimental subjects are less likely to spoil the outcome for an anonymous group than for a single counterpart.**

Despite the results of backwards induction, numerous studies have shown significant levels of both cooperation and defection in 2-player Centipede (*42*). We wondered whether playing in a group – N-player Centipede – would encourage further cooperation and help explain the high levels of altruism seen in the early bitcoin community.

We therefore recruited 832 volunteers to play 8-player Centipede using the online platform Prolific (*44*). Volunteers were divided into groups of eight. (Since the game can end before the $8^{th}$ round, some volunteers did not end up playing.) After each game, players were assigned a new group, for a total of five games per volunteer. Finally, volunteers completed a short exit survey. A volunteer's identity was unknown to the experimentalists and other players. Players also had no way to know who their current group had played with, or what they had done, in earlier games. See Fig 6.

We found that the subgame-perfect Nash equilibrium is almost never played: individuals chose to defect on the first round only 8.7% of the time. Moreover, players cooperated more with each repetition, "learning" to avoid the subgame-perfect Nash equilibrium.

Instead, the average game involved 5-6 rounds of cooperation, followed by a defection, for a total of 7-8 rounds. Strikingly, few players chose to defect in the last round (34.8%), although it was too late for others to reciprocate (*45*).

Quadrupling the payoffs in the experiment did not change the outcome [Supplementary Notes]. The results were similar regardless of a volunteer's income, nationality, or whether they invested in cryptocurrency [Methods].

Despite the fact that the 2-player Centipede enables direct reciprocity – a powerful mechanism for encouraging cooperation (*46*) – the level of defection was consistently lower in the N-player variant. There was much less defection in the first round (2-player: 16.7%; N-player: 8.7%), highlighting that players were more optimistic about cooperation with a group of one-shot, anonymous partners then when taking turns with a single partner. There was also a large difference in the penultimate round (2-player: 35%; N-player: 17.3%). This suggests that players expect less defection in the final round from a random, anonymous player, than from a counterpart who had already cooperated repeatedly (in rounds 2, 4, and 6 of 2-player Centipede). We observed no evidence of learning to cooperate in the 2-player Centipede game.

Taken together, these findings show that playing in an anonymous group encourages cooperation in Centipede: even when their personal payoff does not change, players avoid spoiling the outcome for the group.

**Theoretical analyses predict that larger groups increase the level of cooperation.**

Evolution is commonly invoked to explain altruistic behavior in Centipede and other settings (*47*). To explore the effects of playing Centipede in an anonymous group, we modeled a population of $M$ individuals playing $N$-player Centipede. We considered strategic types $s \in \{0,1,2,3,4,5,6,7,8\}$, with s indicating the last round in which a player will cooperate. Thus the strategy $s=0$ never cooperates, and $s=N$ always cooperates. An individual's payoff $\pi$ is the average of their payoffs in every possible context, e.g., ordered samplings of $N$ players with replacement. Fitness is given by $f=1-w+w\pi$, where $w$ is the strength of natural selection (*48*). Strategies with higher fitness tend to spread. This can reflect either imitation, or a birth-death process in which new individuals are born in a fitness-dependent fashion, replacing random individuals. Occasionally, an individual's type spontaneously mutates to a randomly-selected type. We consider the low mutation limit, whereby a mutant will either fixate or go extinct before another arises. The asymptotic frequency of each strategy can be calculated using a semi-analytic approach (*49*) (*47*).

We began by examining the 8-player Centipede with $b=4$ and $M$=25, varying $w$ and the rate of appreciation $d$. In Fig 6, we plot the strategy with the highest asymptotic frequency for each parameter value. The subgame-perfect Nash equilibrium, $s=0$, is most common only when $d$ is small and selection is very strong. As selection weakens, $s=0$ is replaced by increasingly cooperative strategies. As $d$ increases, the most successful strategies cooperated nearly all the time. These results are robust to variations in $b$ [Methods]. Intuitively, a highly cooperative mutant does worse when others are less cooperative, reducing the chances of successful invasion. But when selection is weak, random drift increases the chance that an invading cooperator will grow to critical mass, enabling it to compete. Increasing *d* reduces the benefit of defection, favoring cooperation.

We then explored the results of varying the group size, *N*. Increasing the group size encourages cooperation, regardless of the choices we made for the other game parameters.

Taken together, anonymous agents entrusted with a rapidly appreciating commodity – whether in the setting of the bitcoin blockchain, a controlled behavioral experiment, or a theoretical analysis – act in ways that are inconsistent with a model based on rational choice. Instead, high levels of cooperation are observed. This effect is enhanced when play involves an anonymous group, which encourages cooperation even more effectively than classic mechanisms such as direct reciprocity. Crucially, these findings are not the result of changes in individual payoff. Instead, they are in line with a model where, even if they derive no personal benefit, players choose to avoid spoiling the outcome for a group.

## Discussion

Since bitcoin was originally designed to be an "anonymous digital currency" (Fig S23), the bitcoin blockchain, a public ledger recording all transactions involving bitcoin from the moment of its launch, has the potential to clarify the economic behavior of an anonymous group in a real-world setting. Here, we use address-linking to associate bitcoin addresses—pseudonymous, random strings—that were controlled by the same agent. We thus catalog many of the agents involved in the early bitcoin community between its launch and when it achieved parity with the US dollar.

Bitcoin's design relies on a decentralized network of agents to keep the currency secure. We observe that Pareto (power-law) distributions in bitcoin mining income, similar to those seen for national economies, emerge within months of bitcoin's launch. This shows that such distributions can form with extraordinary speed in a new economic community, and are associated with disproportionate control of bitcoin's computational network by a small number of agents.

For national economies, it is difficult to study how Pareto distributions emerge and change, because the relevant historical records are limited (*50*). Our catalog of agents allows us to overcome this limitation. We find that changes in the Pareto distribution reflect the emergence of a large class of small bitcoin stakeholders. Unlike bitcoin's founders, these early adopters were much less concerned with the technical challenge of making bitcoin functional, and much more concerned with using bitcoin for actual transactions. In line with this finding, the arrival of these small stakeholders correlates closely with a change in bitcoin's price, from <1¢ to $1. This two-stage process is in line with similar two-stage models that have been proposed in studies of innovation diffusion (*51, 52*), wherein a small group of early influencers learn about an innovation, and then bring about widespread adoption of a new technology by informing their social networks (*53*).

We find that only 64 agents controlled most of the bitcoin mined between when bitcoin launched and when it achieved parity with the US dollar. This is far fewer than estimates from prior studies (75,000) of the number of active agents during this period (*13*). The collective income of these agents accounts for 12.7% of all bitcoin that will ever be created. Together, these agents created a tipping point after which bitcoin became a functional means of exchange for economic transactions worldwide. Interestingly, this contrasts with other social processes, where the tipping point was a significant fraction of the total population (*54*), ranging from 10% in behavioral experiments of agents located at the nodes of a network (*55*), to 25% in pairwise interactions over the web (*56*), to 40% in studies of female influence in the New Zealand Parliament (*57*). Other studies have reported a high level of centralization in the contemporary bitcoin community. Even at present, a large fraction of the computational resources are controlled by a handful of mining collaboratives [Supplementary Notes].

The highly centralized structure of the early bitcoin community has consequences for the security of bitcoin users today. We show that our top 64 agents are extremely central to the contemporary bitcoin transaction network, such that nearly all addresses (>99%) can be linked to a top agent via a chain of less than 6 transactions (*31-33*). These network properties have unintended privacy consequences, because they make the network much more vulnerable to deanonymization using a "follow-the-money" approach. In this approach, the identity of a target bitcoin address can be ascertained by identifying a short transaction path linking it to an address whose identity is known, and then using off-chain data sources (ranging from public data to subpoenas) to walk along the path, determining who-paid-whom to de-identify addresses until the target address is identified.

A key limitation of the follow-the-money approach is the need to identify a known agent who is connected to the target address via a short path. Our results imply that, were the identities of the 64 top agents to become known, it would become easy to identify short transaction paths linking any target address to an already de-identified top agent address. This could adversely affect the privacy of bitcoin transactions. Similar vulnerabilities were identified in a recent preprint studying the Ethereum transaction network (2 million addresses), suggesting that many cryptocurrencies may be susceptible to follow-the-money attacks (*58-60*).

The degree of centralization in the early bitcoin community also made bitcoin vulnerable to 51% attacks, enabling the attacker or attackers to, for instance, fraudulently spend the same bitcoin multiple times. Numerous individuals had the opportunity to perform a 51% attack, but we observe no evidence that such an attack was ever carried out. Why would anonymous agents, possessing enough computational power to attack bitcoin on multiple occasions, choose to act altruistically again-and-again? And how can this be reconciled with that fact that individuals who believe they are acting anonymously are less likely to be altruistic (*1, 61*)?

We model this rational dilemma as an N-player generalization of the classic 2-player Centipede game. In N-player Centipede, a series of anonymous agents make a one-shot decision: to exploit an appreciating resource, or to preserve it for others. We find that experimental subjects cooperate overwhelmingly in this scenario, even though immediate defection is the sole subgame-perfect Nash equilibrium. In fact, players learn to be even more cooperative when they repeat the game.

It is interesting to contrast the N-player and 2-player Centipede games. Both games have identical payoffs and identical choices in each round. They differ only insofar as the latter involves two players taking turns, whereas the former entails a new player making the choice in each round. Direct reciprocity, which the 2-player Centipede game facilitates, is thought to greatly enhance cooperation (*46*). Yet we find that players are more cooperative when playing with a group of anonymous, one-shot players, and that increasing the size of the group amplifies this effect. We hypothesize that players seek to avoid spoiling the outcome for a large group, even when it would have no effect on their personal outcome.

Taken together, our findings show that extraordinary levels of cooperation are possible among groups of anonymous agents, even in high-stakes, real-world scenarios. Of course, we cannot rule out the possibility that other factors also contributed to the choices made by the bitcoin community.

In recent years, there has been a proliferation of novel monetary systems and distributed trust mechanisms, including some that are successful (Ethereum), unsuccessful (Coinye West), in pilot stages (China's digital Yuan) or as of-yet-untested (quantum money; the US digital dollar) (*62, 63*),(*53*). Our findings help clarify how cooperation among groups of anonymous agents can play a role in such mechanisms. These results also raise the possibility that distributed trust mechanisms may depend on the continued appreciation of the associated digital objects, just as the cooperation observed in Centipede games depends on the expectation that payoffs will increase over time. Can participants be relied upon to cooperate if a cryptocurrency stops appreciating?

Our findings also highlight that the detailed design of cryptocurrencies and distributed trust mechanisms has consequences for the security and privacy of their users. It is well-known that systems that are cryptographically secure one day can become completely insecure over time, as analytical methods and computational hardware improve. For an encrypted public database, such as the bitcoin blockchain, these challenges are particularly acute. This is because the sources of information leakage, even once discovered, cannot be retroactively patched. At the time of the first bitcoin transactions, the Bitcoin protocol did not give rise to any obvious information leakage; bitcoin's early users appear to have been either unaware of these leaks or indifferent to them. Despite this, ten years later, we find that it is possible to characterize the population in some detail by aggregating multiple forms of information leakage. Drip-by-drip, information leakage erodes the once-impenetrable blocks, carving out a new landscape of socioeconomic data.

**Acknowledgements:**
We would like to thank Max De Marzi for all of his help and advice with running and working with large graphs and Neo4j. We would also like to mention Jay Brown and John Noplevosky for their help with Neo4j. We'd also like to acknowledge Dimos Gkountaroulis for assistance with the streak equations.

**Competing interests:**
A.B., C.H., Y.E., M.S.S., D.W., and G.S. all hold some amount of cryptocurrency, including, but not limited to, bitcoin. The other authors declare no competing interests.

**Funding:**
This work was supported with funds from an Alfred P. Sloan Foundation grant (2012-3-14TR3), a National Endowment for the Humanities grant (HK-50176-14), and two National Science Foundation Physics Frontiers Center Awards (NSF PHY-2019745 and NSF PHY-1427654).

Conceptualization: AB, GS, ELA
Data curation: AB
Formal analysis: AB, ELA, CH
Methodology: AB, ELA, CH
Investigation: AB, MSS, YE, KK, ELA, DW, CH
Visualization: AB, ELA
Validation: AB, ELA
Funding acquisition: ELA
Resources: ELA
Software: AB, YE
Supervision: ELA
Writing- original draft: AB, ELA
Writing- review & editing: AB, ELA

## Supplementary Notes

### Additional Background

### i. What is Bitcoin?

Bitcoin is a digital cryptocurrency with cryptographic protections against fraud and forgery. Bitcoin can be thought of as a form of digital money. It is based on software designed around mathematical concepts that make it very difficult to forge (counterfeit bitcoin) or commit fraud (spend the same bitcoin more than once). The bitcoin "blockchain" is a permanent, distributed ledger of all bitcoin ever transacted.

A bitcoin transaction is created anytime anybody makes a sale or purchase using bitcoin. A bitcoin transaction can be thought of as two lists of paired values. The first list is of input pairs: an input address that is sending bitcoin and a value for how much bitcoin to send. The second list is of output pairs: an output address that is receiving bitcoin and a value for how much bitcoin to receive. See a toy transaction in Fig S2.

### ii. Mining

A bitcoin "block" can be thought of as a group of transactions bundled together. When an agent is "mining" bitcoin, they take a bundle of these transactions along with a timestamp, a reference to the previous block, an arbitrary integer called a "nonce", and another arbitrary integer called an "extranonce" and feed these as inputs to a hash function in order to get an output lower than a set target number. This target number is akin to handing someone a six-sided die and saying that any number below integer n, where $0<n<7$, will win. If n is 6, the person will win with every roll. If n is 1, they will win, on the average, every six rolls. This target number (called the "difficulty") is recalculated every 2016 blocks (roughly every 2 weeks) to yield one block approximately every ten minutes.

Cryptographic hash functions are non-invertible by design, and their output changes dramatically with tiny changes to their inputs. This means that the output cannot be predicted from the inputs before hashing, so no combination of inputs is more or less likely to yield a winning output. Because miners cannot predict which inputs will yield a winning output, miners that try the greatest number of unique inputs have the best chance of winning. Mining agents with more computational resources will be awarded proportionally more bitcoin. Miners find new input combinations by trivially adjusting their inputs, usually by keeping their timestamp current and incrementing the nonce field after a failed attempt.

When an agent achieves a winning output and "mines" a block, the block gets linked to all the previous blocks mined by virtue of the reference to the previous block that was included as an input to the hash function. The resulting linked chain of blocks is known as the bitcoin blockchain.

In addition, a special transaction paying a mining reward to the miner is added to the bundle of transactions. This special transaction is known as a coinbase transaction, and does not have a list of input pairs, only a list of output pairs reflecting the mining reward. Every 210,000 blocks (roughly every 4 years), the block reward is reduced by a factor of two. For the interval we studied in this paper, the block reward was 50 bitcoin per block.

Today, many individuals in the bitcoin community may own and transact bitcoin without downloading the entire bitcoin blockchain or attempting to mine blocks. However, in the early bitcoin community, the behavior of the default client was to mine bitcoin. In this time period, we are therefore able to use the mining community as a proxy for the total bitcoin community. Our time period, from launch to parity with the US dollar, spans 107,000 blocks. Of these, we only used 106,475 addresses, discarding all coinbase transactions with more than 1 recipient address to avoid the confounding effects of mining pools (situations where individuals form a mining collective and pledge to award all payouts proportionally).

*Note: There is no system-wide time maintained by the blockchain. All block timestamps are self-reported by miners. As such, block timestamps can be noisy.

### iii. Data Leakage and Address Linking: An Overview

Bitcoin's identity masking can be compromised by address-linking—that is, determining that one agent controls multiple addresses. See Figure S1.

Our study made use of two forms of address-linking that are very well-documented in the published literature. The first approach, introduced in Satoshi's whitepaper, notes that all inputs to a transaction are typically controlled by one agent. A second approach exploits the fact that the Bitcoin client typically generated a new address for the sender containing change from a transaction.

Our study also makes frequent use of four other forms of data leakage from bitcoin.

The first form of data leakage exploits how the early bitcoin software generated the extranonce string, one of the two nonsense metadata strings. Specifically, the extranonce string is set to 0 when an agent turns on the mining software, and increases by 1 whenever a block is added to the ledger. Because new blocks are added regularly, the extranonce value associated with a block can be used as a proxy for the length of time that the software had been running on a particular machine when the block was mined. This approximate 'session time' enables us to distinguish between the many machines simultaneously mining bitcoin, frequently making it possible to link all addresses that were generated on a particular computer during a single session of the mining software.

The second form of data leakage exploits the fact that different versions of the early bitcoin software generated the nonce string (the other nonsense metadata string) differently. In particular, certain nonce values were much more common when using one version of the software than when using another. This makes it possible to link addresses associated with similar nonce distributions.

The third form of data leakage exploits an insecure type of transaction that is commonly observed among members of the early bitcoin community. Successful bitcoin mining typically results in the assignment of bitcoin to a large number of addresses controlled by a single agent. In principle, the agent could spend bitcoin assigned to each address separately. In practice, we observe that many agents tend to consolidate all of their bitcoin into one or a handful of addresses, possibly because this precludes the need for keeping track of numerous passwords. (Indeed, we performed an economic experiment in which we generated a fork of bitcoin that was used for transactions among members of a small group of people. Consolidation emerged spontaneously. See the discussion of CO2coin, below.) This behavior can be used to link the addresses involved.

The fourth form of data leakage exploits the fact that agents occasionally perform transactions in highly idiosyncratic ways; for instance, by repeatedly transferring very specific quantities of bitcoin. Performed only once, such transactions would be anonymous. However, when performed over-and-over, they reveal agent-specific patterns that make it possible to link the addresses involved.

By combining these techniques with published methods, we created a list of agents participating in the early bitcoin market (1/3/2009-2/09/2011). See Figure 1.

### iv. Extranonce: Additional Details

A piece of metadata called the "extranonce", illustrated in Figure S3, is preserved in each block. The default bitcoin software available in 2009 and most of 2010 initialized the extranonce at zero and incremented with each new block ("block height rollover") and with exhaustion of the nonce. Since nonce exhaustion was rare in the first 18 months of bitcoin, the extranonce field functions as a slow real time clock of how long the mining client has been running. Critically, when agents paused mining and restarted their computer or mining client, the default software would resume mining with an extranonce of zero. Consequently, if only one machine was mining, we can clearly see mining client stop-and-starts. If we plot the extranonce on the y axis, and timestamps on the x axis, we can see distinct, adjacent, non-overlapping trajectories formed by the behavior of agents' mining clients. See Fig S4.

### v. Co2in

We generated a fork of the Litecoin codebase for internal lab use, dubbed "Co2in". We noted that miners tended to consolidate mining rewards from multiple addresses into a single address (see discussion of mining consolidation, above). A predilection for consolidating round numbers of coins was also immediately evident.

### vi. Review of Previous Address linking efforts

In his first email sharing the whitepaper, Satoshi says "participants can be anonymous," (*4*) and in the white paper, he says "some linking is still unavoidable with multi-input transactions, which necessarily reveal that their inputs were owned by the same owner." (*2*) Thus the concept of address linking has been present from bitcoin's inception.

Address linking approaches have been applied in practice by others, such as Spagnuolo et al. (*14*), who built a systematic framework for address linking. This study also tagged addresses by scraping web forums for addresses that had been publicly associated with online identities. Ober et al. (*13*) and Lischke et al. (*11*) used same-input-based address linking to estimate the total transaction community size and found that the number of addresses attributed to agents followed a power law. Kondor et al. (*10*) uses econophysics models to analyze the degree distribution and wealth distribution of the transaction network and finds the network to be scale-free (*38*) and driven by preferential attachment.

Much of the existing work analyzing data leakage from the nonce and extranonce has been reported by bloggers, where S. D. Lerner has been a key contributor. In particular, his work exploited data leakage from the extranonce to reconstruct the mining activity of Satoshi Nakamoto in detail (*15-17*). The analysis of this so-called "Patoshi" miner has been revisited from time-to-time by others (*64*); in 2014, a blogger reported using a similar analysis to examine roughly two thousand early blocks not mined by Satoshi (*65*).

Other linking approaches that incorporate information not explicitly part of the transaction graph structure, such as logging IP addresses in the peer-to-peer network used to broadcast transactions, have been suggested by Kaminsky (*12*) and formalized in Koshy et al. (*9*) It is worth noting that many of these previous approaches use additional information that is not explicitly part of the blockchain or transaction graph structure: web scraping (*14*), IP logging (*9*), user reported theft data (*7*) (*8, 19*), personal merchant interactions (*19*).

To the best of our knowledge, mining consolidation has not been employed in systematic work on address-linking in the past.

### vii. Fiat Currency Information

All bitcoin prices were from Blockchain.com (*66*) on June 6th, 2022. The estimates for market capitalizations mentioned in the main text date from March 2021 (specifically March 26 when possible), and the sources are as follows: bitcoin, Blockchain.com (*67*); US Dollar, the Federal Reserve (*68*); the Euro, the European Central Bank (*69*); Chinese RenMinBi, an estimate from Trading Economics (*70*). Exchange rates are from the US Federal Reserve (*71*) on March 26, 2021.

Deutsche Bank research reported in March 2021 that the value of circulating bitcoin exceeded all but two fiat currencies (the USD and the Euro), (*72*) but omitted comparison with the Chinese RenMinBi. For completeness, we include a publicly available estimate for the value of circulating RenMinBi.

## Detailed Methods

### i. Acquiring Blockchain Data

The first 500,000 blocks were downloaded as json files using Blockchain's API. In this case, json files were preferable over connecting a node to the network and downloading the entire contemporary blockchain directly due to the human readability and relatively easy parsing offered by the json-formatted data. We note that, although the Blockchain API could be unreliable when used to study extremely recent activity, it is completely reliable for older periods such as the one we examined, since consensus has long been reached. (Of course, contemporaneous versions of the blockchain may differ slightly prior to consensus.)

We specifically note that all popular current forks (Bitcoin, Bitcoin Cash, Bitcoin SV, Bitcoin Gold) share the same chain for the first 250,000 blocks. Bitcoin Cash forked at block 478,558, Bitcoin SV forked at block 556,766, and Bitcoin Gold forked at block 491,407. Until these respective forks, these projects share the same chain as Bitcoin. We note that all the figures in the main text would be unchanged had we used the Bitcoin Cash, Bitcoin SV, or Bitcoin Gold blockchains, with the exception of the analysis of contemporary bitcoin shown in Figure 4.

This is because they depend only on blockchain data prior to any of the above forks. Similarly, we note that the information about the early bitcoin community encoded in the blockchain is not in any way altered, affected, or erased by the process of creating a fork, and as such any security consequences deriving from such analyses could not be fully addressed by the process of creating a new bitcoin fork.

There were a total of 16,176,725 addresses in the first 250,000 blocks, and a total of 346,309,218 addresses in the first 500,000 blocks, where there were 283 million transactions.

### ii. Address Linking Procedure in Detail

Users are represented on the blockchain pseudonymously by one or more alphanumeric public keys. Every bitcoin transaction ever made can be found on the present day blockchain. Therefore, a user with an interest in anonymity and good security hygiene would frequently generate new public/private key pairs and would not reuse public keys. In practice, users frequently reuse key pairs and move bitcoin in easily characterizable ways that disclose personal information or information regarding the nature of the transaction. Metadata from these transaction practices can be used to attribute multiple public keys to the same entity or agent.

Our clustering efforts were geared towards an accurate overall reconstruction of the early bitcoin community, not at deanonymizing individual people or identifying all of a person's bitcoin with complete certainty.

We performed address linking as follows.

a. We began by identifying blocks mined by Satoshi Nakamoto. To do so, we plotted extranonce vs. block height, and looked for linear trajectories with a slope well above unity, the slope formed by mining with the default software. In cases where the assignment of a block to a particular extranonce trajectory was ambiguous (for instance, because multiple extranonce trajectories intersect), we did not assign the block to a trajectory. The resulting set of trajectories, spanning 22171 blocks, was associated with Satoshi Nakamoto. (See *Satoshi Nakamoto aka Agent #1*, below.) These blocks were removed from the dataset during the subsequent agent-identification procedures. See Fig S11. (This step was a refinement of the approach introduced by S. D. Lerner for reconstructing the so-called 'Patoshi' miner.)

b. Next, we identified extranonce trajectories in the remaining data. Each trajectory is a set of blocks that fell along linear contours in the extranonce vs. block height plot. Again, in cases where the assignment of a block to a particular extranonce trajectory was ambiguous (for instance, because multiple extranonce trajectories intersect), we did not assign the block to a trajectory. See Figs S6 and S7 for rationale behind extranonce linking. We note that these block sets were not independently used for address-linking, because we found that doing so was too error-prone. Instead we proceeded as described below, using extranonce data as an additional constraint on our mining consolidation procedure.

c. We defined a graph *G* using the first 250,000 blocks, which corresponds to the blockchain through August 3rd, 2013. The nodes of this graph are 350 million bitcoin addresses that occur in these blocks.

   We further defined two sets of undirected edges, each of which links pairs of nodes (*i,j*) as follows.

   First, whenever *i* and *j* were inputs into the same transaction (input-linking), we added an input-linking edge (represented by the color green) between them.

   Second, we looked for transactions that were consistent with mining consolidation. Two criteria were used. One criterion was that the transaction transferred a large number of coinbase rewards into a single address *j*. The second criterion was that at least three of the input coinbase rewards had been annotated as lying along a single extranonce trajectory in step (b) above. Once a mining consolidation event had been identified, we added a mining consolidation edge (represented by the color purple) between *i* and *j* whenever *i* was an input address into the mining consolidation transaction. We found that the integration of extranonce constraints into the mining consolidation step yielded a much less error-prone set of links then would be obtained by using either the extranonce or mining consolidation on their own.

d. We constructed a list of connected components of the graph *G*. These connected components are sets of bitcoin addresses that correspond to candidate agents in an address-linking scheme employing only same-input (green) and mining consolidation (purple) edges. We note that these connected components also benefit from some of the data leakage due to the extranonce, as described in step (c) above.

e. We plotted the extranonce against block height for each agent, and identified candidate agents that exhibited distinct extranonce trajectories that overlapped in time. This behavior (dubbed 'overlapping extranonce trajectories' below) is extremely unlikely for a single machine mining bitcoin, and we therefore subjected such agents to additional scrutiny. Specifically, the mining-consolidation edges associated with these agents were set aside for subsequent manual curation (see step g). After these edges were removed, the candidate agents (i.e., the connected components of G) were recalculated.

f. Next we looked for pairs of putative agents that shared extranonce trajectories, in the sense that both agents were associated with addresses that were part of the same extranonce trajectory. When two putative agents exhibited many addresses on shared extranonce trajectories, they were combined into a single agent.

g. We manually curated the mining-consolidation edges set aside in step e, including mining-consolidation edges when (i) the putative agents they linked shared extranonce trajectories, and (ii) the combined agent did not exhibit overlapping extranonce trajectories. Note, edges removed in step (e) are selectively reintroduced here provided they meet conditions (i) and (ii).

h. We identified putative agents that mined exactly 20 or 40 blocks and who transferred all the corresponding coinbase rewards to a single address, consistent with mining consolidation (although not meeting the specific criteria we employed in step c). When multiple putative agents transferred to the same address, we combined them into a single agent, unless either (i) the nonce profiles of the individual agents were incompatible, or (ii) the combined agent exhibited overlapping extranonce trajectories.

See Fig S5 for a schematic of the pipeline, and Figs S8, S9, and S10 for agents put together with this pipeline.

### iii. Transaction Linking Heuristics - Pseudocode

In this section we provide pseudocode for key address linking heuristics that are triggered by particular types of transactions.

In our notation {input_addresses} represents the set of input addresses associated with a transaction, {output_addresses} represents the set of output addresses associated with a transaction, {old_addresses} represents the set of all addresses that exist prior to the transaction in question, and || represents the cardinality of a set.

The notation{enonce_annotation} represents a set of blocks that correspond to a single extranonce trajectory, annotated in step (b) above.

a. **Same Input - See figure S1D.**
if $|$ {input_addresses} $|$ > 1,
$\Rightarrow$ add a link (i, j) $\forall$ i, j $\in$ {input_addresses}

b. **Change Rule - See figure S1E.**
if $|$ {output_addresses} $|$ = 2,
and if $output_address_1 \in$ {old_addresses} and $output_address_2 \notin$ {old_addresses},
and if $output_address_1 \notin$ {input_addresses} && $output_address_2 \notin$ {input_addresses},
$\Rightarrow$ add a link (i, $output_address_2$) $\forall$ i $\in$ {input_addresses}

c. **Mining Consolidation - See figure S1F.**
if $|$ {input_addresses} $|$ $\geq$ 3 and $|$ {output_addresses} $|$ = 1,
and if $\exists$ {enonce_annotation} such that |{input_addresses} $\cap$ {enonce_annotation}| $\geq$ 3,
$\Rightarrow$ add a link(i, $output_address_1$) $\forall$ i $\in$ input_addresses

*Note: The mining consolidation heuristic helps capture runs of the mining software and the corresponding extranonce trajectories while minimizing error.

### iv. Manual Cluster Grading

We again plotted extranonce against block height in order to validate the resulting candidate agents.

a. Convincing agent: A candidate agent exhibiting a series of dense extranonce trajectories, typically with a slope of one. The trajectories do not overlap. One trajectory often begins shortly after the previous fragment has ended along the time axis (showing mining client stop and restart).
b. Ambiguous agent: A candidate agent with few extranonce trajectories and no overlap, or with no extranonce trajectories. No clouds of points.
c. Separable agent: Many extranonce trajectories are seen, each with a slope of one. The trajectories clearly overlap in time. This indicates over-consolidation, or that an agent has more than one machine mining simultaneously.
d. Poor agent: No dense line fragments, only clouds of points.

*Note: It is certainly possible for one agent to have more than one mining machine running at a time. If there seemed to be sufficient visual evidence for more than one mining machine, the cluster was not broken up.

### v. Tagging

Address-linking is prone to two types of errors. First, it is possible that some links are erroneous (i.e., false positives). Second, because we cannot guarantee that all true address links are detectable, it is possible that multiple agents in a given list are all a single person (i.e., false negatives).

To validate our list of agents, a number of bitcoin discussion forums, most notably bitcointalk.org, were scraped in search of posts in which a user publicly indicated ownership of an address. An online identity publicly associated with a bitcoin address will be hereafter referred to as a "tag," as in an address was "tagged" with an online identity. Bitcointalk.org was crawled up until the end of 2013 to avoid more contemporary speculation on old addresses. In the very early community, it was common for users to post "rich lists" of community users and publicly discuss addresses with very high balances. Popular "rich lists" were also collected for tagging. Additionally, screenshots of users mining rewards were very useful for tagging. For example, at one point, the user Theymos published a screenshot of a number of their recently mined blocks. See Fig S12. All of the addresses shown in screenshots of this type could be tagged as belonging to Theymos. We also compiled tags from Bitiodine (*14*), bitcoin-otc.com (*73*), archived IRC logs, and Blockchain.com's submitted tags. All known Satoshi correspondence to early users was reviewed, including but not limited to, emails to Mike Hearn, which allowed the identification of the Mike Hearn-Satoshi transaction (*74*).

Tags from web crawling can have a very high error rate as forum users often discussed the transactions of others, especially if they were very early or very large transactions. To circumvent this error rate, only agents with more than 150 bitcoin mined were tagged and all tags associated with agents were manually curated to be as accurate as possible. A very high number of positive pairwise intra-agent comparisons comes from several users sharing a large number of their addresses (as in the case of Theymos' or Hal Finney's screenshots). However, this problem is endemic to the method of tagging and every literature heuristic would exhibit this true positive "boost."

This yielded a list of 50,323 addresses, each labelled with value corresponding to their online identity. This list was used as a proxy for ground truth data. Note that, since our purpose was not to de-anonymize individuals, these numbers almost certainly underestimate overall levels of participation on bitcoin forums.

From these tags, all possible pairwise comparisons were examined. If the tags matched and the agent id matched, it was counted as a true positive. If the tags did not match and the agent ids did not match, it was counted as a true negative. If the tags matched and the agent ids did not match, it was counted as a false negative. If the tags did not match and the agent ids matched, it was counted as a false positive.

We compared our address-linking strategy to a strategy using only links derived from same-input linking. We found that the specificity was similar for both approaches, but the rate of false negatives was >25-fold higher using same-input linking alone. We then compared our strategy to same input linking combined with only change linking (change linking was not used in our methods), yielding lower specificity (~95%) and a similar false negative rate as same input linking alone. This is consistent with the fact that many agents cannot be reliably reconstructed only on the basis of same-input linking.

See Fig S13.

As an additional sanity check, we compared the timestamps of sequential blocks. A negative time delta between blocks mined sequentially by the same agent would indicate a clustering mistake. There are no negative time deltas in any of our agents, nor are there any negative time deltas in the dataset in general. In other words, time stamps increase monotonically for the first 107,000 blocks.

### vi. Majority Attacks

The largest agents were ranked and the top n agents required to mine >50% of the total bitcoin between January 3rd 2009 and February 9th 2011, became our 64 top agents. To explore the number of colluders required to attack the network via majority attack at any given time, small time intervals between 3 blocks and three days were examined. It became clear that there were several intervals where only one agent (Agent #2) controlled most of the computational resources used by the bitcoin network. This would have been readily apparent to Agent #2 at the time. See Fig S14.

Next, we calculated the frequency of unilateral and bilateral streaks, which are sequences of consecutive blocks all mined by only one or two agents. We also calculated the frequency of majority streaks, in which a single agent mines most of the blocks. The results are in Figures 5 and S15.

For the purpose of comparison, we analytically determined the expected number of streaks of various types in a population of $M$ mining agents, assuming that each agent has identical computational resources, distributed uniformly over time. We obtain the following equation for the expected number of unilateral streaks, $X_{unilateral}$:

$$E[X_{unilateral}] = \frac{N-k+1}{M^{k-1}}$$

Where $k$ is the length of the streak in blocks and $N$ is the total number of blocks. We obtain the following equation for bilateral streaks:

$$E[X_{bilateral}] = \frac{N-k+1}{M^{k-1}}\left(1+\left(\frac{2^k-2}{M}\right)\right)$$

Finally, we obtain the following equation for majority streaks:

$$E[X_{majority}] = \frac{N-k+1}{M^{k-1}} \sum_{j=\left\lceil\left(\frac{k+1}{2}\right)\right\rceil}^{k} \binom{k}{j} (M-1)^{k-1}$$

For the theoretical curves shown in the figures, we set $N$ as the number of blocks mined by non-Satoshi miners during our period of study.

To calculate effective population size as reported in the paper, we used bilateral streaks with a length of between 3 and 12 blocks, inclusive. We obtain best fits for the following values of $M$: 5.0522, 5.6121, 5.59224, 5.37115, 5.18127, 4.91012, 4.62591, 4.44368, 4.27408, and 4.25033, suggesting a total effective population size of 4.93 agents. The results are similar using unilateral and majority streaks.

We note that there remains a considerable amount of centralization among the top mining pools even at present (*75*).

### vii. Follow-The-Money

The structure of the bitcoin transaction network has specific consequences not seen for other networks. When encountering criminal activity associated with bitcoin, it is not uncommon for law enforcement to seek to identify agents who control a particular target address. One approach, which we dub the "follow-the-money" strategy, is to begin at an address controlled by a publicly known agent who is several steps away from the target address. By using off-chain metadata, it is possible to trace a path from the publicly known agent to the address-of-interest. For instance, the office of Special Prosecutor Robert Mueller was able to determine that a group of Russian hackers known as Fancy Bear were responsible for the theft and dissemination of documents from the US Democratic National Committee by connecting them with a target bitcoin address (*6*).

Thus, our findings imply that, should a law enforcement agency determine the identity of the 64 top agents, the agency could de-anonymize nearly any target bitcoin address using at most 6 steps. Similarly, the follow-the-money strategy could be employed by a rogue entity to compromise the privacy of all bitcoin users.

We began by defining a graph whose nodes are all bitcoin addresses. We then added directed links between a pair of nodes *i,j* whenever there was a transaction in the first 500,000 blocks in which address *i* had sent bitcoin to address *j*. The resulting graph had 350 million address nodes.

All of the addresses belonging to the top 64 agents were labelled. Finally, for every node in the network, we calculated the length of the shortest path between connecting that node to one of the labelled nodes. For instance, all addresses belonging to top 64 agents yielded a path length of 0; any immediate transaction partners who were not themselves owned by top 64 agents had a path length of 1, and so on. The above graph calculation was performed using a custom variant of Dijkstra's algorithm (*76*) and run on an AWS node with 1 terabyte of RAM to accommodate the large network size, as the register of visited nodes was too large to be kept in memory on a smaller machine. The algorithm was implemented in java and run on a Neo4j graph database.

To visualize the results, we selected a subset of 5,000 nodes at random and showed the shortest path between each of the chosen nodes and a labelled node (Figure 4 in the main text). Most (91.9%) addresses are separated from one of the labelled nodes by a chain of four or fewer transactions. Nearly all (99.3%) are separated by a chain of six or fewer. The very large outliers may be the result of a deliberate effort to obscure the flow of bitcoin through transactions, rather than true examples of inter-agent exchange.

## viii. Pareto Distributions

The time period between January 3rd 2009 and February 9th 2011 was divided into six equal intervals. In each interval, agent size, defined as the percentage of coin mined by an agent in that interval, was fit with a power law using the package "poweRlaw" in R, which uses maximum likelihood estimation to fit the complementary cumulative distribution function (*77*).

Agents with exactly 2000 or 4000 blocks mined were suppressed for the Pareto fits. While a monotonically increasing Pareto exponent is observed from Interval 1 to Interval 6 without suppressing these artifacts, the anomalous number of agents of this size is an artifact of users tending to move bitcoin in round numbers and does not depict a true spike in the number of agents with these amounts of bitcoin mined.

The strengths of the clustering methods used in this paper are in putting together large to very large agents. Finding fractile income or capital ownership proportions to study evolving inequality is not feasible in a situation like this one, where we have a good idea of the mining income of a few dozen top agents, but the mining income of the bottom fractiles is very noisy. Consequently, we study the evolution of the Pareto exponent, which is fit to the complementary cumulative distribution function, making it well suited to a study of well reconstructed, large agents.

Fits of Intervals 1-4 yield Pareto exponents of less than one. A Pareto exponent of less than one does not make sense in a real-world scenario. To wit, this would mean the distribution has infinite variance, which is impossible, but the Pareto exponent estimate is broadly indicative of levels of inequality (granted, above a certain threshold). See Fig S16.

## ix. Topic Modeling

To understand how community discussion changed between the two stages (Intervals 1-4 vs Intervals 5&6), we scraped all posts from the official bitcoin forum (bitcointalk.org) between its first post (November 22, 2009) and the date of dollar parity (February 9th, 2011). There were 146 threads in stage 1 (November 22nd, 2009 – May 29th, 2010) and 2770 threads in stage 2 (May 29th, 2010 – February 11th, 2011).

BeautifulSoup was used for text scraping and top2vec (*78*) was used for topic modeling. Top2vec was ultimately chosen for topic modeling over the more established methods of non-negative matrix factorization (NMF) and latent Dirichlet analysis (LDA) (*79*), due to four main reasons: LDA is ill suited to very short texts (many forum posts are short); both LDA and NMF require picking the number of topics before clustering, information that is not available a priori; both LDA and NMF are highly sensitive to preprocessing; and both LDA and NMF use a bag-of-words (BOW) approach to represent texts, which might not be sufficient to separate semantically similar text documents within a specialized domain.

Top2vec is a relatively new NLP approach, and so a short description is as follows. First, a joint embedding of both words and documents is trained. Topics are represented by vectors lying in this same semantic space. UMAP (*80*) is then performed on this space for dimensionality reduction. Document vectors are then clustered by HDBSCAN (*81*), a density-based clustering algorithm. Topic vectors are finally picked by choosing the centroids of these clusters. Topic keywords are the closest words by cosine similarity to each topic vector. Top2vec requires almost no preprocessing (save for punctuation/capitalization standardization), as stopwords (high frequency, less meaningful words) will be equidistant from all topic vectors.

All forum posts had punctuation and capitalization removed. Initial topic modeling runs identified three topics: English posts, French posts, and Russian posts. While this does reflect the growing audience of bitcointalk.org (the foreign language posts were only seen in the second stage, after the arrival of early adopters), these results were less useful in identifying the change in topic discussion over time.

Subsequently, non-English documents were filtered out using the module langdetect (*82*). Top2vec was run on all documents from both stages (Intervals 1-6), with a total of 2765 documents total. Note that this number of topics is greater than the total number of threads. If a thread spanned both stages, it was split into two documents: comments posted in stage 1 (Intervals 1-4) and comments posted in stage 2 (Intervals 5&6). Likewise, threads with comments posted after the conclusion of Interval 6 were truncated so all data comes only from Intervals 1-6. Otherwise, one thread constituted one document.

Top2vec is an unsupervised, stochastic method, and as such, is sensitive to initial state conditions. To get the numbers cited in the main text, we ran 20 trials with one set of hyperparameters. Coherence score ($C_v$) (*83*) was assessed for the top 10 words for each topic in all 20 trials. The model with the highest coherence score was chosen for main text Fig 3.

### x. Controlled Experiment: 8-player Centipede Game

*Experimental design*

In the 8-player Centipede game experiment, subjects are randomly matched into groups of N=8 players, which participate in the following sequential move game with an exponentially appreciating commodity. The game resembles a centipede game, in which a new, anonymous player is added to play in each new round. We note that other N-player generalizations of the Centipede game are also possible; this generalization is meant to recapitulate the dilemma faced by bitcoin's founders (*84*). The eight players take turns in choosing between two actions – to cooperate or to defect – and take their decisions one after another. Each player's position in the sequence is randomly determined at the beginning of the game and is retained across all repetitions. If a player chooses to cooperate, the round ends, the value of the commodity increases by a factor of d, and the next player in the sequence is selected to make the same choice. If a player chooses to defect, that ends the game, and all players who have played must sell off their commodity, gaining a payoff $\pi = d^{n-1}$, where n is the round in which the defection took place. The defecting player benefits from this decision by gaining $b*\pi = b*d^{n-1}$. Players who have not made a choice because an earlier player in the sequence has defected receive no payoff. The game also ends after a pre-specified number of rounds, N=8, even if no one has defected. If so each of the N players gain the payoff, $\pi = d^N$. Each subject takes part in five repetitions of the game, in each of which they are assigned to a new group of eight.

All instructions are phrased in abstract and neutral terms without any mentioning of terms like "custodian", "cooperate", "defect", or "commodity". The full experimental instructions including all game parameters, as well as a screenshot of the decision screen are available on the following Open Science Framework (OSF) repository: https://osf.io/u5qg3/

*Recruitment and data collection using Prolific*

832 unique subjects, who gave informed consent, were recruited using the online crowd-working platform Prolific (*44*) in November 2021. Subjects were required to be fluent in English, have a minimum Prolific approval rate of 95%, and were not allowed to have had previously participated in the same study.

792 subjects completed all eight rounds in all five repetitions of the 8-player Centipede game; the remainder also read the experimental instructions and passed five comprehension questions, but discontinued the study or got timed out during one of the five repetitions. Games in which at least one of the eight players was timed out were excluded from the analysis. After five repetitions of the game, subjects completed a short exit survey, in which we elicited their risk preferences, financial literacy, cognitive reflection abilities, as well as some basic demographics. The experiment was programmed and conducted using an adaption of the bubble game application for oTree (*85, 86*).

All subjects who completed the study received a fixed reward of GBP 2.00 as well a bonus payment depending on the outcome of one randomly determined repetition of the 8-player Centipede game. Bonus payments varied between GBP 0.00 and GBP 1.28 (mean: 0.36; std. dev.: 0.31).

54.8% of subjects were male. Their age ranged from 18 to 86 years with a mean of 29 years. The most common participating nationalities were British (15.4%), Polish (14.8%), Portuguese (10.8%), U.S.-American (9.7%), and South African (8.3%). 32.0% of subjects reported to have had invested in one or more cryptocurrencies in the previous five years.

*Preregistration*

The experimental study was preregistered on OSF using the AsPredicted.org protocol: https://osf.io/tvuyf

*Code and data availability*

The oTree code of the experimental software as well as all collected data and the R script used for statistical analyses are available on OSF: https://osf.io/u5qg3/

*Supplementary results and statistical analysis*

- Defecting on the first round:
  38 out of 438 games ended at the first node (8.7%), i.e. individuals chose to "defect" in the first round; this is significantly different from the subgame-perfect Nash equilibrium of 100% ($p < 0.001$, CI = [0.062, 0.117], two-sided binomial test).

- Defecting on the last round:
  In 78 out of 224 games that lasted the full 8 rounds (34.8%), the last individual in this game chose to "defect" in the last round; this is significantly different from 100% ($p < 0.001$, CI = [0.286, 0.415], two-sided binomial test).

- Players tend to become more cooperative with each repetition:
  Implied conditional probabilities of defection in the first round are 13.0%, 10.3%, 8.0%, 7.2%, and 5.0% for repetitions 1 to 5.
  In repetitions 1 and 2, 21 out of 179 games (11,7%), and in repetitions 4 and 5, 10 out of 171 (5.8%), ended at the first node ($p = 0.053$, CI = [2.349e-05, 1.177e-01], two-sided test for equality of proportions)

- Results hold for different cultures/nationalities and hold equally well for cryptocurreny investors as for those who had made no such investments:
  Among the 10 nationalities of whom at least 10 subjects participated in the experiment, average cooperation rates range from 82.4% (United Kingdom, N=128) to 91.1% (United States, N=81) and are not statistically different ($p = 0.505$, two-sided Kruskal-Wallis test).
  In the exit survey, 32.0% subjects also answered the question "Have you invested in bitcoin or any other cryptocurrency in the last 5 years?" with "Yes". On average, they cooperate in 90.4% of their decisions, compared with 85.7% cooperation among those who made no such investments ($p = 0.013$, CI = [-0.082, -0.010], two-sided t-test).

- Comparison with centipede game:
  An additional 86 unique subjects were recruited from Prolific to take part in a 2-player centipede game experiment, applying identical procedures than for the 8-player Centipede game experiment outlined above. None of the additional subjects had previously participated in the 8-player Centipede game. The 2-player centipede game uses the identical game parameters than the 8-player Centipede game, except insofar as the 2-player Centipede game has only two players who alternate making choices over up to eight rounds.
  35 out of 209 2-player centipede games ended at the first node (16.7%), i.e., individuals chose to "defect" in the first round; this is significantly different from the corresponding proportion in the 8-player Centipede game (p = 0.004, CI = [-0.141, -0.020], two-sided test for equality of proportions).

In 25 out of 67 games that lasted the full 8 rounds (37.3%), the last individual in this game chose to "defect" in the last round; this is not significantly different from the corresponding proportion in the 8-player Centipede game (p = 0.819, CI = [-0.166, 0.116], two-sided test for equality of proportions).

In 36 out of 103 games that lasted at least 7 rounds (35.0%) in the 2-player centipede game, and in 47 out of 271 games that lasted at least 7 rounds (17.3%) in the 8-player Centipede game, the individual in the penultimate round 7 chose to "defect". This difference is statistically significant ($p < 0.001$, CI = [-0.285, -0.067], two-sided test for equality of proportions). See Fig S17, S18, S19.

- *Comparison with other game variants*

  Another 246 subjects from Prolific were recruited to take part in one of three similar variants of the 8-player Centipede game. None of the additional subjects had previously participated in the 8-player Centipede game. Variant LR uses the identical game parameters than the 8-player Centipede game, but with one difference: if a player chooses to defect, all players who have cooperated in rounds 1 to n-2 gain a payoff $\pi = d^{n-2}$, and the player who cooperated in round n-1 gains a payoff $\pi = d^{n-1}$, where n is the round in which the defection took place. In this variant, a player in round n cannot directly affect (and thereby reciprocate to) the player who cooperated in round n-1. Variant HS uses the identical game parameters than the 8-player Centipede game but with all payouts multiplied by ten to induce higher stakes. Variant 4P uses the identical game parameters than the 8-player Centipede but with only four players and four rounds instead of eight.

  6.7%, 17.8%, and 6.1% of games in the LR, HS, and 4P variants, respectively, ended at the first node, i.e., individuals chose to "defect" in the first round (Differences to 8-player Centipede game: LR: p = 0.857, CI = [-0.070, 0.110]; HS: p = 0.087, CI = [-0.218, 0.036]; 4P: p = 0.735, CI = [-0.058, 0.109], two-sided tests for equality of proportions).

  In 50%, 40%, and 50% of games that lasted until the last round in the LR, HS, and 4P variants, respectively, the last individual in this game chose to "defect" in the last round (Differences to 8-player Centipede game: LR: p = 0.389, CI = [-0.459, 0.155]; HS: p = 0.769, CI = [-0.276, 0.172]; 4P: p = 0.266, CI = [-0.407, 0.103], two-sided tests for equality of proportions).

  In 10.5%, 7.4%, and 38.0% of games that lasted until the penultimate round in the LR, HS, and 4P variants, respectively, the individual in the penultimate round chose to "defect" (Differences to 8-player Centipede game: LR: 0.653, CI = [0.173, 0.105]; HS: p = 0.291, CI = [-0.030, 0.228]; 4P: p = 0.007, CI = [-0.383, -0.027], two-sided tests for equality of proportions).
- *Power analysis*

  Our analysis contains 438 8-player centipede games and 209 2-player centipede games in the first round as well as 224 and 67 games in the last round, respectively. This allows us to detect small- to medium-sized effects between $h = 0.236$ (first round) and $h = 0.390$ (last round) with 80% power at the 5% significance level in testing for differences in the proportion of defections between the two games (where $h = \left|\left(2 \arcsin \sqrt{p_1}\right) - \left(2 \arcsin \sqrt{p_2}\right)\right|$ with $p1, p2$ representing the two proportions; e.g., a small effect of $h = 0.2$ translates to a 5 to 10 percentage points difference (*87*)). The respective minimum detectable effect is even smaller for the one-sample tests of defecting in the 8-player centipede game alone (between $h = 0.133$ (first round) and $h = 0.187$ (last round)). In comparing defection rates between the first and last two rounds in the 8-player centipede game, our sample size also allows us to detect a small- to medium-sized effect $h = 0.300$.

Note, however, that due to the smaller sample sizes, statistical power is lower in the comparisons with other game variants (LR, HS, and 4P). For those comparisons we are only able to detect medium- to large-sized effects between $h = 0.422$ (first round) and $h = 0.772$ (last round).

**xi. Human Subjects Approval**

We consulted with the Baylor College of Medicine Institutional Review Board, which determined that our analysis of the bitcoin blockchain does not constitute human subjects research, and does not fall under the regulations for IRB review of human subjects research, as the information examined is not considered private information.

Approval for the Centipede game experiments was provided by the Competence Center for Experimental Research at Wirtschaftsuniversität Wien, the Vienna University of Economics and Business (Reference Number WU-HSRP-2021-017), and all participants gave informed consent.

**Extended Legend for main text Figure 1.**

**i. Satoshi Nakamoto aka Agent #1**

Trajectories with a slope of one are formed by the default behavior of the original mining client: the extranonce increments by one every time a block is mined. There are also many trajectories with a slope much greater than one. These runs are almost completely unspent, do not overlap, and start at the very beginning of the bitcoin blockchain. For months at a time, these trajectories are of approximately uniform length, formed by an agent turning on their mining software, which initializes the extranonce at zero, mining continuously for a set period of time, then stopping the mining software, backing up their wallet, and starting again. We attribute the blocks following this pattern to Satoshi Nakamoto, the inventor of bitcoin.

All in all, there are 22,171 blocks (₿1,108,550, or $34,707,591,950 USD) following this pattern. We believe this is the most accurate estimate to date.

We also attribute three transactions where mining rewards were moved to Satoshi. The first bitcoin transaction recorded was a payment from Satoshi to Hal Finney in Block 170 on January 11, 2009, moving coin mined in Block 9 on January 9, 2009. The second Satoshi transaction moving a mining reward is from Satoshi to Mike Hearn in Block 11408 on April 18, 2009, moving coin mined in Block 5326 on February 23, 2009. Satoshi disclosed his public key via email to Mike Hearn, telling Hearn that if Hearn sent Satoshi ₿32.51, Satoshi would return the coin plus an additional ₿50. Mike Hearn made these emails public in August 2017 (*74*). The third Satoshi transaction takes place in Block 56173 on May 17, 2010 and moves the rewards from ten blocks (Blocks 1760, 3479, 9443, 9925, 10645, 14450, 15817, 19093, 23014, 28593) on ten different

extranonce trajectories. There has been a good deal of speculation on Satoshi Nakamoto's ultimate fate (*88*) but this transaction shows he did still have possession of early private keys at this later date. Notably, this is just before we see him or her exit the community for good on May 3, 2010. See Fig S11 and S20.

**ii. "Dr. Evil" aka knightmb aka Michael Mancil Brown aka Agent #19**

Michael Mancil Brown, known as knightmb on bitcointalk.org, was federally convicted for attempting to ransom Mitt Romney's tax returns (*89*).

**iii. Agent #6**

Agent #6 is the first agent to exhibit mining trajectories consistent with GPU mining. See Fig S21.

Until the summer of 2010, bitcoin was mined only on CPUs. Miners discovered GPUs could do the work required to mine bitcoin much faster than CPUs and GPU mining quickly became widespread. As GPUs can increment extranonce values much faster with regard to session time, this rate increase manifests as a change in slope when we plot extranonce value vs time. See Fig S21.

**iv. "Dread Pirate Roberts" aka Ross Ulbricht aka altoid aka Agent #67**

Ross Ulbricht was the founder of the Silk Road, a criminal marketplace on the dark web which used bitcoin for payment and which was online from January 2011 to October 2013. He was federally convicted of "distributing narcotics, distributing narcotics by means of the Internet, conspiring to distribute narcotics, engaging in a continuing criminal enterprise, conspiring to commit computer hacking, conspiring to traffic in false identity documents, and conspiring to commit money laundering" and was sentenced to life in prison and ordered to forfeit 186 million dollars (*90*). He made a bitcointalk.org profile under the username "altoid."

Main text figure one was created with a Moore curve (*91*) under a Shirley-Chiu transform (*92*). See Fig S22.

**Packages and Software Used**

python- numpy, pandas, scipy, networkx, bokeh, nltk, gensim, matplotlib, beautifulsoup
java
R- poweRlaw
Neo4j
Gephi

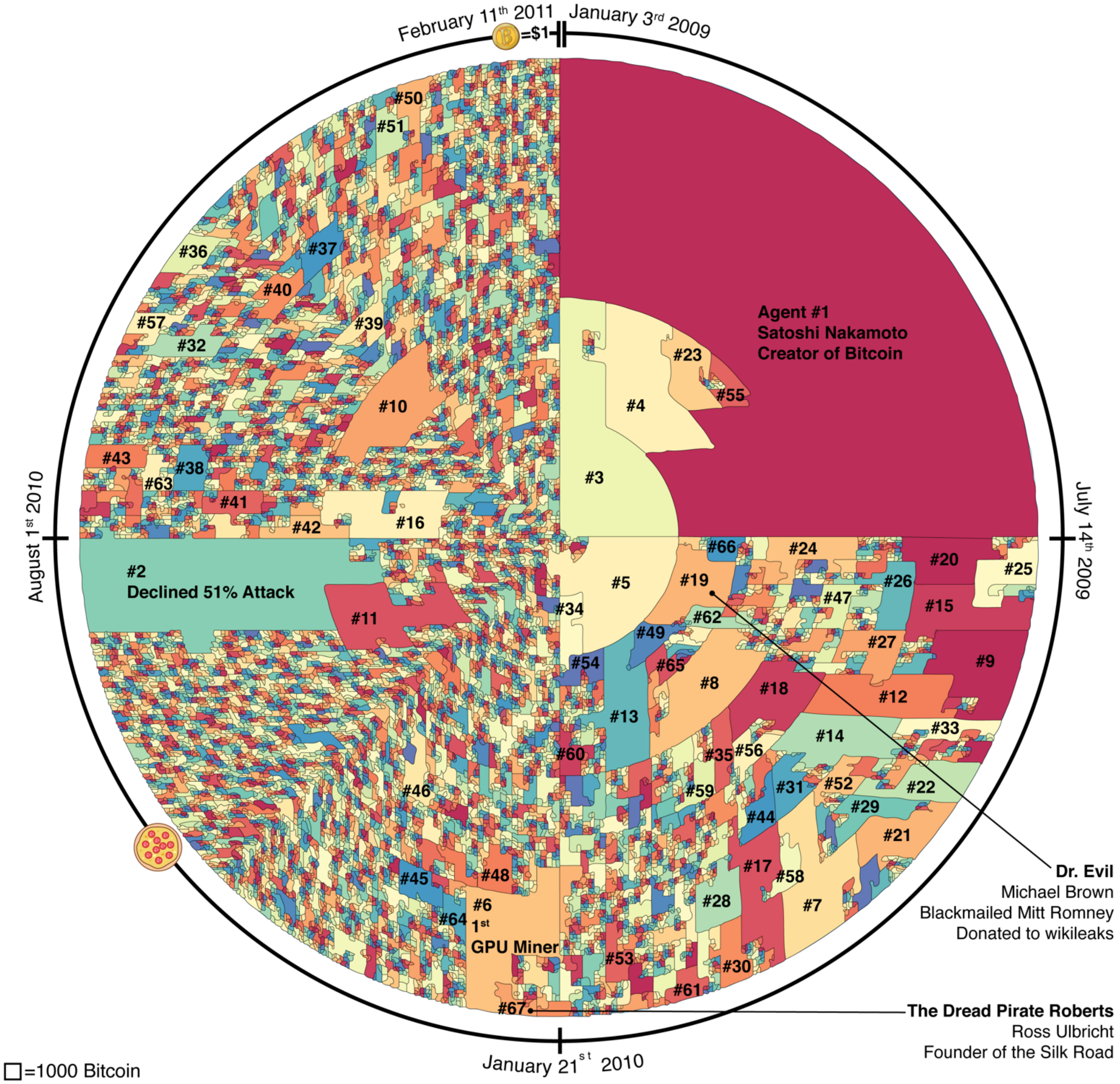

**Fig 1. Sixty-four agents mined most of the bitcoin between bitcoin's launch and when it achieved price parity with the US dollar.** We exploited data leakage to construct a map of the blockchain in early 2011, in which bitcoin are arranged according to the agent that mined them. The map was generated as follows: we sorted bitcoin by the agent that mined them; sorted agents by the date on which they first mined a bitcoin; and then arranged the bitcoin in 2D using a disc-filling curve. Consequently, each agent corresponds to a single map tile, whose area is proportional

to the quantity of bitcoin mined by the agent. The position of the tile along the curve indicates roughly when the agent was active, moving clockwise from the launch of bitcoin (January 3rd, 2009) to the moment that bitcoin achieved parity with the US dollar ($1=Ƀ1, February 9th, 2011). Agents are labelled in descending order of size, beginning with the largest agent (#1, Satoshi Nakamoto), and ending with Agent #67 (Dread Pirate Roberts). Except when otherwise specified, the resulting ordering is used to enumerate agents throughout the text. Smaller agents are shown but not labelled. Collectively, 64 top agents mined most of the bitcoin generated during this period. The earliest bitcoin users, the community's "founders," were able to mine a disproportionate fraction of the currency. Later agents, the "early adopters," correspond to smaller tiles, as they had to compete for bitcoin with an increasingly large community. Early adopters of GPU-based methods are an exception, exploiting a technological advantage to achieve extraordinary size despite a later entry date. As GPU mining becomes more widespread, this advantage, too, dissipates.

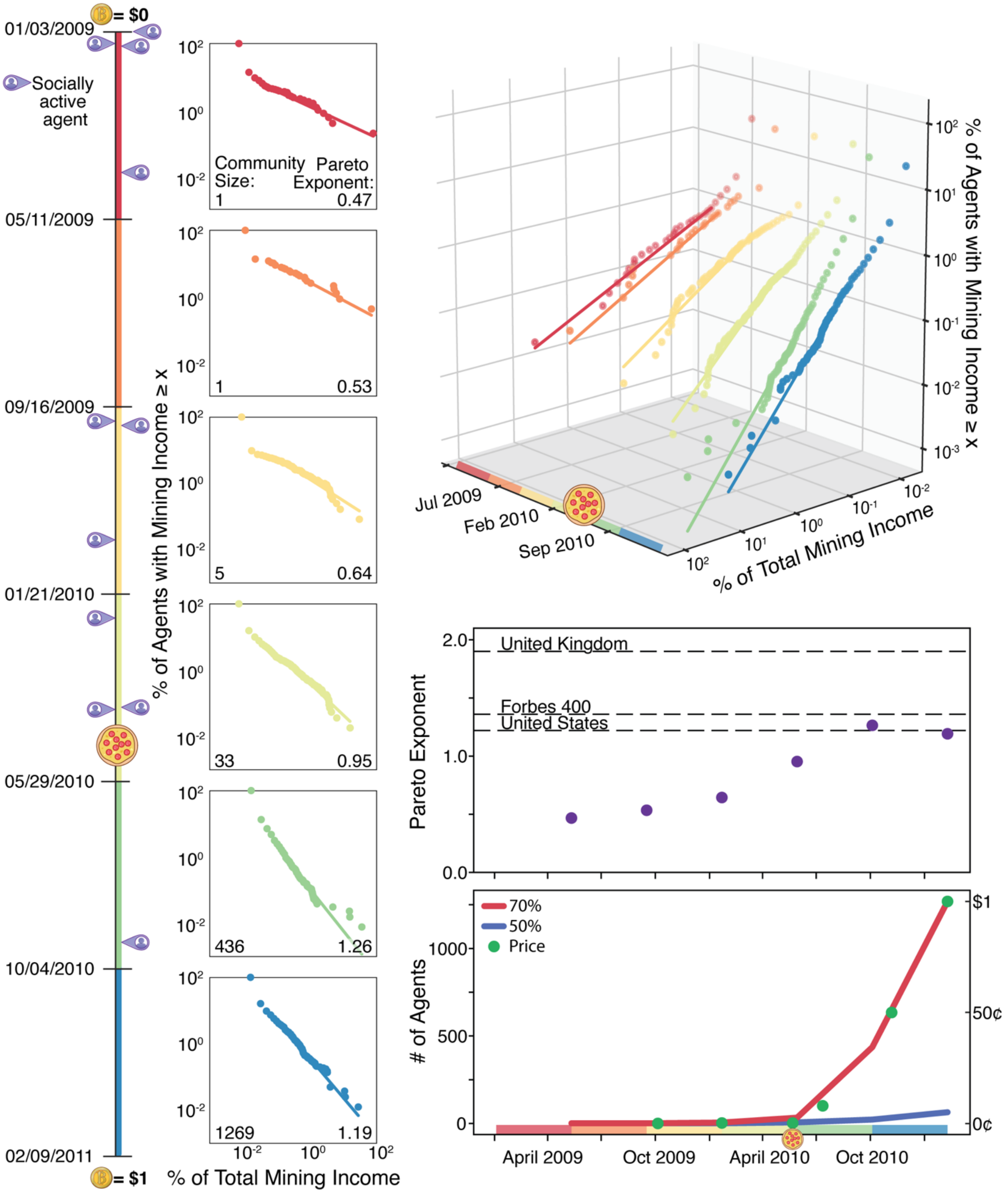

= $0
01/03/2009
Socially active agent
05/11/2009
09/16/2009
01/21/2010
05/29/2010
10/04/2010
02/09/2011
= $1
% of Agents with Mining Income ≥ x
Community Size:
Pareto Exponent:
1
0.47
1
0.53
5
0.64
33
0.95
436
1.26
1269
1.19
% of Total Mining Income
Jul 2009
Feb 2010
Sep 2010
% of Total Mining Income
% of Agents with Mining Income ≥ x
Pareto Exponent
United Kingdom
Forbes 400
United States
# of Agents
70%
50%
Price
$1
50¢
0¢
April 2009
Oct 2009
April 2010
Oct 2010

**Fig 2. Pareto distributions of income inequality emerge rapidly, and their dynamics reflect a two-stage process of technological adoption.** On the left, we plot the entry of socially active agents into the bitcoin community. All but one of these agents enter during the first four intervals (sixteen months). In the middle, we plot the fraction of agents whose bitcoin mining income was larger than a certain fraction of the community's total income. These values fall on a line in a log-log plot, indicating that incomes in the bitcoin community, like those in other economies, follow a power law, known in this context as a "Pareto distribution." The negation of this line's slope minus one is known as the "Pareto exponent." During the first four intervals (sixteen months), the power laws are extremely shallow, indicating extraordinarily high levels of income inequality not observed in ordinary economies: nearly all income is distributed among a tiny group of founders. The power laws become steeper over time. During the final two intervals, income inequality declines greatly, falling in line with the levels seen in typical economies. This corresponds to the emergence of a new class of small stakeholders, dubbed "early adopters." The process is illustrated in a succession of 2D plots (left), with respect to a timeline indicating the launch of bitcoin, the pizza sale event (after roughly sixteen months), and the parity event (after roughly two years). Bitcoin's value appreciates rapidly between the latter two events. We also exhibit this phenomenon in a single 3D plot (upper right). We also show the dynamics of the Pareto exponent (right, middle). Finally, we exhibit a comparison between the market price of bitcoin and the number of agents required to account for 50% (blue) and 70% (red) of bitcoin mined. Although the early adopters tended to mine very little bitcoin per capita, their arrival correlates with the rapid transition in bitcoin's price from <1¢ to $1 (right, bottom).

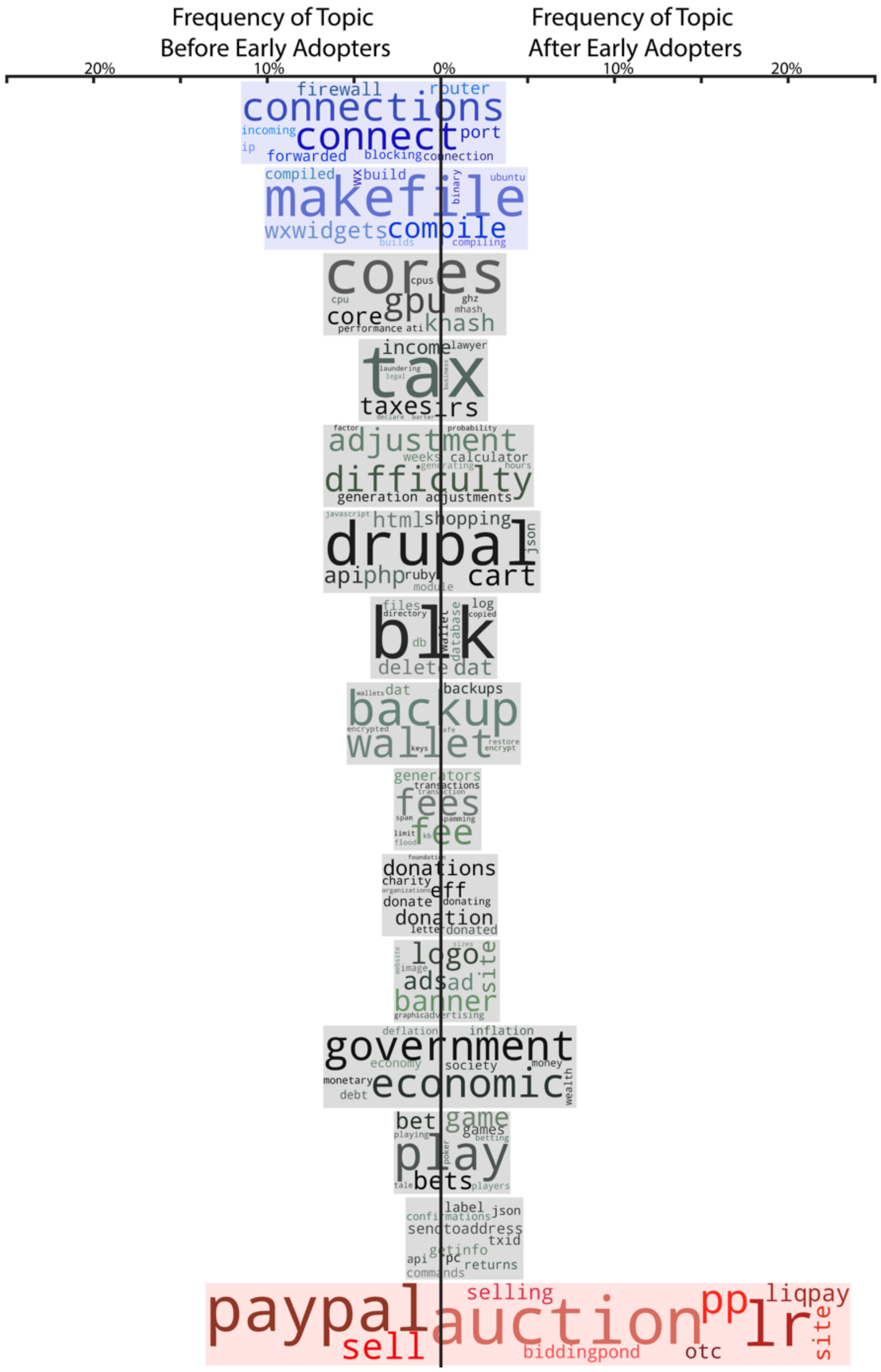

**Fig 3. Bitcoin's founders were focused on the technical challenge of getting bitcoin to work, whereas the much larger population of early adopters were focused on using bitcoin for transactions.** Here we exhibit the top 15 topics for bitcointalk.org during our period of study. Topics were obtained using top2vec. The bar chart displays topic frequency during two periods, November 22nd, 2009 – May 29th, 2010 (left) and May 29th, 2010 – February 9th, 2011 (right). The 10 most characteristic words for each topic are superimposed on the bars. Topics are ordered from top to bottom by the difference in frequency between period 1 and period 2. The two topics that are most biased towards the founder period (shown in blue) relate to the implementation of the core software and peer-to-peer network. The topic most biased towards the early adopter period (red) relates to the use of bitcoin for transactions.

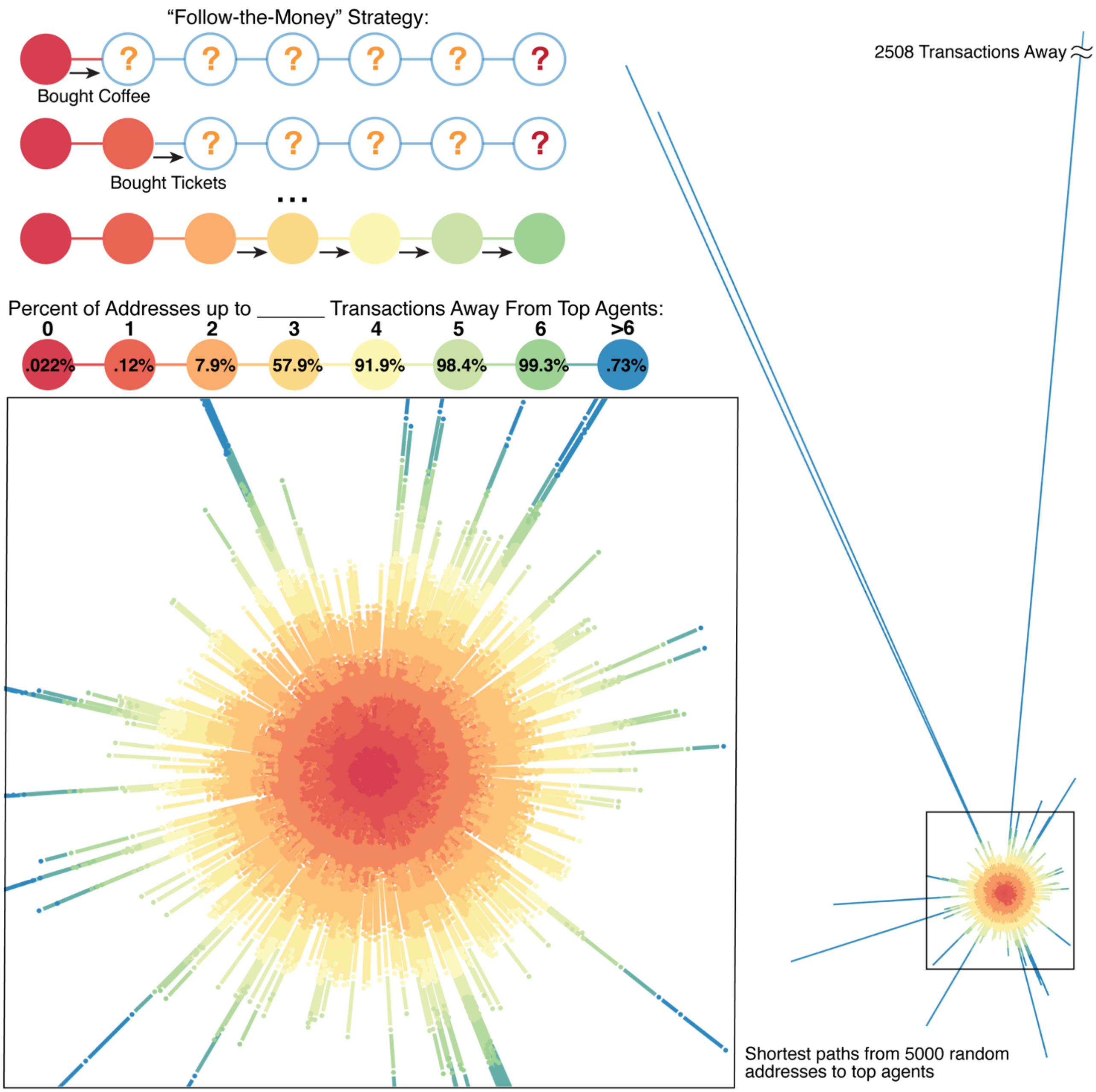

**Fig 4. Over 99% of bitcoin addresses fall within 6 transactions of bitcoin's 64 top agents.** Sixty-four top agents collectively mined most of the bitcoin generated prior to parity ($1=Ƀ1). We examine the centrality of these agents in the contemporary bitcoin transaction network (spanning 1/3/09 through 12/18/17) by constructing a network in which the nodes are addresses and the edges are transactions. (This network is too large to show.) *Upper-left*: Short paths in the bitcoin transaction network can introduce vulnerabilities to the privacy of bitcoin users, making it easier to "follow-the-money" from a known node (red, left) through a chain of transactions involving intermediate agents (orange question mark) in order to determine the identity of a target address (red question mark). *Right*: We select 5000 bitcoin addresses at random. For each address, we identify the shortest path between that address and an address controlled by one of the 64 top agents. The resulting 5000 paths in the bitcoin transaction network are arranged, radially, around

the top agent addresses (bright red). In this visualization, each step along the network (i.e., a transaction) is roughly uniform in length, and nodes and edges are colored based on distance to a top agent address. A handful of exceptional addresses cannot be linked to a top agent address by means of a short path. In fact, one address required a chain of over 2500 transactions. *Lower-left*: Zooming in on the network, we find that the overwhelming majority of bitcoin addresses (>99%) are within 6 steps of a top agent address. The table atop the panel shows the fraction of addresses that can be linked to at least one top agent using a path whose length is a given number of transactions. Nearly all addresses can be linked to a top agent by a path of 6 transactions or fewer.

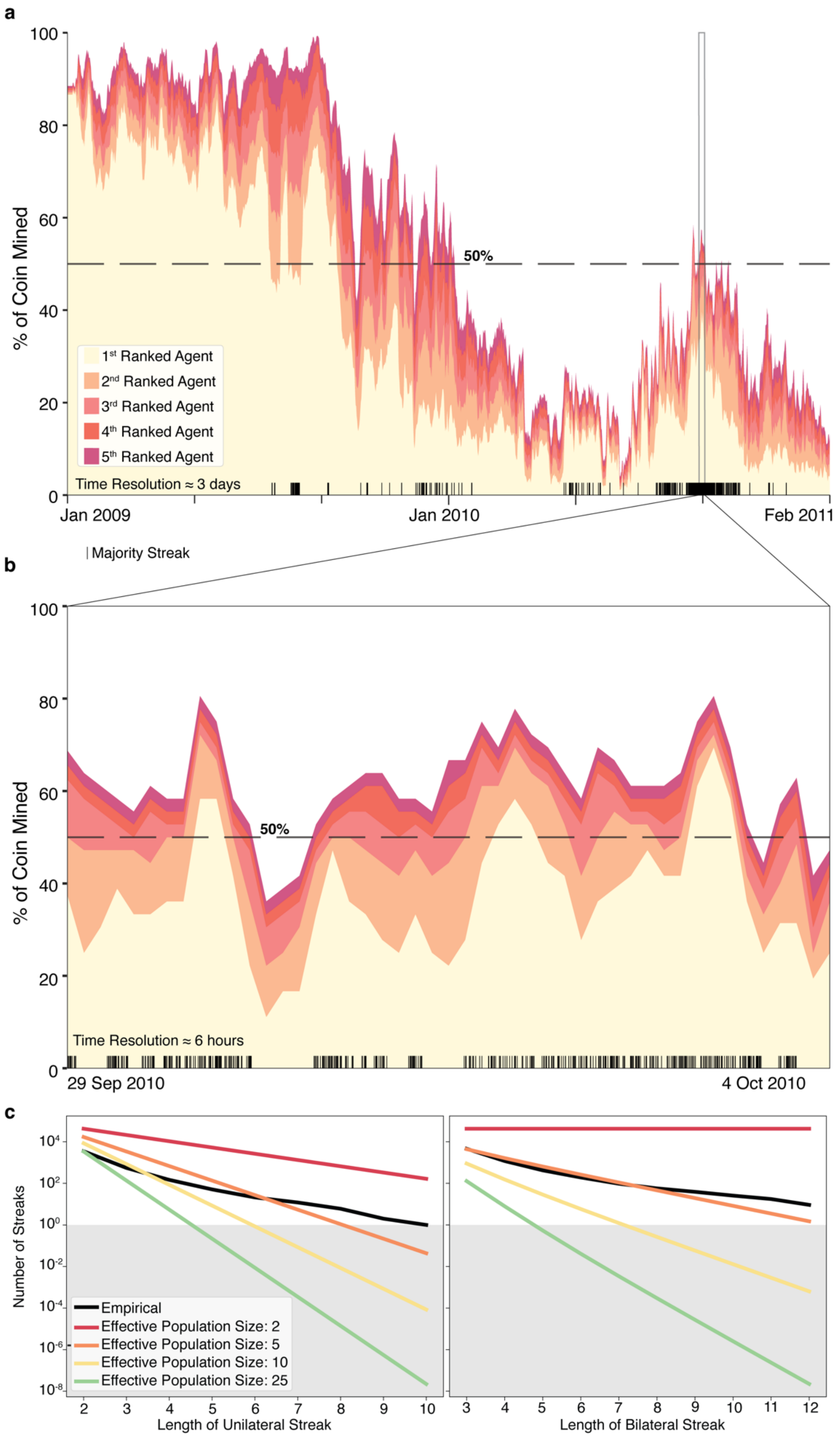
a
100
80
60
40
20
0
% of Coin Mined
50%
1st Ranked Agent
2nd Ranked Agent
3rd Ranked Agent
4th Ranked Agent
5th Ranked Agent
Time Resolution ≈ 3 days
Jan 2009
Jan 2010
Feb 2011
Majority Streak
b
100
80
60
40
20
0
% of Coin Mined
50%
Time Resolution ≈ 6 hours
29 Sep 2010
4 Oct 2010
c
Number of Streaks
Empirical
Effective Population Size: 2
Effective Population Size: 5
Effective Population Size: 10
Effective Population Size: 25
Length of Unilateral Streak
Length of Bilateral Streak

**Fig 5. Agents repeatedly decline to perform 51% attacks despite having sufficient computational resources.** A well-known vulnerability of the Bitcoin protocol is that, if any one agent has over 50% of the computational power of the bitcoin mining network, the agent can unilaterally enrich themselves by validating fraudulent transactions. This is called a 51% attack. (A) Here we show the 5 agents in the bitcoin network employing the most computational power at a given time, and the fraction of computational power held by each. Each datapoint reflects an interval of 432 blocks, or roughly 3 days. Note that the specific agent with a particular rank may not be the same in different time intervals. Until December 2009, the agent with the most computational power (at the time this was Agent #1, Satoshi Nakamoto) had sufficient resources to perform a 51% attack. Subsequently, resources are more evenly distributed, until, roughly a year later, the use of GPUs enables massive accelerations in bitcoin mining. Black tick marks at the bottom of each plot indicate streaks of six blocks or more in which a single agent mined most of the bitcoin, suggesting that the agent would have been able to perform a majority attack. Even during periods when the blockchain was – overall – relatively decentralized, such streaks are frequently seen. (B) During a weeklong period from September 29, 2010 and October 4, 2010, the agent with the most computational power (no longer Satoshi Nakamoto, but instead Agent #2) has enough resources to perform a 51% attack during several 6+ hour long windows. Agent #2 declines to perform a 51% attack, and instead continues to exhibit cooperative behavior. Each datapoint reflects an interval of 36 blocks, or roughly six hours. In principle, a 51% attack can be performed as long as the miner knows, with high probability, that they will successfully mine the next block. Again, black tick marks at the bottom of the plot indicate streaks of six blocks or more in which a single agent mined most of the bitcoin. (C) We estimate the effective population size of the decentralized bitcoin network by counting the frequency of streaks in which all blocks are mined by one agent (bottom-left) or two agents (bottom-right). These are compared to the expected values for idealized networks comprising $P$ agents with identical resources. The comparisons suggest an effective population size of roughly 5, a tiny fraction of the total number of participants. The grayed out region corresponds to an expected value of less than one streak; for instance, given an effective population of 25 agents, a unilateral streak of length 6 should never be observed in the two-year interval we studied. In fact, we observe 21 such streaks. Bitcoin is streakier than a fair die, but not as streaky as a fair coin.

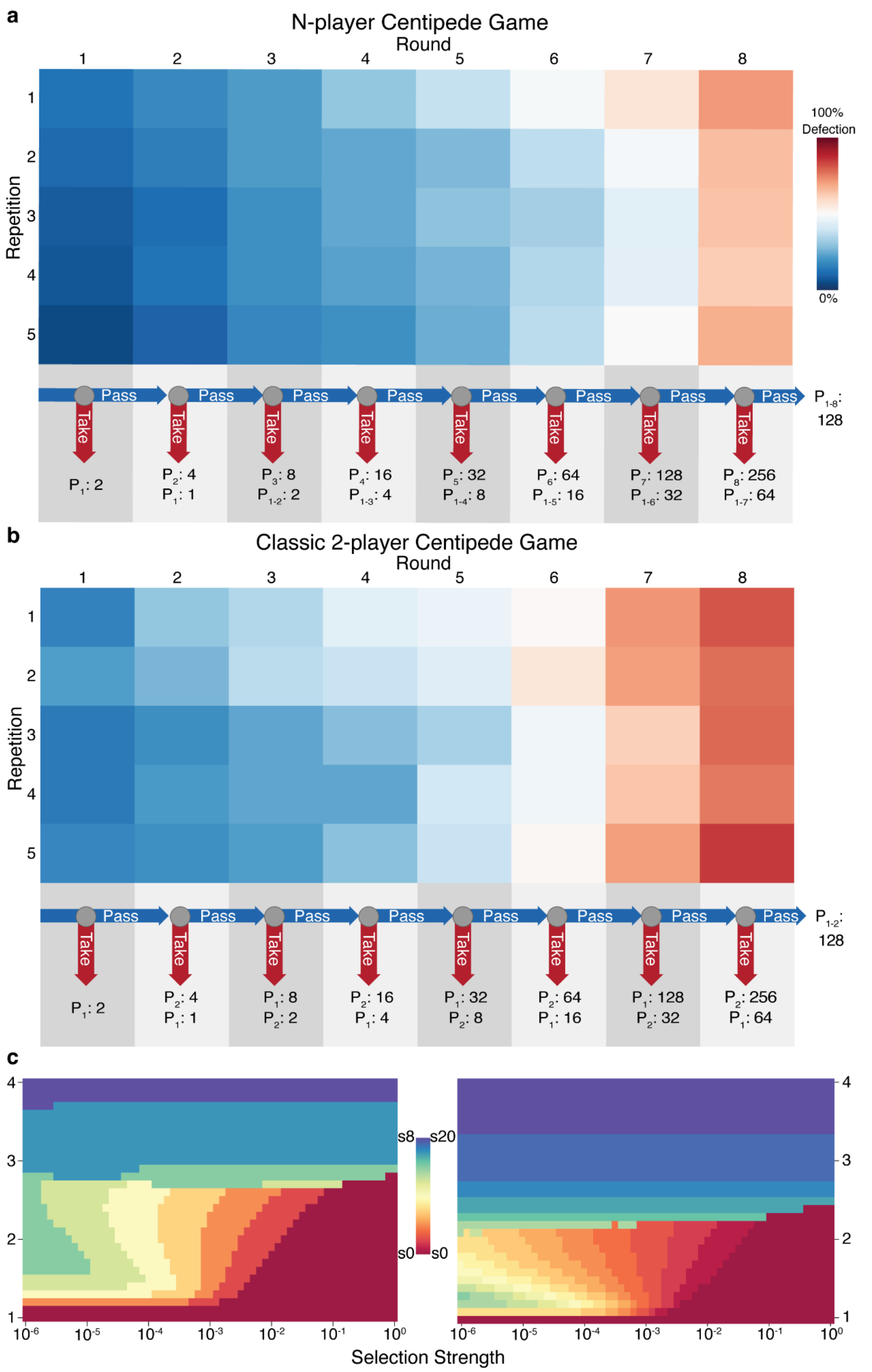
a
N-player Centipede Game
Round
1 2 3 4 5 6 7 8
Repetition
1 2 3 4 5
100% Defection
0%
Pass
Take
$P_{1-8}$: 128
$P_1$: 2
$P_2$: 4 $P_1$: 1
$P_3$: 8 $P_{1-2}$: 2
$P_4$: 16 $P_{1-3}$: 4
$P_5$: 32 $P_{1-4}$: 8
$P_6$: 64 $P_{1-5}$: 16
$P_7$: 128 $P_{1-6}$: 32
$P_8$: 256 $P_{1-7}$: 64
b
Classic 2-player Centipede Game
Round
1 2 3 4 5 6 7 8
Repetition
1 2 3 4 5
Pass
Take
$P_{1-2}$: 128
$P_1$: 2
$P_2$: 4 $P_1$: 1
$P_1$: 8 $P_2$: 2
$P_2$: 16 $P_1$: 4
$P_1$: 32 $P_2$: 8
$P_2$: 64 $P_1$: 16
$P_1$: 128 $P_2$: 32
$P_2$: 256 $P_1$: 64
c
1 2 3 4
s8 s20
s0 s0
$10^{-6}$ $10^{-5}$ $10^{-4}$ $10^{-3}$ $10^{-2}$ $10^{-1}$ $10^{0}$
Selection Strength

**Fig 6. We model cooperation among bitcoin's agents as a variant of the classic Centipede game, and find that playing in an anonymous group enhances the level of cooperation.** (A) We modeled the social dilemma faced by bitcoin's founders as an 8-player Centipede game, whose game diagram is shown below. In each round, an anonymous new player enters the game, and can choose to cooperate ("pass"), doubling the payoff for all players, or defect ("take"), quadrupling their personal payoff while ending the game. The last player is incentivized to take, and has no possibility of reciprocity; similarly, by backwards induction, the only subgame-perfect Nash equilibrium is to take on the first round. Above, we show the likelihood that an 8-player Centipede game stops before the $N^{th}$ pass. Although game theory predicts that everyone should defect in the first round, we observe high levels of cooperation, even in the very last round of the game. More cooperation is seen as the game is repeated. The fraction of people who take on the first round declines as the game is repeated, indicating that players learn to avoid the subgame-perfect Nash equilibrium. (B) We compare the 8-player Centipede game to a classic two player Centipede game. Again, the game diagram is shown below. The payoffs are identical, but instead of 8 players, there are only two, alternating players. Levels of cooperation are consistently lower in every round of every repetition. Players are more likely to cooperate with an anonymous new player than with someone who has been consistently cooperative. (C) Left: We used evolutionary game theory to explore the behavior of a finite population playing an 8-player Centipede game under the Moran (birth-death) process. The possible strategies are to never cooperate (the subgame-perfect Nash equilibrium, s0), to cooperate till round 1 but defect thereafter (s1), and so on; strategy s8 always cooperates. We plot the strategy that is most commonly observed in the long run as a function of selection strength and cooperation benefit. The subgame-perfect Nash equilibrium dominates when selection is strong and the benefit of cooperation is small. As selection becomes weaker and cooperation becomes more beneficial, more cooperative strategies ascend. When the benefit of cooperation is sufficiently high, the most common strategy is highly cooperative, regardless of selection strength. Right: The results for a 20-player Centipede game. As the anonymous group gets larger, the tendency towards cooperation becomes increasingly strong. Even when the benefit of cooperation is not as large, highly cooperative strategies are the most common.

## Supplemental Figures:

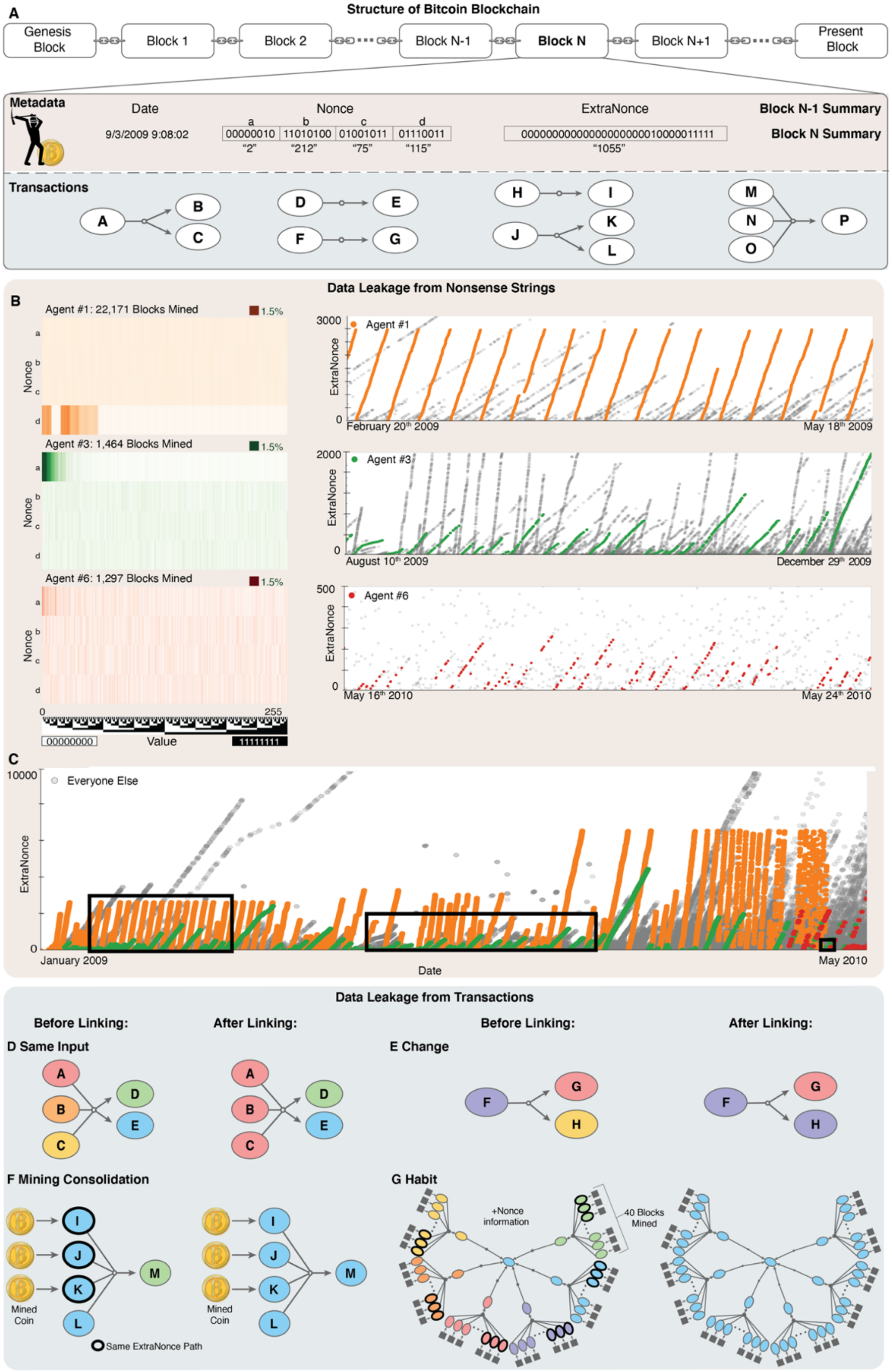

**Fig S1. Data leakage from the blockchain compromises bitcoin's anonymity.** (**A**) The blockchain is a permanent ledger of every bitcoin transaction that has ever taken place. It is made up of individual blocks, or bundles of transactions (bottom, representing as graphs linking sender and receiver addresses), coupled with associated metadata (top), such as the block date. To make the blockchain forgery-proof, each block refers to the contents of prior blocks. New blocks can only be appended when a particular random hash function (which is run on the contents of the block) yields an extremely small value. To find acceptable blocks, bitcoin "miners" iterate through possible values of two nonsense string fields: the "nonce" and the "extranonce". Both the transactions and the nonsense fields can give rise to data leakage by which it becomes possible to link addresses controlled by the same individual. (**B**) The way an agent generates their nonce and extranonce fields is an identifying characteristic. (Left) The nonce heat maps show the distribution of nonce values over every block mined by an agent. Some agents use the field as a counter, exhibiting a pattern like Agent #3. Other agents exhibit a more uniform distribution, like Agent #6. Still others, like Agent #1, employ a limited range of seemingly arbitrary nonce values. (Right) Extranonce scatter plots showing how the extranonce values for an individual agent increment over time. The short line fragments of positive slope beginning on the x-axis show mining sessions. Notice how these lines do not overlap. This is due to an agent mining for a while, stopping and restarting their client, and then starting to mine again with their extranonce back at 0. The agent of interest is shown in color; other agents are shown in gray. (**C**) Here, we see the extranonce over time for all agents during the first 17 months of bitcoin mining. The three agents from the previous panel are highlighted in color. (**D**) The bitcoin transaction graph, in which the nodes are bitcoin addresses and the edges represent bitcoin transactions, also facilitates multiple forms of address-linking. Here, addresses A, B, and C are nominally different agents (shown using different colors). But, because they are all inputs to the same transaction, we can deduce that they correspond to a single agent (and are therefore shown in the same color after the address-linking procedure has been carried out, as shown on the right). (**E**) In a different transaction, a new address, H, is created to hold change (indicated using green outline). This implies that a single agent controls G and H. (**F**) Analysis of extranonce data reveals that addresses I, J, and K correspond to a single bitcoin mining session. Consolidating the results of such sessions into a single address is extremely common. Thus, when the contents I, J, K, and L are transferred to M, we deduce that a single agent controls all 5 nodes. (**G**) An idiosyncratic mining consolidation pattern. This agent repeatedly transfers the results of 40 successful mining events into an intermediate address. Then, the contents of the intermediate addresses are combined into a single address.

| Inputs | | Outputs | |
|---|---|---|---|
| Addresses | Values | Addresses | Values |
| Address A | 2 BTC | Address D | 6 BTC |
| Address B | 3 BTC | Address E | 4 BTC |
| Address C | 5 BTC | | |

**Fig S2. A Toy Transaction**
In this toy transaction, addresses A, B, and C, are sending two, three, and five bitcoin, respectively. Addresses D and E will receive six and four bitcoin, respectively. The sum of input values must be equal to or greater than the sum of the output values. Every input pair must reference a previous unspent output pair, with the notable exception of "coinbase" transactions, which create bitcoin. This means that address A must have been sent exactly two bitcoin in a past transaction. The same is true for addresses B and C, they must have been sent exactly three and five bitcoin, respectively, at some point in the past.

```
01000000010000000000000000000000000000000
000000000000000000000000000000000ffffffff
4d04ffff001d0104455468652054696d657320303
32f4a616e2f32303039204368616e63656c6c6f72
206f6e206272696e6b206f66207365636f6e64206
261696c6f757420666f722062616e6b73ffffffff
0100f2052a01000000434104678afdb0fe5548271
967f1a67130b7105cd6a828e03909a67962e0ea1f
61deb649f6bc3f4cef38c4f35504e51ec112de5c3
84df7ba0b8d578a4c702b6bf11d5fac00000000
```

**Fig S3. The raw genesis block**.
The first bitcoin block is shown in hexadecimal, beginning with the block header. The extranonce is shown in the blue box.

| Block # | Bitcoin Mined | Date Mined (UTC) | Extranonce |
|---|---|---|---|
| 1 | 50 | Jan 3rd 2009, 18:15:05 | 00000100 (binary data) = 04 (hex) = 4 (decimal) |
| 2 | 50 | Jan 9th 2009, 02:54:25 | 00000100 (binary data) = 04 (hex) = 4 (decimal) |
| 3 | 50 | Jan 9th 2009, 02:55:44 | 00001011 (binary data) = 0b (hex) = 11 (decimal) |
| 4 | 50 | Jan 9th 2009, 03:02:53 | 00001110 (binary data) = 0e (hex) = 14 (decimal) |
| 5 | 50 | Jan 9th 2009, 03:16:28 | 00011010 (binary data) = 1a (hex) = 26 (decimal) |
| 6 | 50 | Jan 9th 2009, 03:23:48 | 00100000 (binary data) = 20 (hex) = 32 (decimal) |
| 7 | 50 | Jan 9th 2009, 03:29:49 | 00100011 (binary data) = 23 (hex) = 35 (decimal) |
| 8 | 50 | Jan 9th 2009, 03:39:29 | 00101011 (binary data) = 2b (hex) = 43 (decimal) |
| 9 | 50 | Jan 9th 2009, 03:45:43 | 00101100 (binary data) = 2c (hex) = 44 (decimal) |
| 10 | 50 | Jan 9th 2009, 03:54:39 | 00110100 (binary data) = 34 (hex) = 52 (decimal) |

**Figure S4. The first 10 bitcoin blocks together with some associated metadata.**
This table shows the block number, date mined, and decimal extranonce value for each of the first ten blocks mined. Note we use the term “block #” in place of block height as a more general term; Block number is block height plus one. For the extranonce, the raw blockchain data is shown in binary; its interpretation in hexadecimal and decimal is also shown, for convenience.

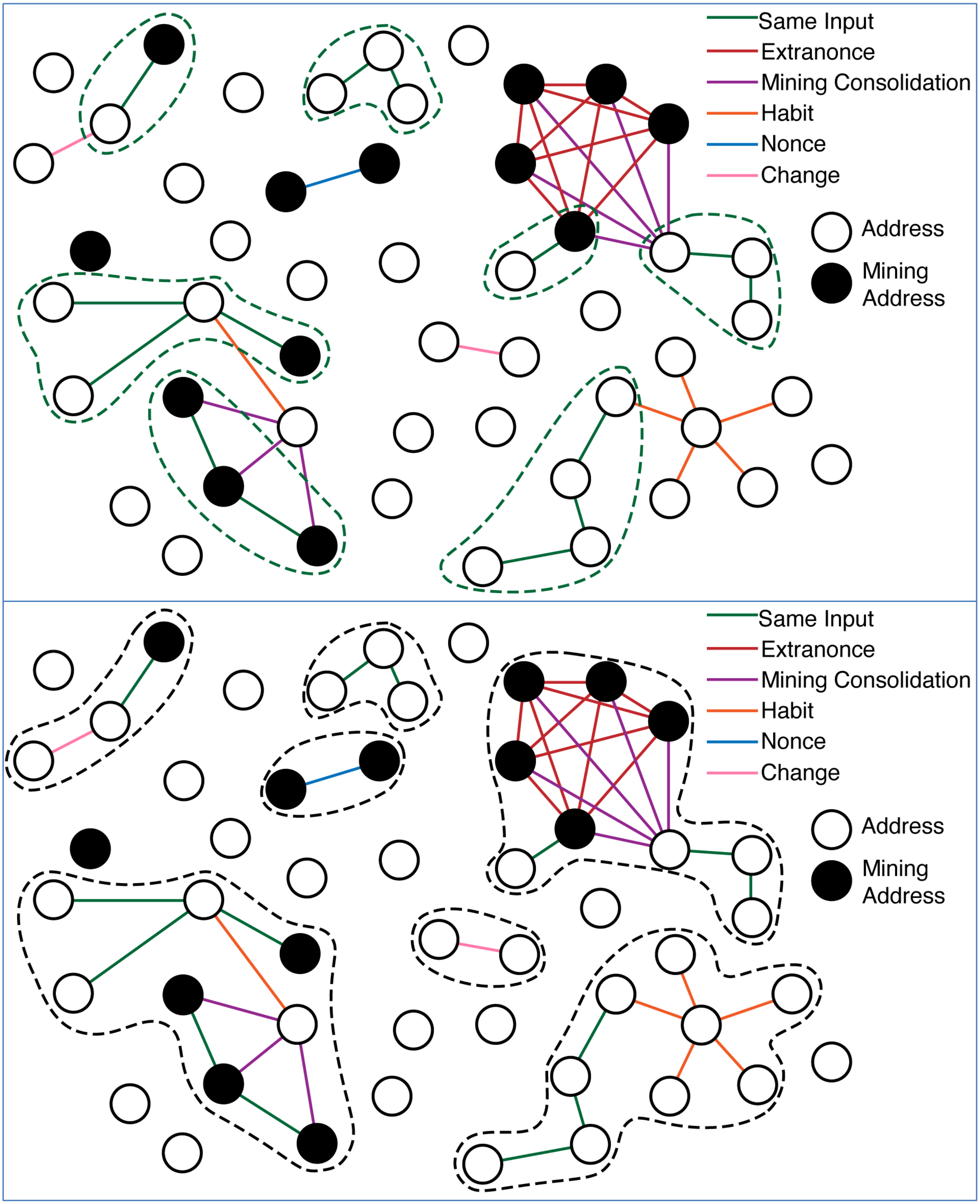

**Fig S5. Address-linking is a strategy for identifying individual agents in the bitcoin blockchain.** This schematic shows how address linking can be used to compromise the identity masking features of the bitcoin blockchain. Bitcoin is held by addresses, illustrated here as nodes of a graph. In both panels, we show two different types of nodes. Users of bitcoin, referred to as agents throughout this manuscript, can generate as many addresses as they want using the bitcoin software (white nodes). These addresses, which function as pseudonyms for the user, enable them to mask their activity. Mining addresses, shown in black, are generated by the bitcoin protocol itself when a block is successfully mined. Control of mining addresses is given to the miner of that block; as such, the mining address serves as still-another pseudonym. Address-linking strategies undermine this pseudonymity by identifying pairs of addresses that are held by the same user. Here, we illustrate address-links as colorful edges, where the color indicates the type of evidence that supports the link. Examples include the use of two addresses as an input to a transaction ("same input" links, shown in green) or the fact that two addresses lie along the same extranonce trajectory ("extranonce" links, shown in red). Consequently, connected components in this graph correspond to putative agents, insofar as every address in the connected component can be identified with every other address via an evidence chain. Of course, depending on which evidence is employed (i.e., using edges with certain colors, rather than all edges), a different agent list will result. In the top panel, agents formed with only same input linking are shown by dashed green line. In the bottom panel, agents formed via our methods are shown in dashed black line.

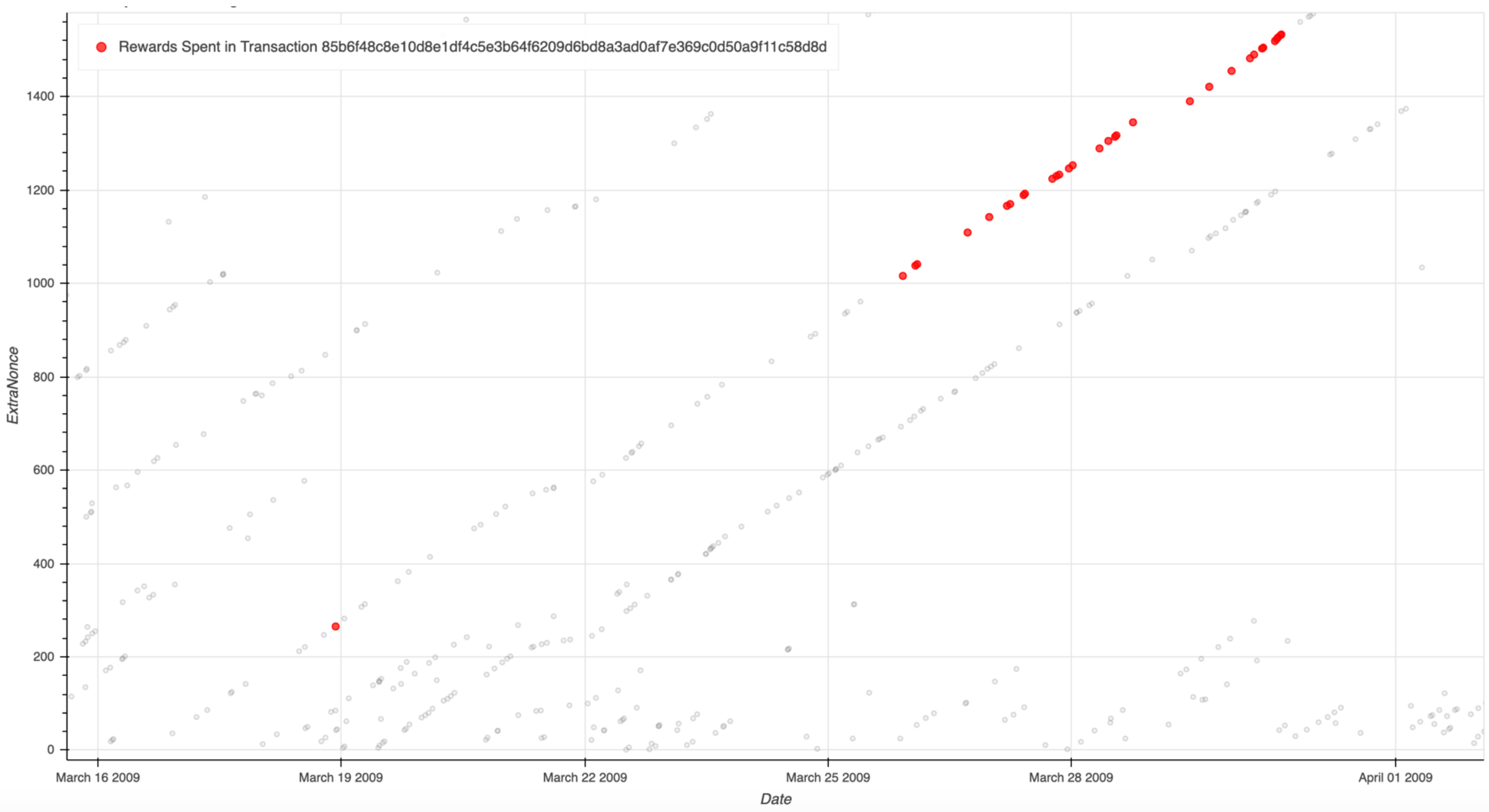
Rewards Spent in Transaction 85b6f48c8e10d8e1df4c5e3b64f6209d6bd8a3ad0af7e369c0d50a9f11c58d8d
1400
1200
1000
800
600
400
200
0
ExtraNonce
March 16 2009
March 19 2009
March 22 2009
March 25 2009
March 28 2009
April 01 2009
Date

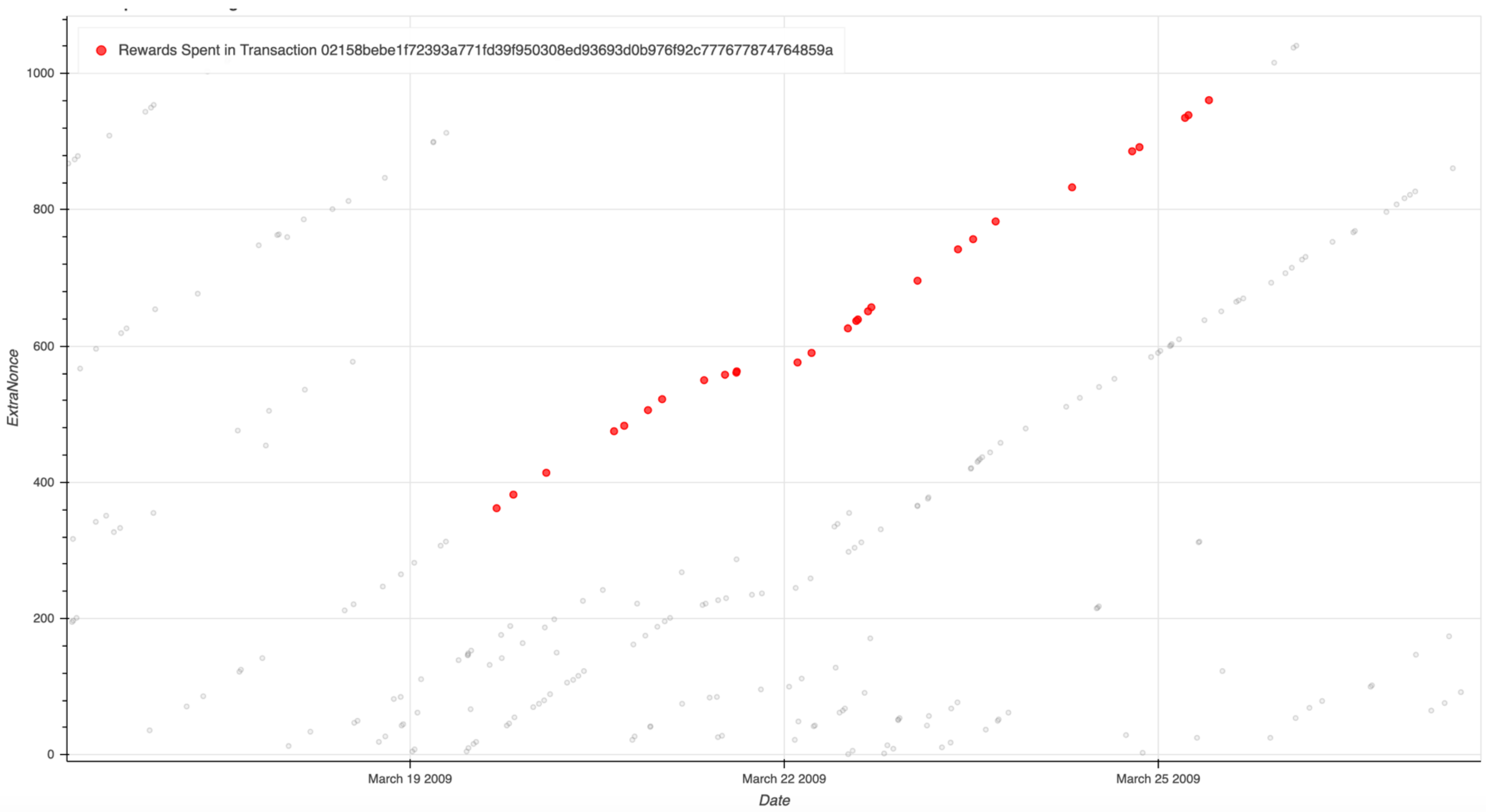
Rewards Spent in Transaction 02158bebe1f72393a771fd39f950308ed93693d0b976f92c777677874764859a
1000
800
600
400
200
0
ExtraNonce
March 19 2009
March 22 2009
March 25 2009
Date

**Fig S6. When we plot extranonce vs. time, mining rewards spent in the same transaction are often colinear.** Both plots show blocks mined by date and their extranonce values for a limited time interval. In the top plot, we select one transaction, and color all of the mining rewards spent in this transaction in red. Other mining rewards mined around the same time are shown in gray. The red rewards are all colinear. This is consistent with a model wherein co-linear blocks derive from the activity of a single agent. The bottom plot shows the same information for a different transaction. Blocks associate with Satoshi Nakamoto are removed from these plots.

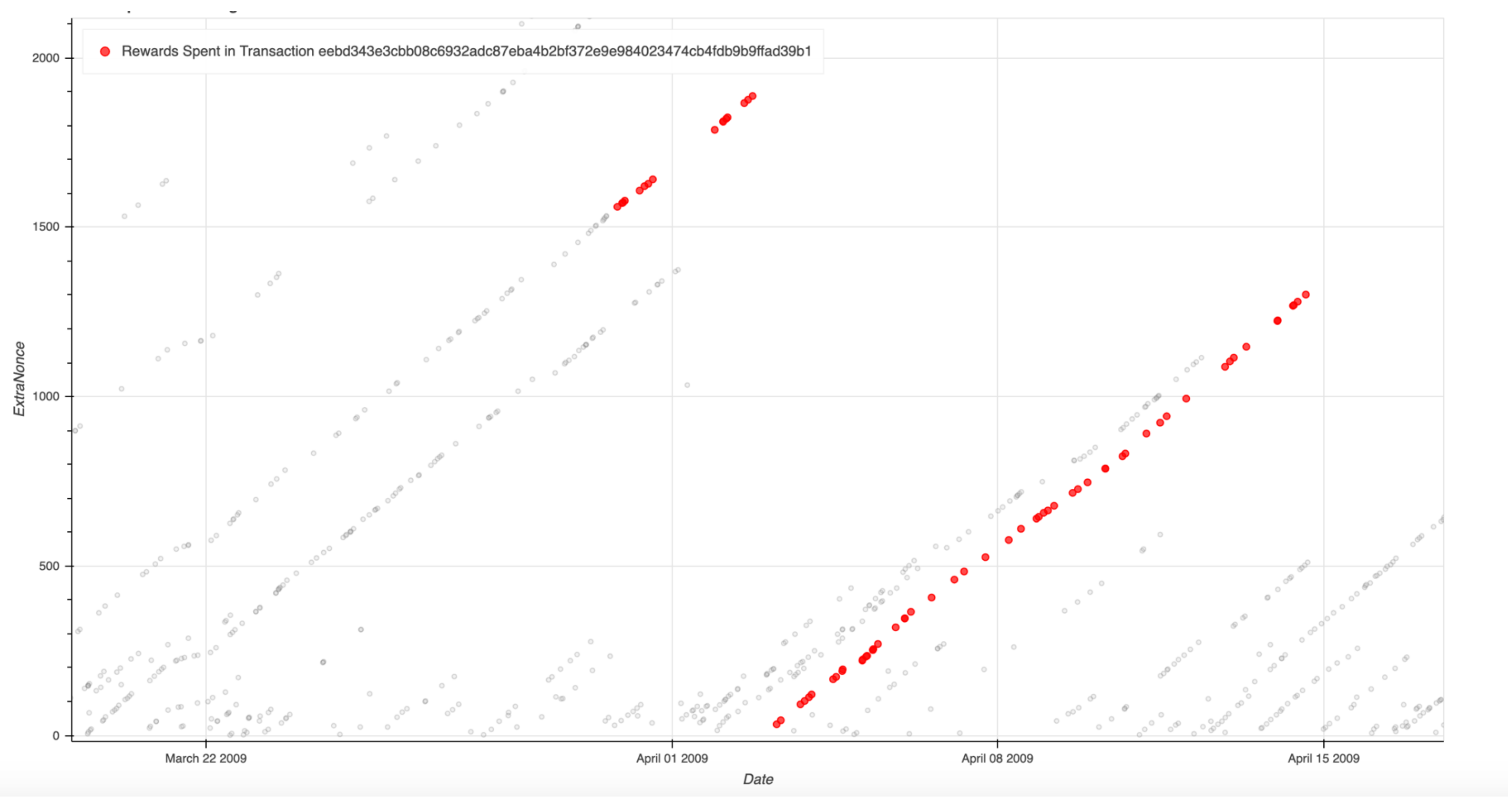
Rewards Spent in Transaction eebd343e3cbb08c6932adc87eba4b2bf372e9e984023474cb4fdb9b9ffad39b1
ExtraNonce
2000
1500
1000
500
0
March 22 2009
April 01 2009
April 08 2009
April 15 2009
Date

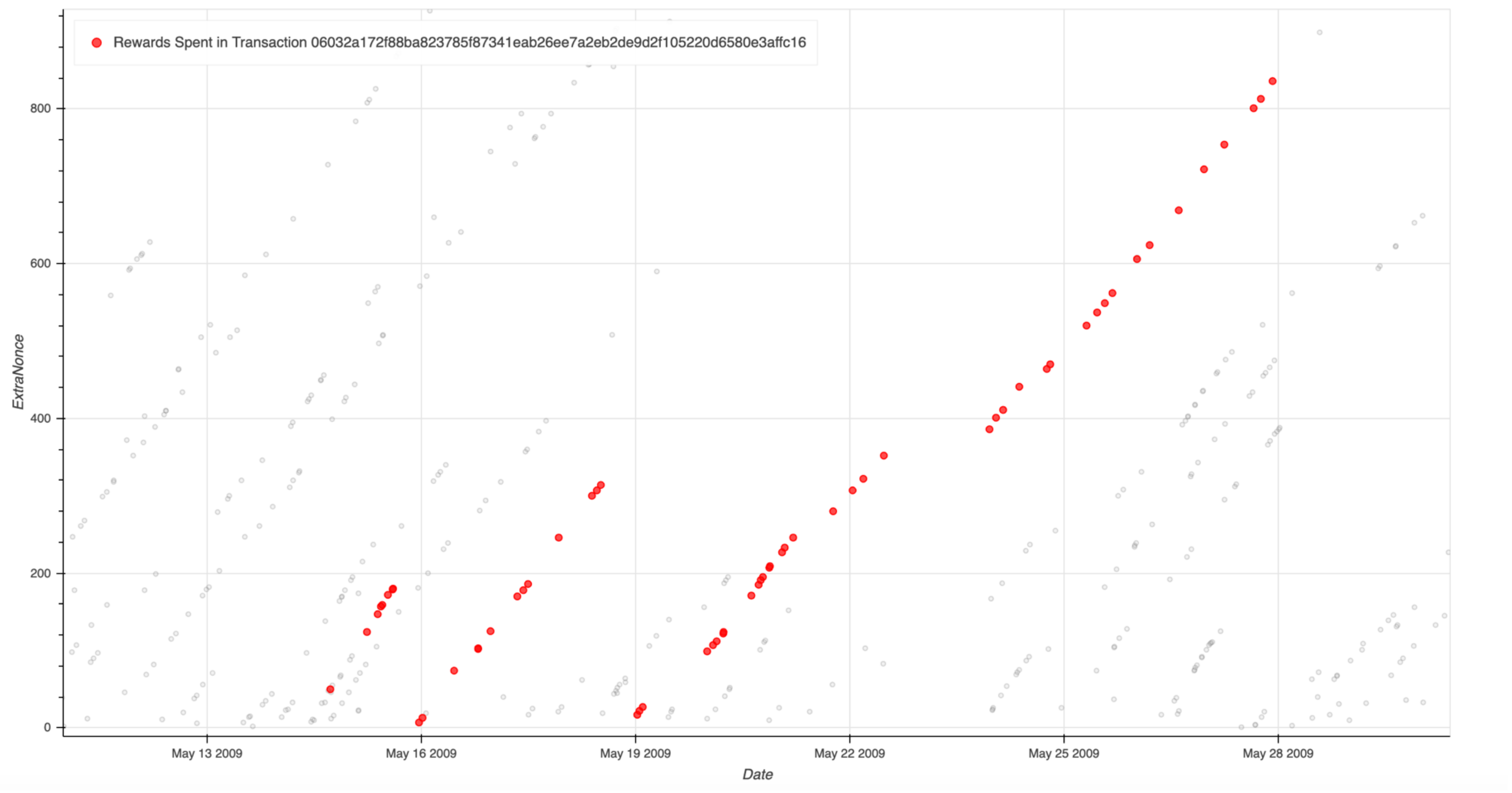
Rewards Spent in Transaction 06032a172f88ba823785f87341eab26ee7a2eb2de9d2f105220d6580e3affc16
ExtraNonce
800
600
400
200
0
May 13 2009
May 16 2009
May 19 2009
May 22 2009
May 25 2009
May 28 2009
Date

**Fig S7. Mining rewards spent in the same transaction often fall in adjacent linear trajectories.** For the top plot we select one transaction and show all of the mining rewards spent in this transaction in red. The blocks fall on adjacent linear trajectories, in that when one trajectory stops, the next one starts. This is consistent with adjacent linear trajectories belonging to a single user. Specifically, the data is consistent with a scenario wherein the user is spending mining rewards that span the end of one mining session and the beginning of the next mining session. When the user stops mining, we see one extranonce trajectory end, and shortly afterward, another trajectory starts near the x axis as the user resumes mining with an extranonce of zero. Other mining rewards mined around the same time are shown in gray. In the bottom plot, the same thing is shown for a different transaction. This plot shows three mining sessions forming three extranonce trajectories. Again, Satoshi Nakamoto is removed from these plots.

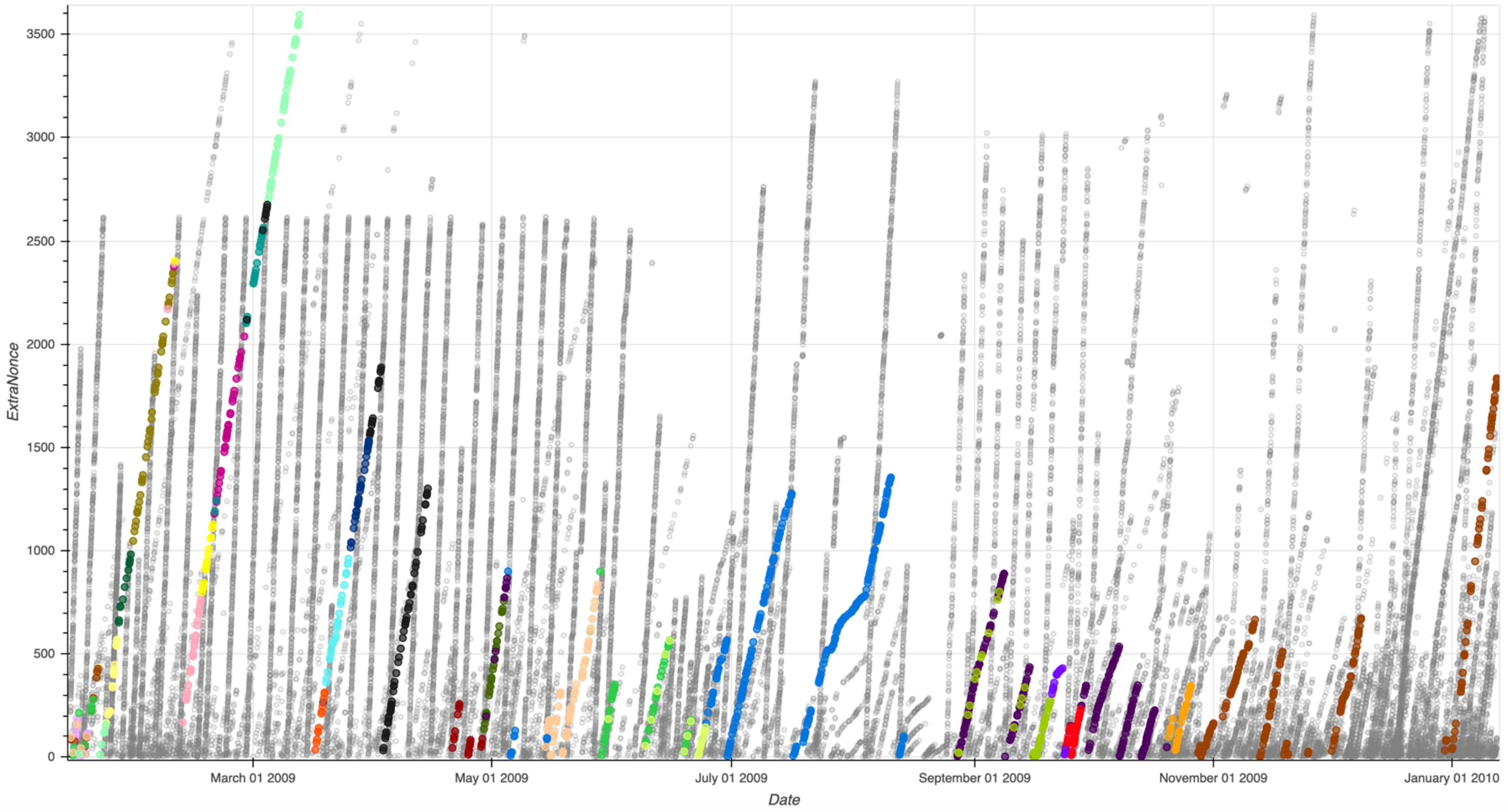

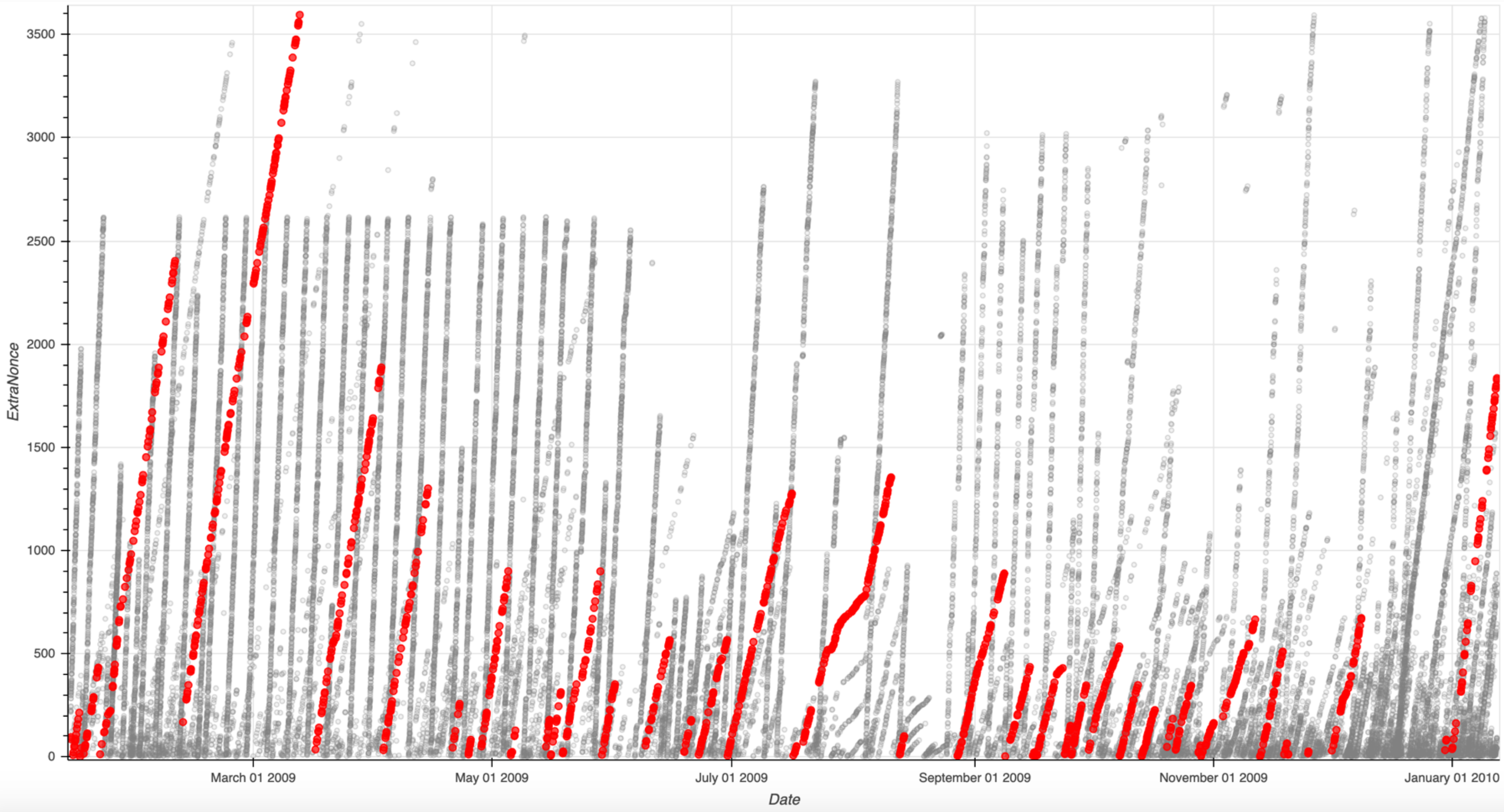

**Fig S8. Agent #4 in our list corresponds to 32 agents obtained using same-input-linking alone.** The top plot shows connected components formed by same-input linking only. Different colors represent different components. The bottom plot shows how these components come together to form our Agent #4, shown in red. For clarity, Satoshi Nakamoto (Agent #1) is removed from these plots. The plots highlight the advantages of address linking using multiple forms of evidence, rather than same-input-linking only.

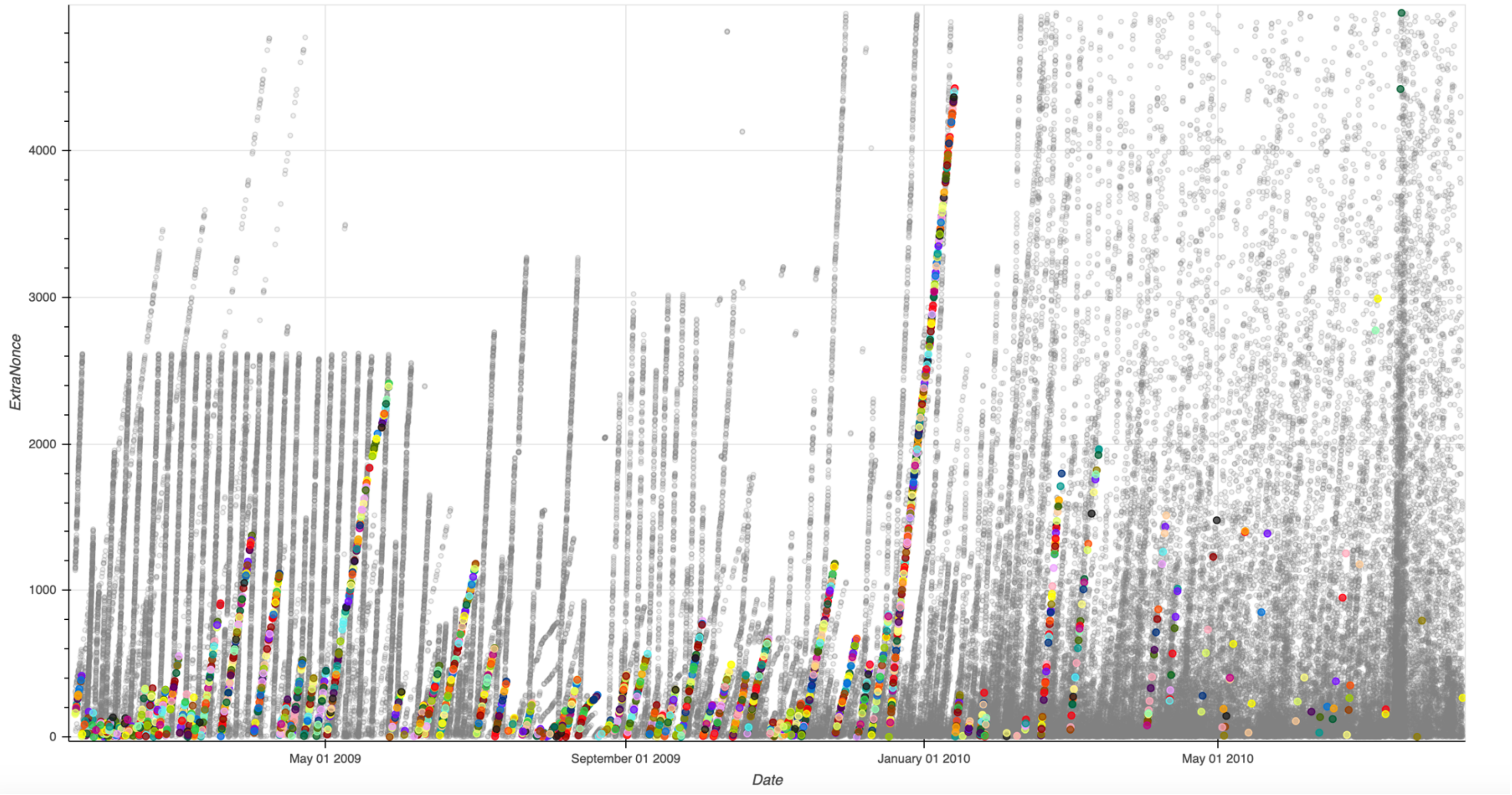

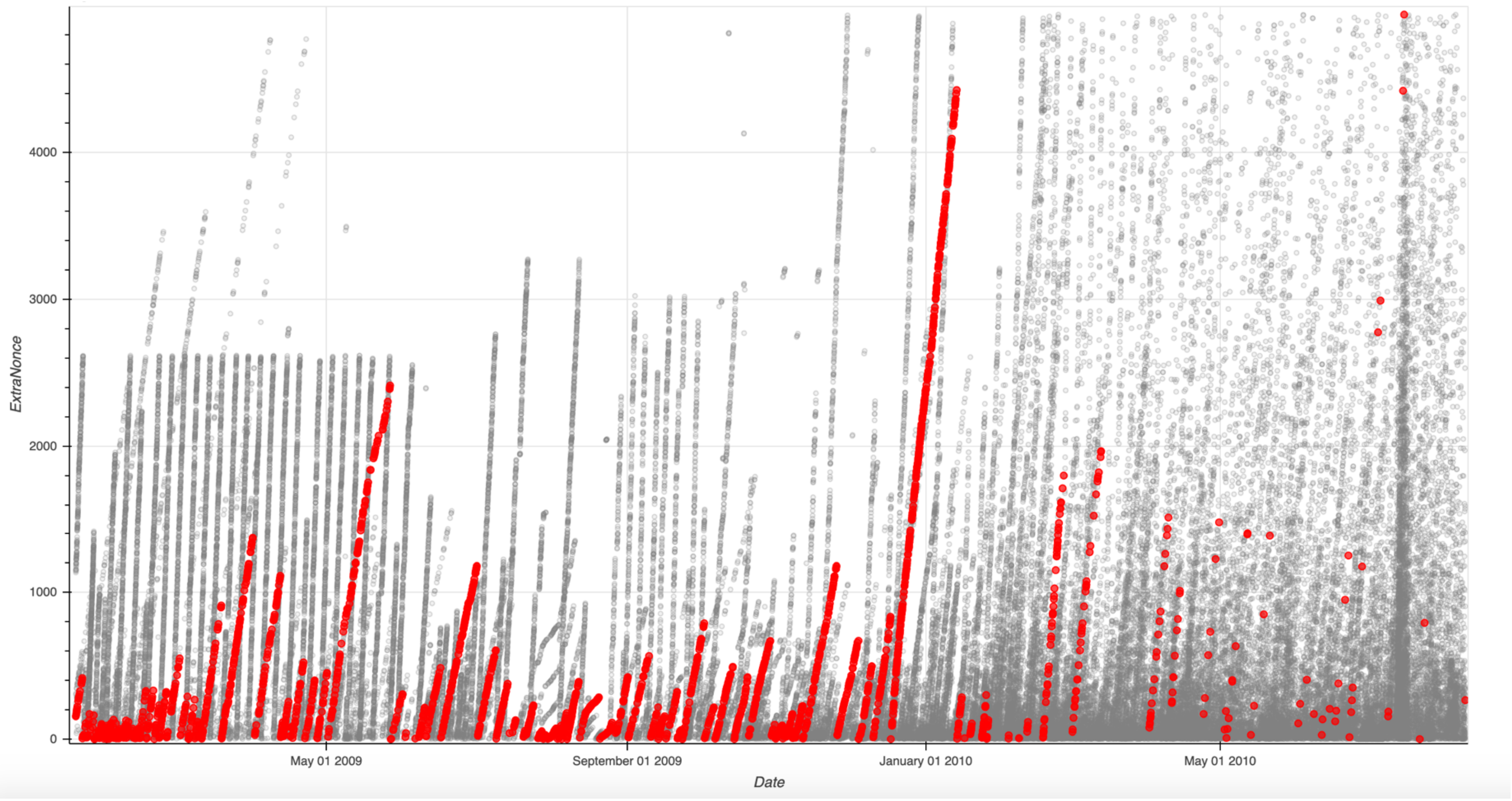

**Fig S9. Agent #3 in our list corresponds to 62 agents obtained using same-input-linking alone.** The top plot shows connected components formed by same input linking only. Different colors represent different components. The bottom plot shows how these components come together to form our Agent #3 shown in red. For clarity, Satoshi Nakamoto (Agent #1) is removed from these plots. The plots highlight the advantages of address linking using multiple forms of evidence, rather than same-input-linking only.

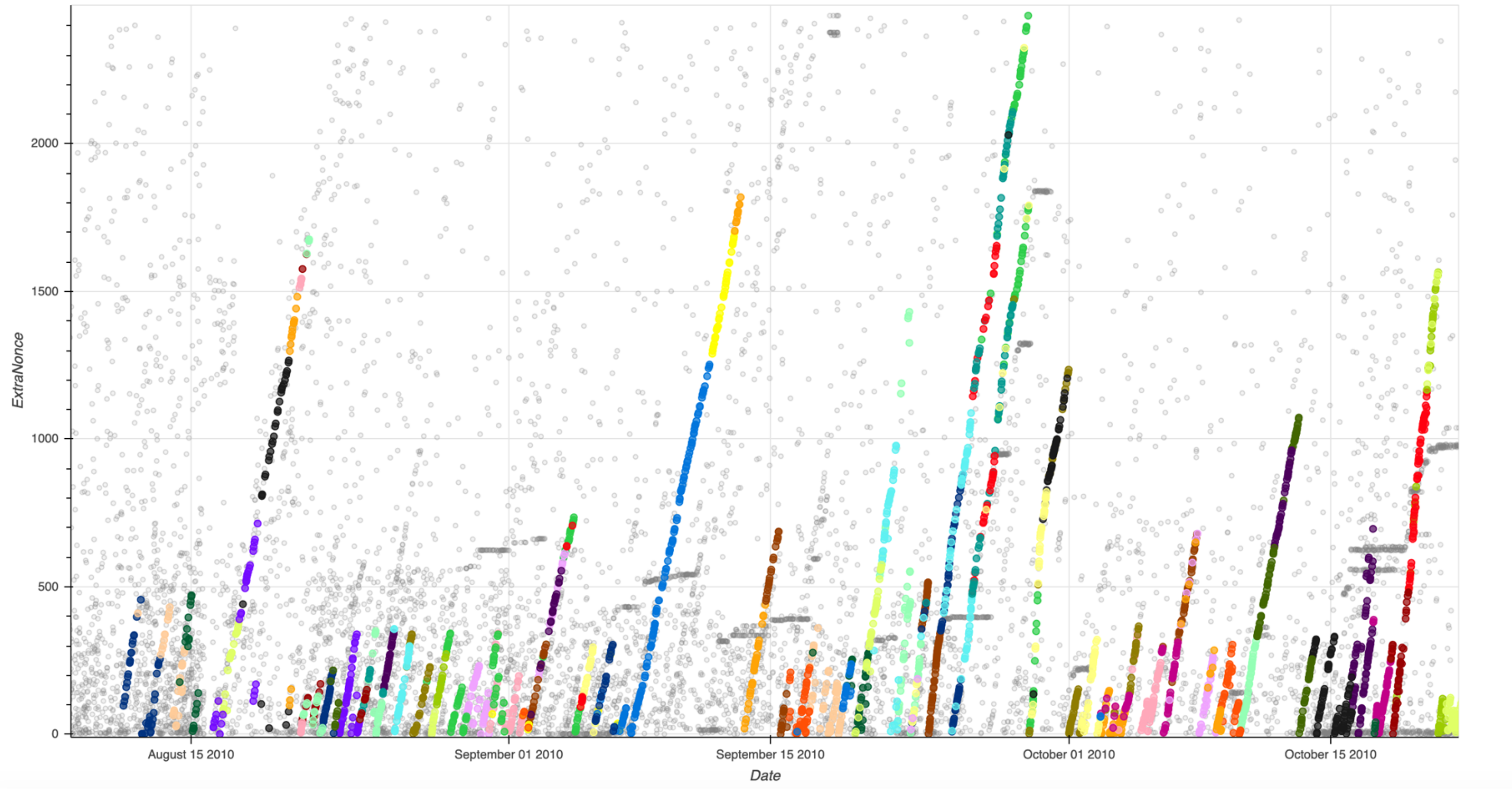

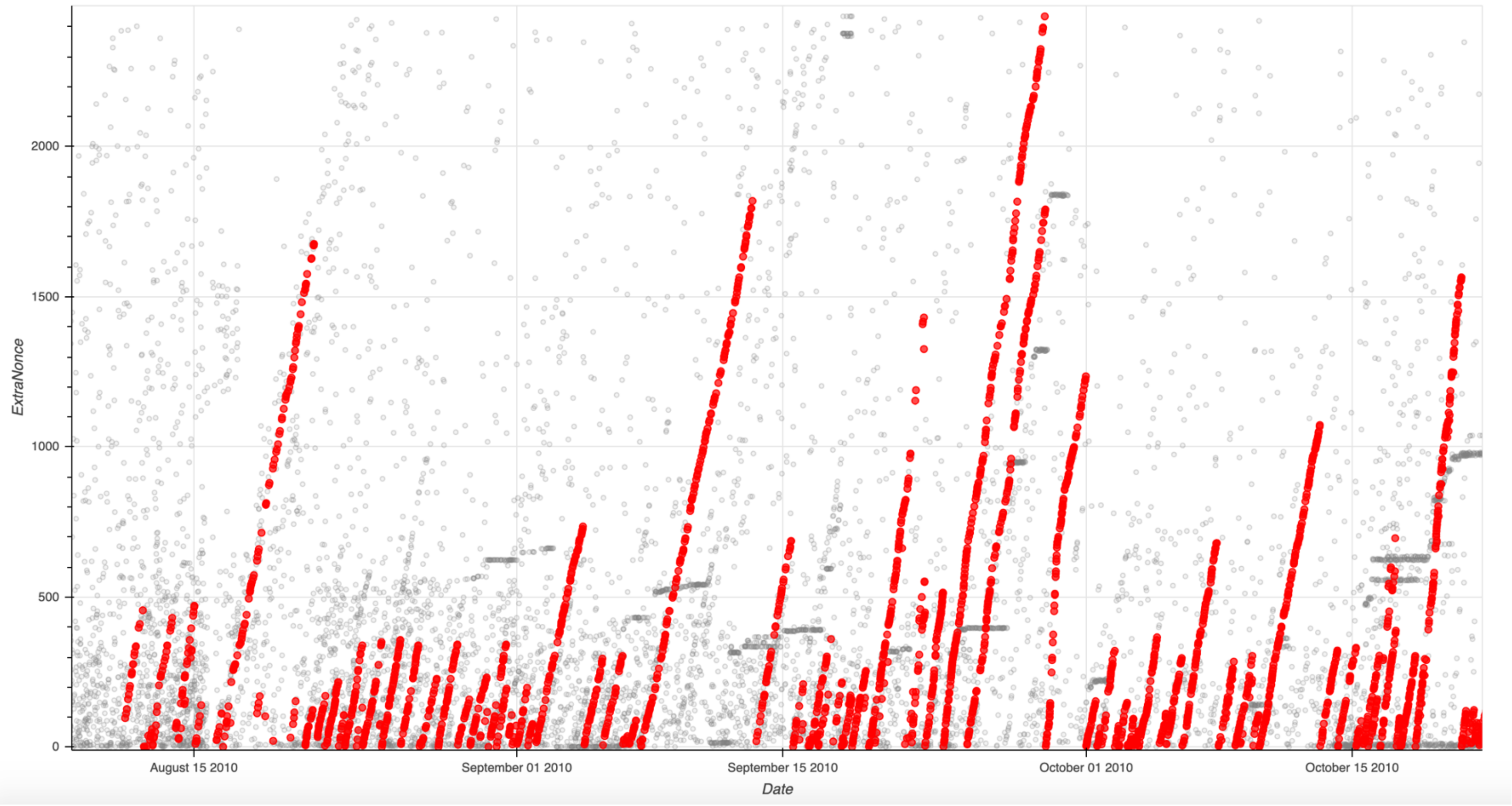

**Fig S10. Agent #2 in our list corresponds to 103 agents obtained using same-input-linking alone.**
The top plot shows connected components formed by same input linking only. Different colors represent different components. The bottom plot shows how these components come together to form our Agent #2, shown in red. For clarity, Satoshi Nakamoto (Agent #1) is removed from these plots. The plots highlight the advantages of address linking using multiple forms of evidence, rather than same-input-linking only. Note, Agent #2 is not shown in its entirety.

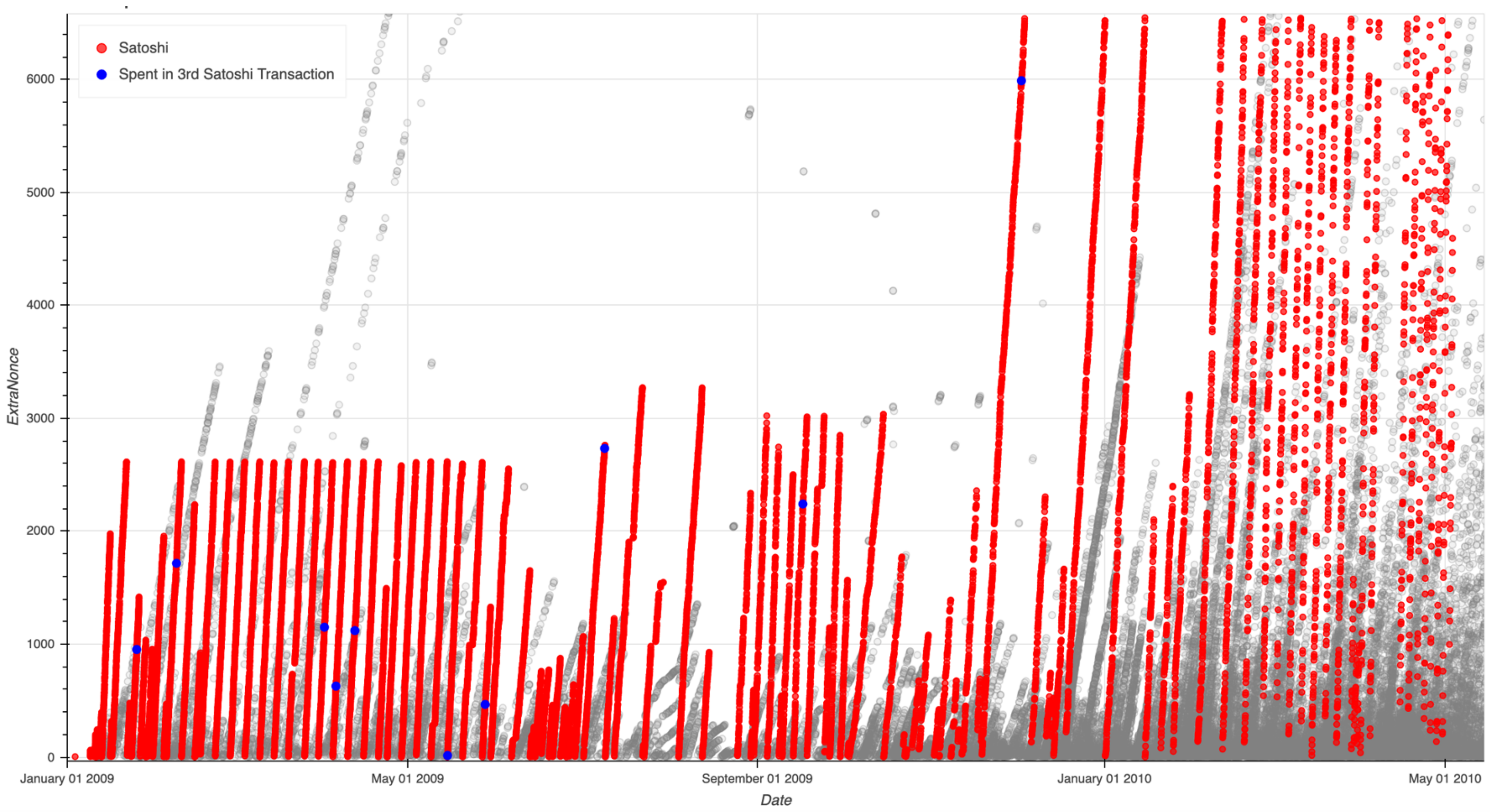

**Fig S11. Satoshi spent ten early mining rewards in a transaction on May 17, 2010.**
Blocks mined by Satoshi are shown in red or blue. Blocks mined by other agents are shown in gray. The Satoshi miner ceases to mine on May 3, 2010. Fourteen days later, Satoshi spends an early ten blocks. These ten blocks are shown in blue and lie across ten different extranonce trajectories. While another agent may mine one block at a similar time and extranonce as Satoshi, it is highly improbable another agent would be able to do this ten times if they did not have knowledge of Satoshi's extranonce values.

| Date | Type | Address | Amount |
|---|---|---|---|
| 2/14/2010 06:05 | Mined | (1GvpZ7hwrMqVKbGLfs7gZiSh642mRfKsae) | 50.00 |
| 2/14/2010 05:18 | Mined | (19ePheuntFU6FGsd1Mc2gRJBP1NhLobTFw) | 50.00 |
| 2/13/2010 20:34 | Mined | (1GNekSjZMMyWfcQCijXexXNFfmGdrbbAhD) | 50.00 |
| 2/13/2010 19:28 | Mined | (1GQ8ALq2x3DszEBPhA5ky1MMj2HRoe7DMD) | 50.00 |
| 2/13/2010 14:31 | Mined | (1QBHnS6wDfMbw1LXu8CDw6uEQTi7F19AAc) | 50.00 |
| 2/13/2010 12:50 | Mined | (1DAVBCzbbm9dBTQgJ8xkdGoLUSP7yn6Pt1) | 50.00 |
| 2/13/2010 10:14 | Mined | (1Ar9TkY4Rt8x1bBDs7rHuHYo4ypS88JxjY) | 50.00 |
| 2/13/2010 02:23 | Mined | (1PWL3D4BcTAhnf73vc9B9vyYFgKpNzGpLn) | 50.00 |
| 2/13/2010 01:51 | Mined | (1PNEjPmmWZTVML1uB3FmFvbzpzTwv7RCcA) | 50.00 |
| 2/13/2010 01:17 | Mined | (17GyhbDFi7RQMPfxR3h1VnPa6vFiVHVhK1) | 50.00 |
| 2/12/2010 22:24 | Mined | (112pJXp3VvjS9ECMKFQtHSgbxi79wchDzK) | 50.00 |
| 2/11/2010 17:21 | Mined | (1FcEwmLSRNZ57PTeZYianPM6pWy5eRZGFJ) | 50.00 |
| 2/10/2010 17:25 | Mined | (1MQj9gL5xR1wRRSMXkHS1x1oThtaqZnFQ) | 50.00 |
| 2/9/2010 20:12 | Mined | (142bpcSfcbYTwsoNf1MUCbeBaijpJQUs1J) | 50.00 |
| 2/9/2010 12:32 | Mined | (1HxQavcwsYnLbntANzj1V1yC5ckxZiRELg) | 50.44 |
| 2/8/2010 21:00 | Mined | (1Cu42YDpoZRdmfipnSKVJokSF7h7QKgKfF) | 50.00 |
| 2/8/2010 19:28 | Mined | (19qLedpeHpK3oncugU3BdzVKbJNEfR7Jot) | 50.00 |
| 2/8/2010 12:38 | Mined | (1DNvckqZmnfSWkKz4AEHncNtHCqaQv4P8Y) | 50.00 |
| 2/8/2010 01:14 | Mined | (1N5TKZ82GqZ9e1jwokEvF5wt726Er9DhXM) | 50.00 |
| 2/7/2010 23:51 | Mined | (1GB4NvJsHo32QC9bmj8a7a5bR9JeK6Tijd) | 50.00 |
| 2/7/2010 19:58 | Mined | (13f3nyvfo3pm3E6EucUJi2EHKrFrGLeJQW) | 50.00 |
| 2/7/2010 18:02 | Mined | (19QSH7vr8ysD495ztPXnPsdaoXUizbKZve) | 50.00 |
| 2/7/2010 15:48 | Mined | (1BKAXj2WR56zNXRLZeVzG4xQBnqkCAW34v) | 50.00 |
| 2/7/2010 12:30 | Mined | (1EHBjax1YoCBiVTq1XusVQtGZXcBtC67u4) | 50.00 |
| 2/7/2010 01:08 | Mined | (195zWsxn9cVESmSkpB3fv5pRYN5hNkanqa) | 50.00 |
| 2/6/2010 21:18 | Mined | (1N5FJKJg9TJr36vccWNPcjgh9NbX19J7Zp) | 50.00 |
| 2/6/2010 20:40 | Mined | (19BFPLqRvevYcgkRzCCRo5HW14CcZMYXmL) | 50.00 |

**theymos**
Administrator
Legendary

Activity: 4060
Merit: 8158

**Re: Satoshi's Fortune lower bound is 100M USD** #25
April 14, 2013, 04:19:19 AM

**Quote from: Sergio_Demian_Lerner on April 14, 2013, 03:53:30 AM**
> So I can tell you with confidence that you mined very few blocks (e.g. 10 blocks) during that time and you're not millonaire, or you are part of the Satoshi group, period.

Do I get to be Satoshi too? I was off by only a few days... https://i.imgur.com/w57rtbs.png

I know first-hand that there were several different people who mined before January 2010. It's kind of funny that history I've lived through is being questioned...

1NXYoJ5xU91Jp83XfVMHwwTUyZFK64BoAD

**Fig S12. A publicly posted screenshot of a user's mining rewards.**
On top, this user posted a screenshot of their mining rewards to bitcointalk.org. We tag all of the addresses shown in this screenshot as belonging to this user. On bottom, the original message posted to the forum, linking the screenshot. This information is publicly accessible here: https://bitcointalk.org/index.php?topic=175996.20

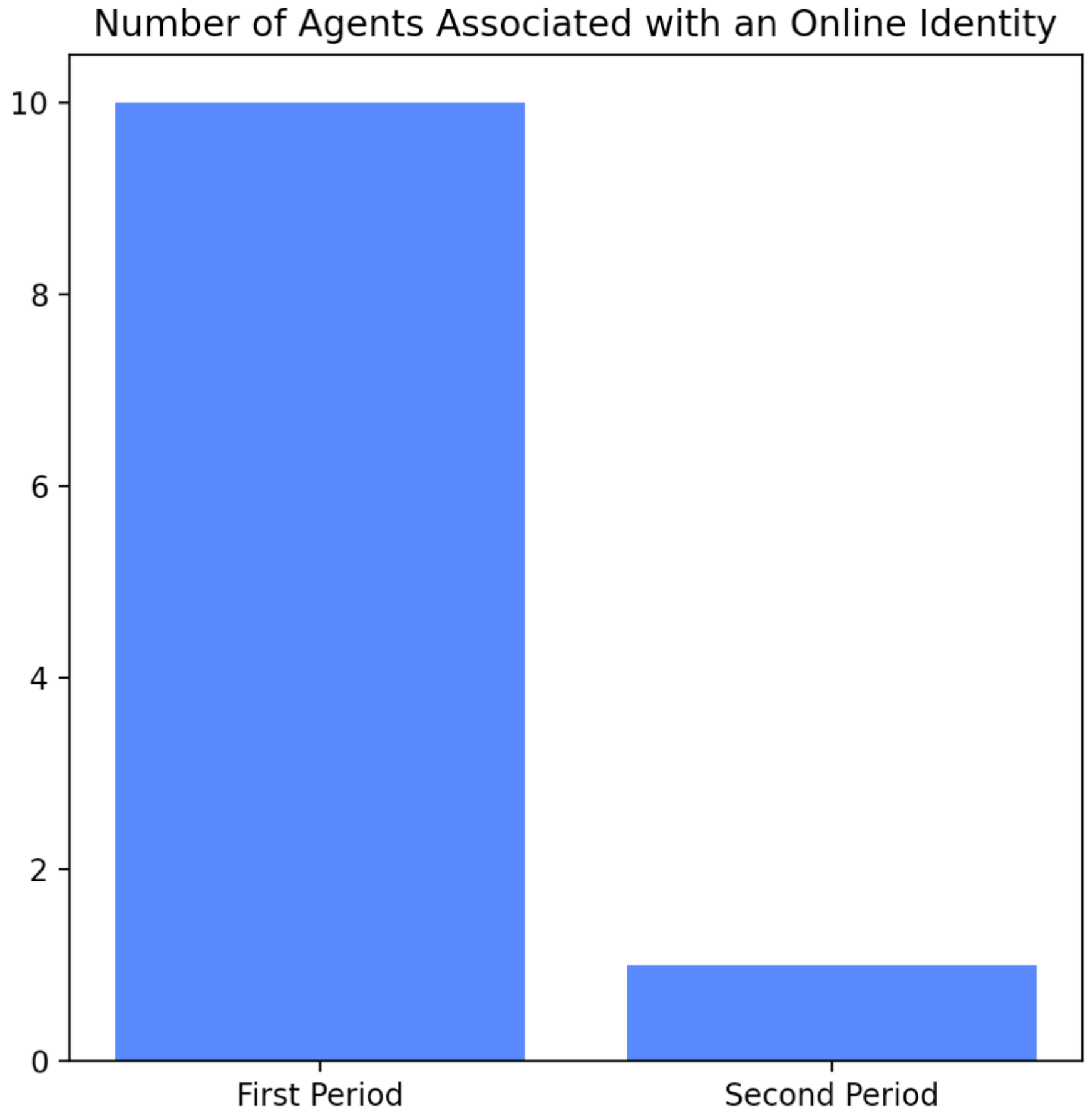

**Fig S13. In the first time period, large agents were more likely to be associated with an online identity.**
For both time periods discussed in the main text, we ranked the top 100 largest agents who began mining in both time periods and attempted to associated them with our tags associated with publicly known online entities. We were able to tag 10 agents in the first period (Intervals 1-4) and 1 agent in the second period (Intervals 5-6). This plot includes Satoshi.

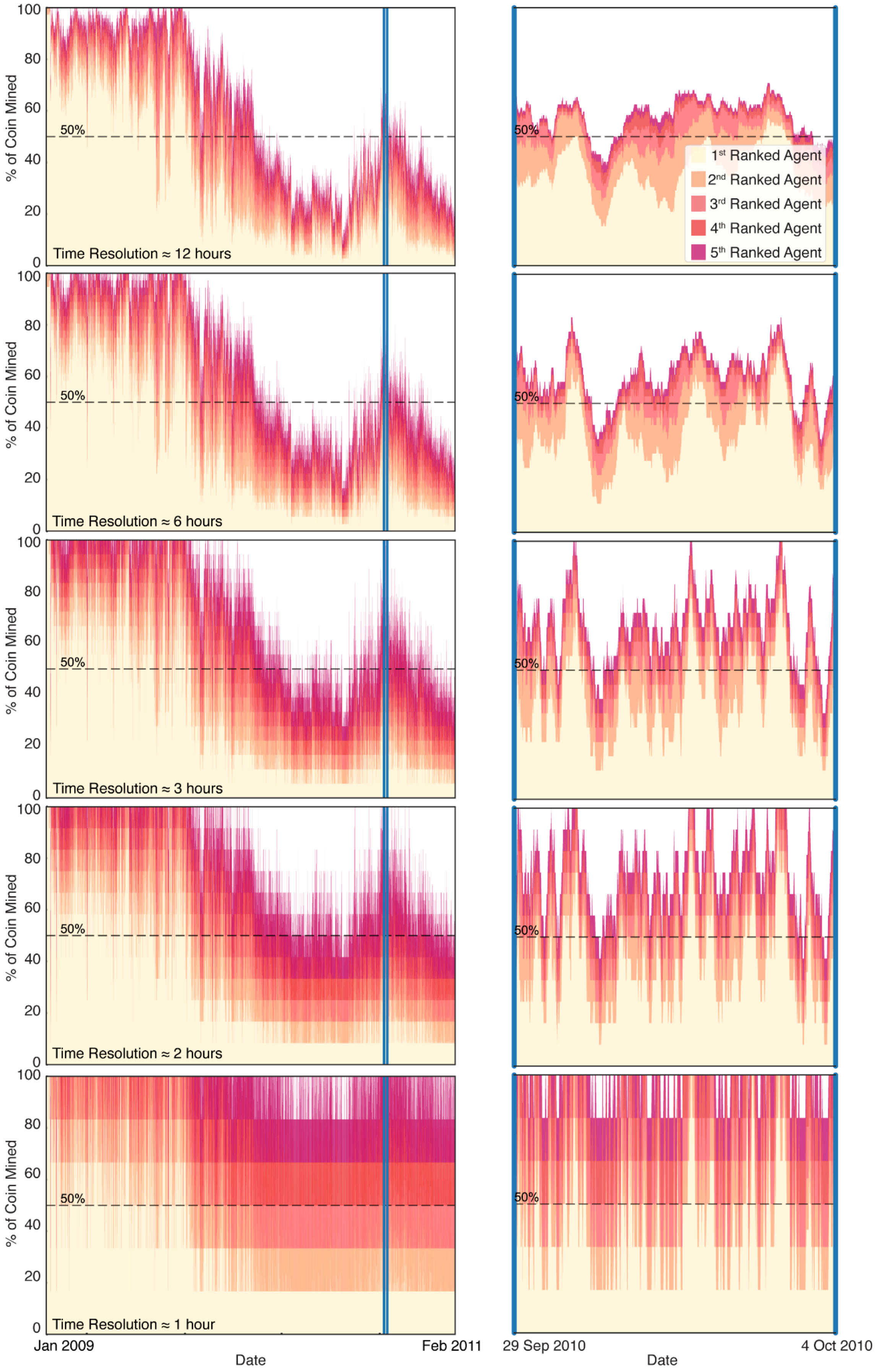

100
80
60
40
20
0
% of Coin Mined
50%
Time Resolution ≈ 12 hours
Time Resolution ≈ 6 hours
Time Resolution ≈ 3 hours
Time Resolution ≈ 2 hours
Time Resolution ≈ 1 hour
1st Ranked Agent
2nd Ranked Agent
3rd Ranked Agent
4th Ranked Agent
5th Ranked Agent
Jan 2009
Feb 2011
Date
29 Sep 2010
4 Oct 2010
Date

**Fig S14. The likelihood of having a dominant agent increases when shorter time-windows are examined.** In a variant of Figure 5, we here show the 5 agents in the bitcoin network employing the most computational power at a given time, and the fraction of computational power held by each. *Left:* Each datapoint in the top row reflects an interval of 72 blocks, or roughly 12 hours. Note that the specific agent with a particular rank may not be the same in different time intervals. Subsequent rows are identical, but the datapoints reflect narrower time windows: 6 hours (36 blocks), 3 hours (18 blocks), 2 hours (12 blocks) and 1 hour (6 blocks). As the time resolution of the data points grows finer, the plot becomes increasingly jagged. This reflects the fact that, even if there is no dominant agent over a long interval, such as 12 hours, there are often dominant agents during the shorter subintervals. (B) A weeklong period from September 29, 2010 and October 4, 2010, in which the agent with the most computational power (no longer Satoshi Nakamoto, but instead Agent #2) has enough resources to perform a 51% attack during several 6+ hour long windows. Again, the blockchain is more decentralized when longer intervals are examined than when shorter intervals are examined. We find that, despite the fact that 64 agents mined a majority of bitcoin during the two-year period we studied, differences in computational resources among the agents over time lead to a scenario in which, for short time intervals, the blockchain can be extremely centralized.

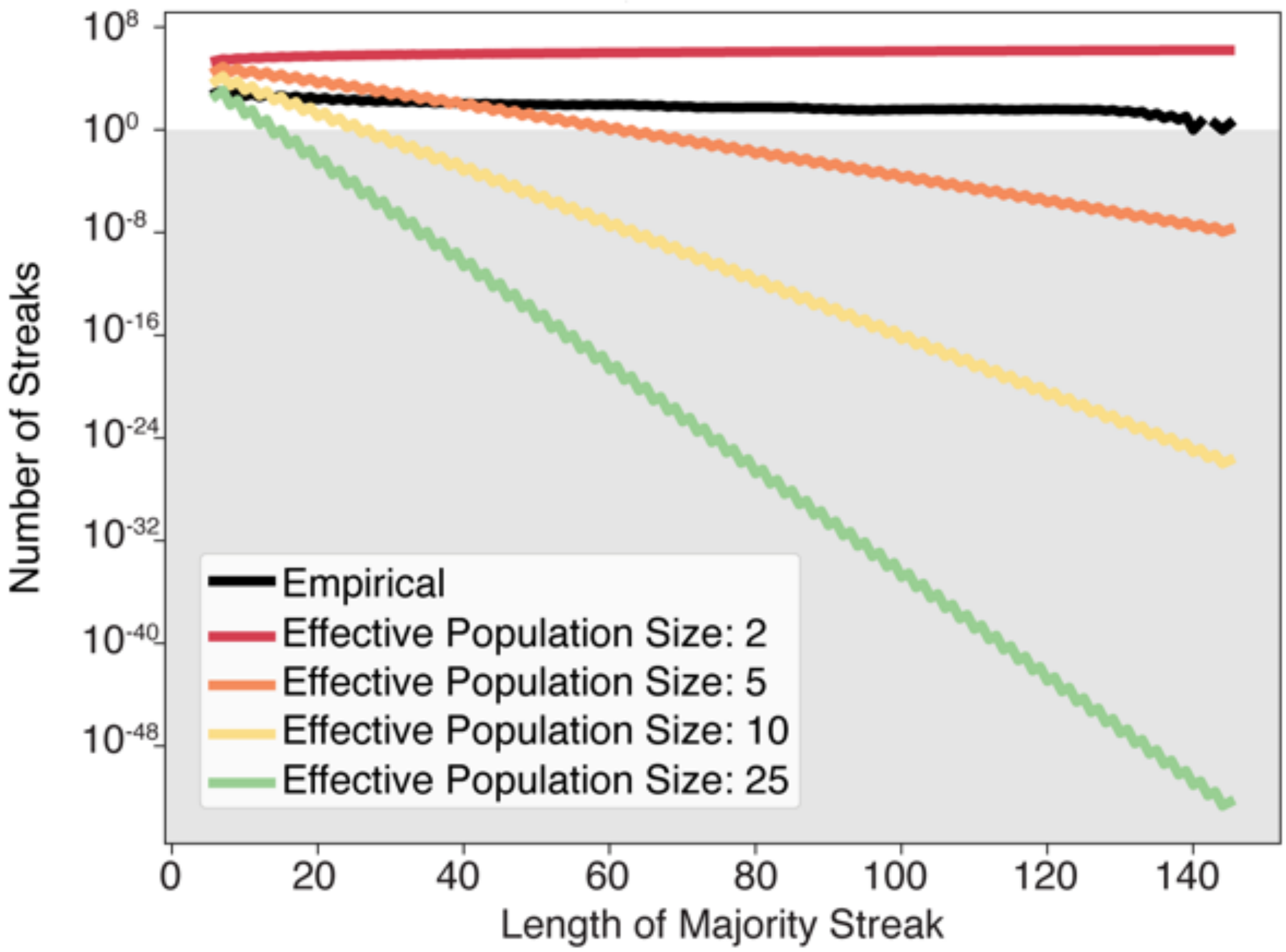

**Fig S15. We estimate the effective population size of the decentralized bitcoin network by counting the frequency of streaks in which most blocks are mined by a single agent.** These are compared to the expected values for idealized networks comprising $P$ agents with identical resources. The comparisons suggest an effective population size of between two and five. The grayed-out region corresponds to an expected value of less than one streak.

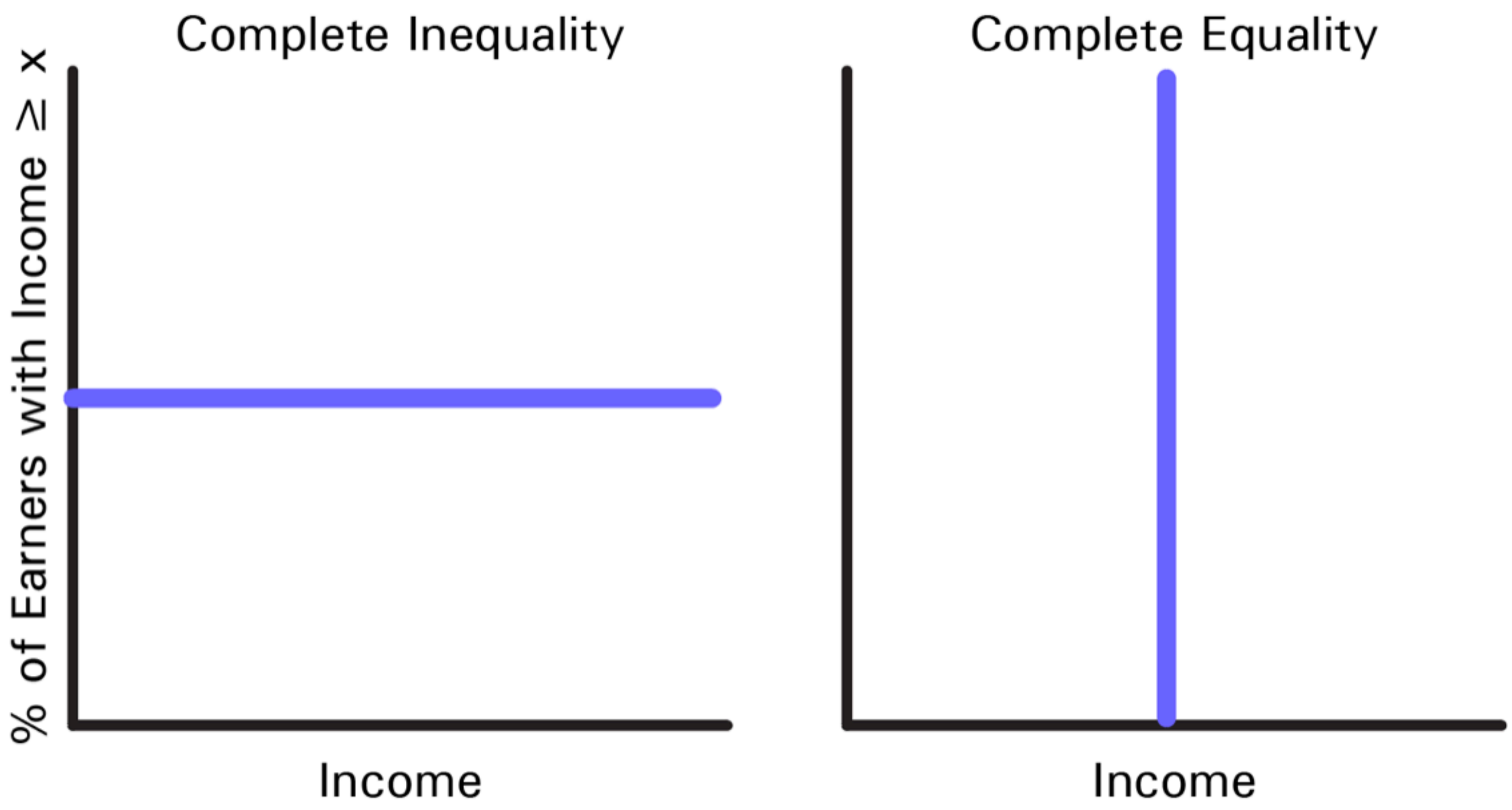

**Fig S16. The theoretical limits of the Pareto distribution as the Pareto exponent approaches -1 and infinity.**

On the left, one individual earns 100% of all income. The slope of the fit is 0 and the Pareto exponent is -1. On the right, every individual earns exactly the same income. The slope of the line is undefined and the Pareto exponent is infinite. The integral of both of these is infinite, implying an infinite money supply, and is clearly not feasible in a real life scenario.

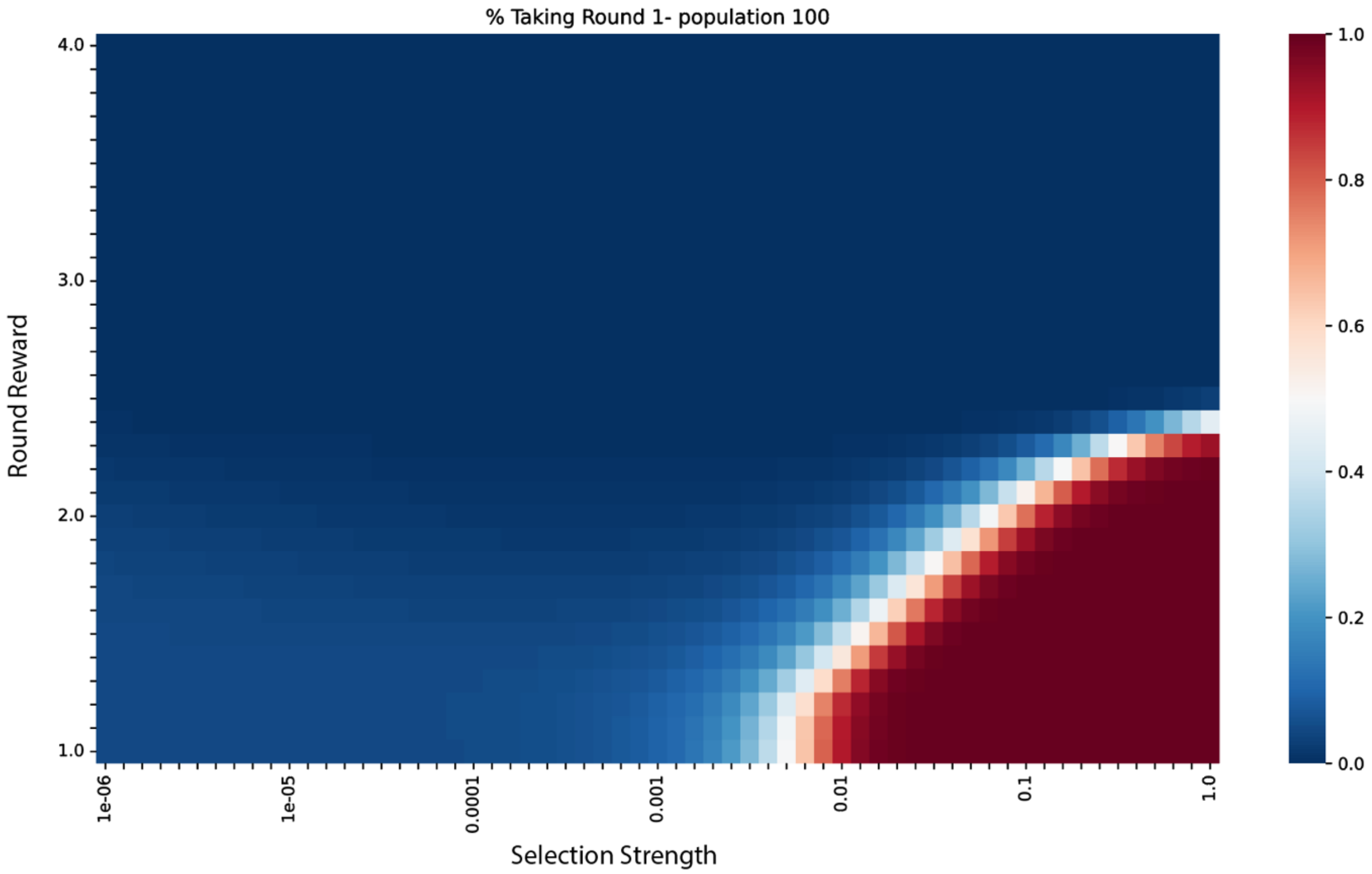

**Fig S17. Theoretical likelihood of passing on round 1 in 8-player Centipede game (n=100 population).** The fraction of players who defect on the first round in our evolutionary model. Again, the subgame-perfect Nash equilibrium is rare except when selection is strong and the benefits of cooperation are weaker.

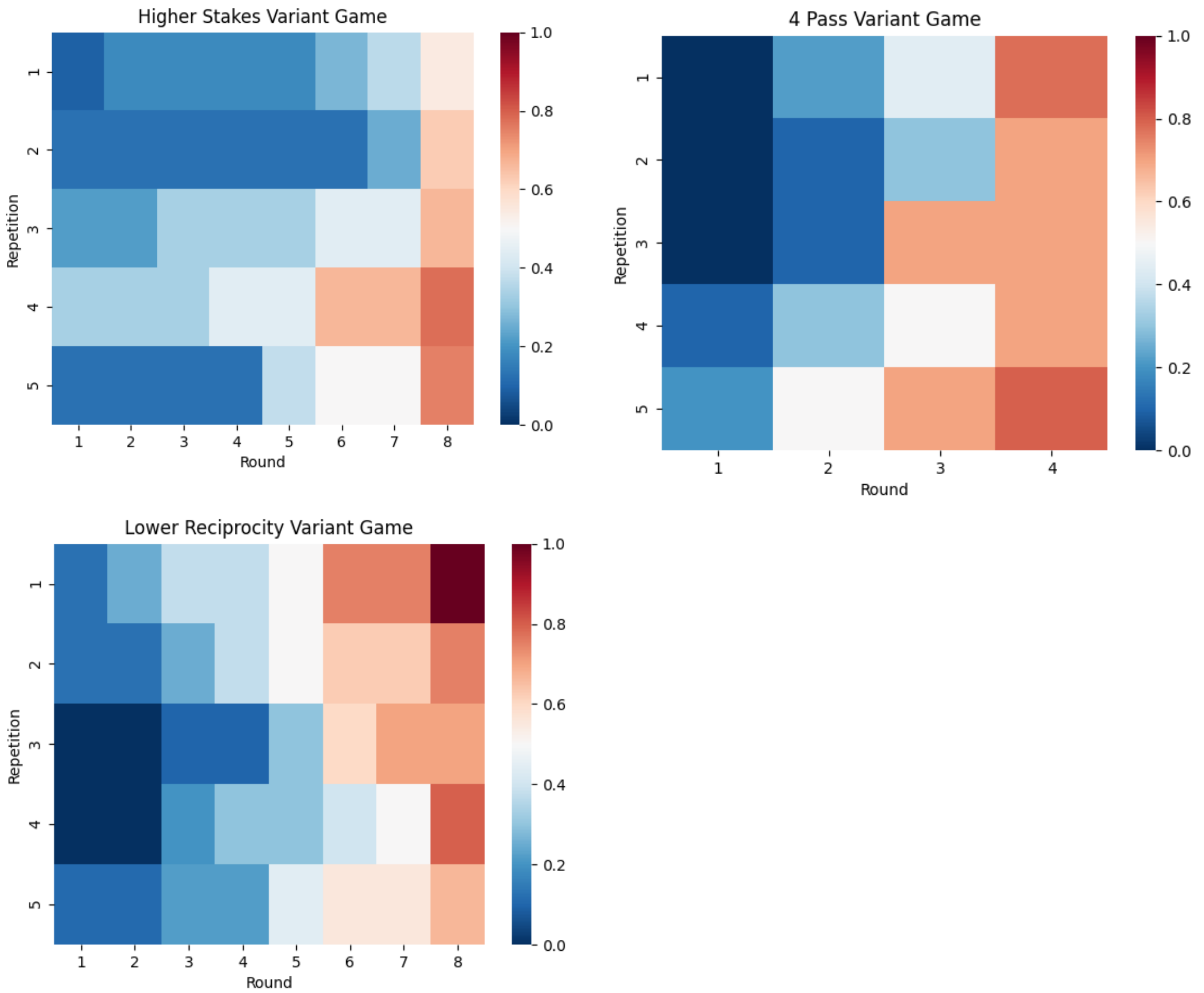

**Fig S18. Empirical likelihood that variant games stop before the N$^{th}$ pass.**
As in the 8-player Centipede game, game theory predicts that everyone should defect in the first round.

| 8-player Centipede Game | | | | | | |
|---|---|---|---|---|---|---|
| | Repetition | | | | | |
| | 1 | 2 | 3 | 4 | 5 | Total |
| Mean Number of Passes | 5.0223 | 5.5130 | 5.6574 | 5.6873 | 5.7099 | 5.5118 |
| Actual Passes/Possible Passes | 62.8% | 68.9% | 70.7% | 71.1% | 71.4% | 68.9% |
| Mean Length of Game | 5.7390 | 6.1570 | 6.3110 | 6.3040 | 6.3880 | 6.1730 |
| Actual Length / Possible Length | 71.7% | 77.0% | 78.9% | 78.8% | 79.9% | 77.2% |

| 2-player Centipede Game | | | | | | |
|---|---|---|---|---|---|---|
| | Repetition | | | | | |
| | 1 | 2 | 3 | 4 | 5 | Total |
| Mean Number of Passes | 4.2326 | 4.2683 | 4.9286 | 4.8810 | 4.6585 | 4.5284 |
| Actual Passes/Possible Passes | 52.9% | 53.4% | 61.6% | 61.0% | 58.2% | 56.6% |
| Mean Length of Game | 5.0465 | 5.0488 | 5.7143 | 5.6429 | 5.5122 | 5.3923 |
| Actual Length / Possible Length | 63.1% | 63.1% | 71.4% | 70.5% | 68.9% | 67.4% |

**Fig S19. Comparison Between 2- and 8- player Centipede games**

Across every repetition, participants in the 2-player game exhibited less passing (cooperation) and more taking (defection). Moreover, participants in the 2-player centipede game did not exhibit the same "learning to cooperate" trend as seen in the 8-player game.

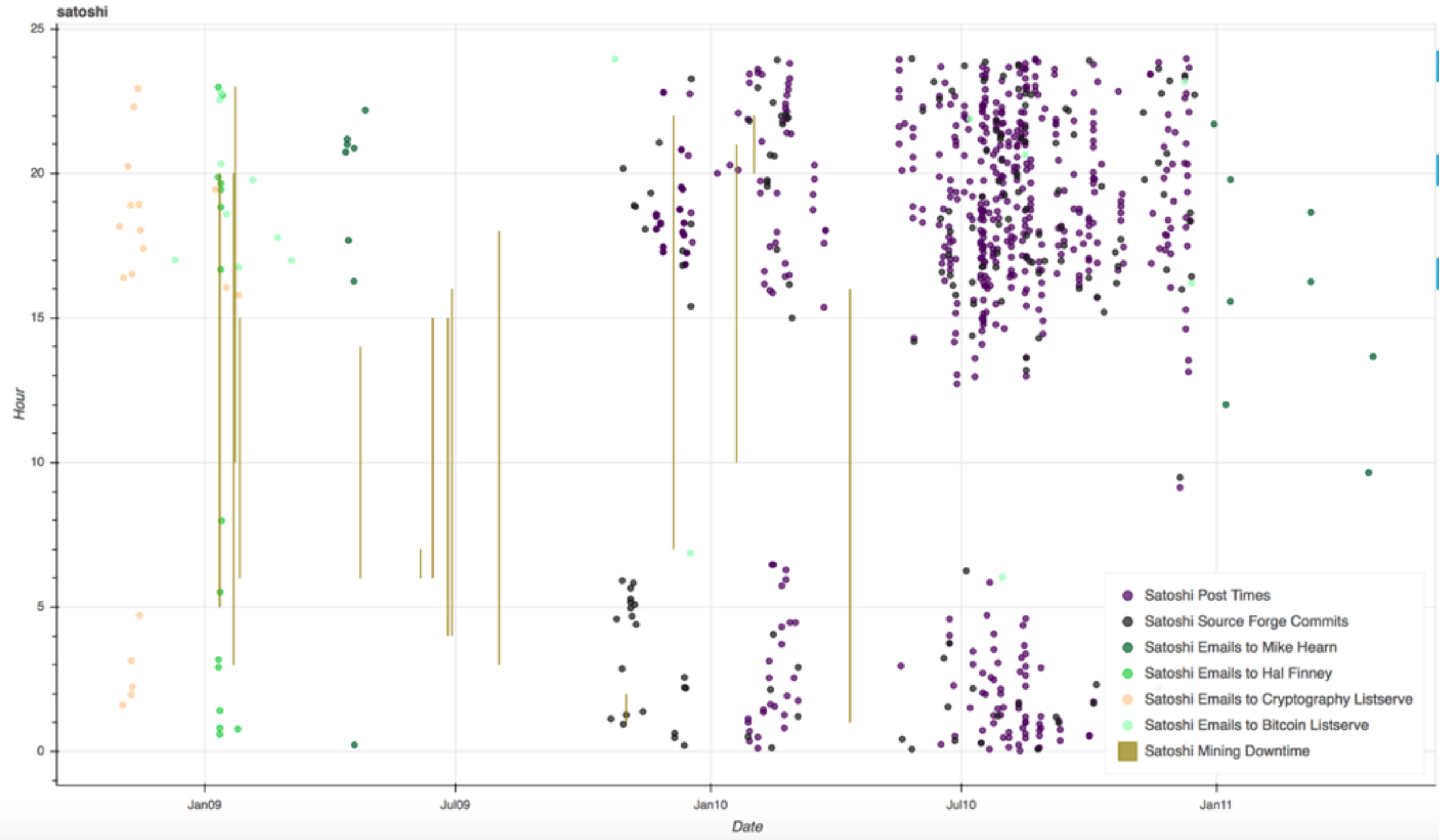

**Fig S20. Activity of Satoshi Nakamoto by date and hour.** This plot shows the date and time of known activity by Satoshi Nakamoto, including the UTC (Universal Time Coordinated) timestamp of emails sent to Mike Hearn (dark green); to Hal Finney (bright green); to the Cryptography listserv (tan) where bitcoin was first announced; and to the Bitcoin listserv created shortly thereafter (neon cucumber). It also includes the timestamp of posts by Satoshi to bitcointalk.org (purple) and of Sourceforge code commits (grey). All of these activities suggest that Satoshi is awake. We also plot prolonged periods of time in which the Satoshi miner appears to be inactive. This is consistent with the possibility that Satoshi is asleep, and cannot address the outage. Taken together, these datasets suggest that Satoshi is typically inactive (and perhaps asleep) for 8 hours between 6 and 14 UTC, which corresponds to 10PM – 6AM Pacific Standard Time, or 1AM – 9AM Eastern Standard Time. These data are in line with the possibility that Satoshi Nakamoto was living in either North or South America. Of course, we cannot rule out the possibility that Satoshi Nakamoto tended to sleep at anomalous times with respect to his time zone.

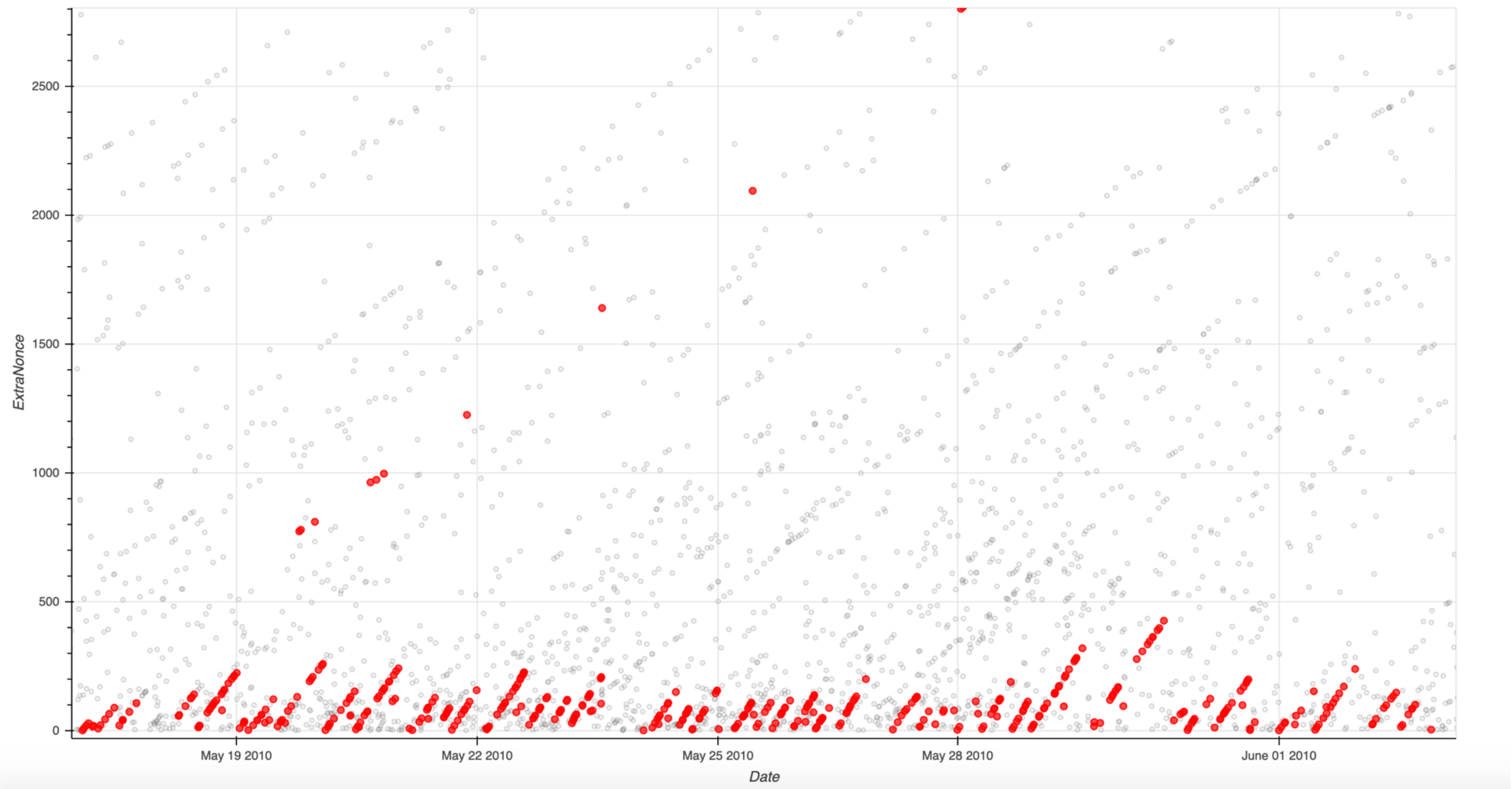

**Fig S21. Agent #6 exhibits mining trajectories consistent with GPU mining.**
Agent #6 is the first agent to exhibit short dense lines with a steeper slope and frequent restarts, consistent with mining on GPUs vs CPUs. The increased slope is formed by the faster hashing speed of GPUs, which exhausts the nonce field faster, causing the extranonce to increment more frequently.

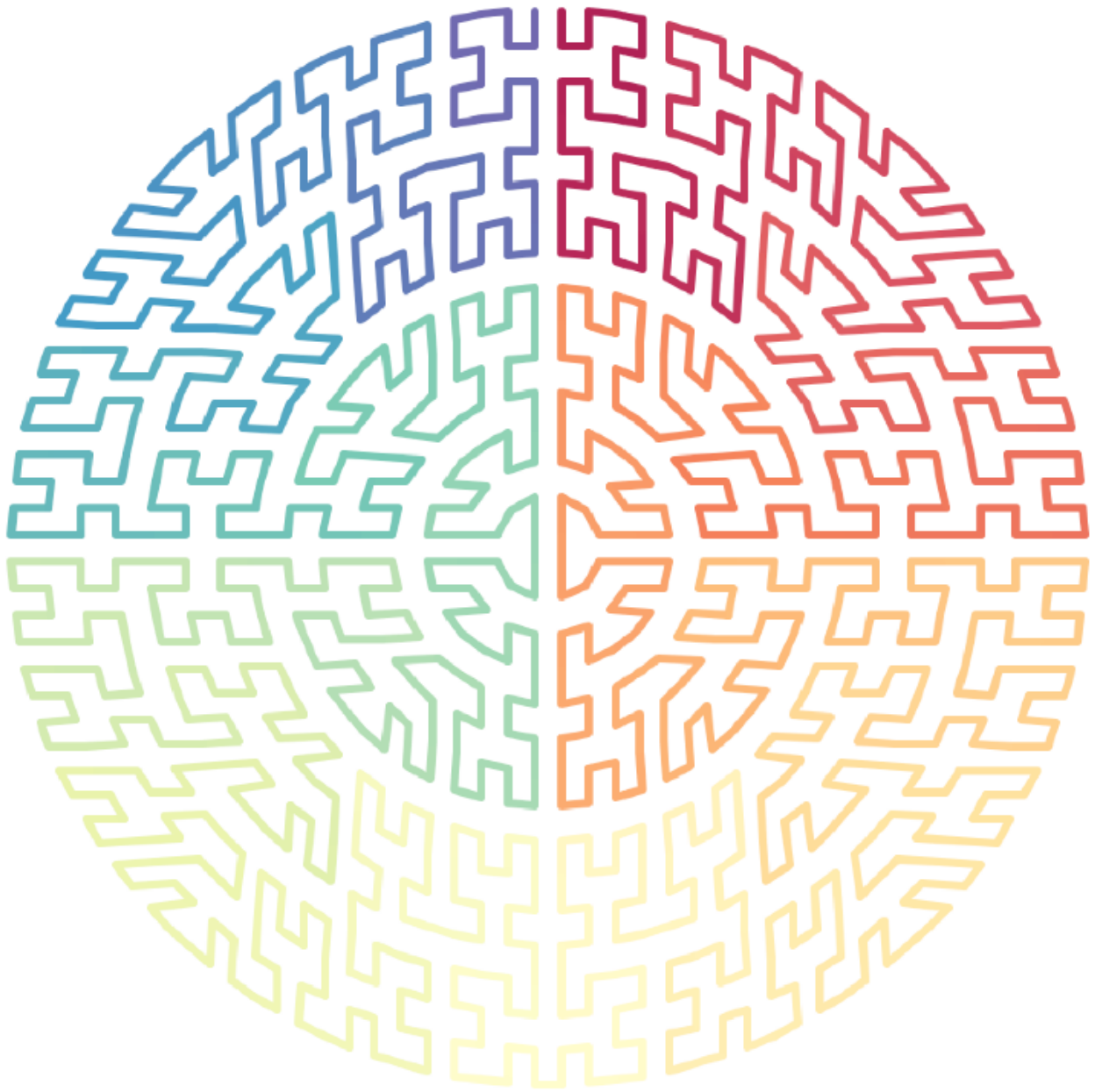

**Fig S22. We used the Shirley-Chiu transformation of the Moore curve to lay out the blocks and generate Figure 1.** In Figure 1, we sought to lay out the blocks mined by each agent to satisfy several desiderata. First, that the blocks mined by each agent form a contiguous region. Second, that the size of each region be proportional to the number of blocks. Finally, that the position of each region along the outer perimeter of the circle correspond roughly to when the corresponding agent joined the bitcoin community. This was achieved by combining the equiareal Shirley-Chiu transformation with the space-filling Moore curve. The Moore space-filling curve, introduced by E.H. Moore in 1900, used iterative application of an underlying motif to construct a curve that, in the infinite limit, densely fills the unit square. This was a modification of earlier work on such curves by Giuseppe Peano and David Hilbert. The total area covered by a line segment in a finite iteration of the Moore curve is proportional to the length of the segment. We then used the Shirley-Chiu transformation to map the unit square to the unit disc in an area-preserving fashion. The figure above shows the results of applying an equiareal Shirley-Chiu transformation of a finite iteration of the Moore Curve. Here, color indicates 1d position along the curve, from the beginning (red) till the end (blue). We used such curves to help generate Figure 1, as follows. First, agents are sorted by when they first mined a block. Blocks are first sorted by agent, and then by block height. Blocks are colored by agent. We then apply a Moore curve transformation to the 1D block color array, mapping it to the unit square. Finally, we apply the Shirley-Chiu transformation to map the resulting regions to the disc. The resulting image of mining activity, shown in Figure 1, achieves all of the desiderata: consecutive blocks mined by the same agent are adjacent in the plot; the area occupied by an agent is linearly proportional to the number of blocks mined by that agent; and position along the perimeter corresponds roughly to when an agent began mining relative to other agents. The exact point in time that an agent began can be estimated more precisely by comparison to the above figure.

Bitcoin

# Bitcoin P2P Cryptocurrency

## Bitcoin P2P Cryptocurrency

Bitcoin is a peer-to-peer network based anonymous digital currency. Peer-to-peer (P2P) means that there is no central authority to issue new money or keep track of transactions. Instead, these tasks are managed collectively by the nodes of the network. Anonymity means that the real world identity of the parties of a transaction can be kept hidden from the public or even from the parties themselves. Advantages:

- Transfer money easily through the internet, without having to trust third parties.
- Third parties can't prevent or control your transactions.
- Be safe from the unstability caused by fractional reserve banking and bad policies of central banks. The limited inflation of the Bitcoin system's money supply is distributed evenly (by CPU power) throughout the network, not monopolized by the banks.

Bitcoin is an open source project created by Satoshi Nakamoto, and is currently in beta development stage. Get the latest version of Bitcoin at the Sourceforge project page.

Screenshots:

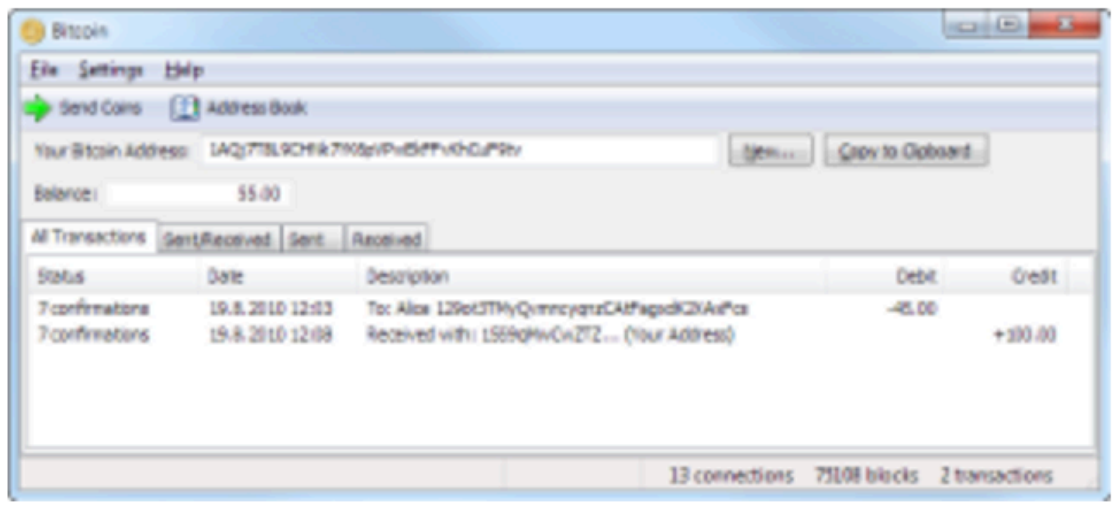

**Fig S23. A screenshot of bitcoin.org on January 6, 2010.**
Early materials written by Satoshi advertised bitcoin as being anonymous, free from third party control, and decentralized. This can be viewed courtesy of the Internet Archive's Wayback Machine: https://web.archive.org/web/20100106082749/http://www.bitcoin.org:80/ (*93*).